\documentclass[twocolumn]{aastex7}
\usepackage{amsmath,mathtools,mathrsfs,cleveref,graphicx,float}
\usepackage[caption=false]{subfig}
\usepackage{threeparttable}
\usepackage[T1]{fontenc}
\usepackage{makecell}
\usepackage{multirow}
\usepackage{hyperref}
\usepackage{color,soul}
\usepackage{nicefrac}
\usepackage{xspace}
\usepackage{enumitem}
\usepackage{tabularx}
\usepackage{url}
\usepackage{epstopdf}
\usepackage{gensymb}
\usepackage{multirow}
\usepackage{ragged2e}
\usepackage{hyperref}
\usepackage{url}
\usepackage{booktabs}
\usepackage{caption}
\usepackage{comment}
\usepackage{afterpage}
\usepackage{mathtools}
\usepackage{bold-extra}
\usepackage{relsize}
\usepackage{colordvi}
\usepackage{braket}

\hypersetup{
  colorlinks=true,        
  linkcolor=blue,         
  citecolor=magenta
}

\definecolor{forestgreen}{RGB}{34,180,34}

\newcommand{\sgra}{Sgr~A$^*$\xspace}

\newcommand{\bhspin}{a_*\xspace}

\newcommand{\kharma}{kharma\xspace}
\newcommand{\ipole}{\texttt{ipole}\xspace}
\newcommand{\charm}{\texttt{CHARM}\xspace}

\defcitealias{eht_sgra_6}{Paper~VI}

\begin{document}

\nocite{medeiros_2023_primo1}
\nocite{medeiros_2023_PRIMO_M87}

\title{Prospects for Improving the Theoretical Uncertainty for Tests of General Relativity with the EHT}

\author[0000-0003-2342-6728]{Lia~Medeiros}
\email{lia2@uwm.edu}
\altaffiliation{These authors contributed equally to this work.}
\affiliation{Center for Gravitation, Cosmology and Astrophysics, Department of Physics, University of Wisconsin–Milwaukee, P.O. Box 413, Milwaukee, WI 53201, USA}
\affiliation{Department of Astrophysical Sciences, Peyton Hall, Princeton University, Princeton, NJ, 08544, USA}
\author[0000-0001-6952-2147]{George~N.~Wong}
\email{gnwong@ias.edu}
\altaffiliation{These authors contributed equally to this work.}
\affiliation{Princeton Gravity Initiative, Princeton University, Princeton, NJ 08544, USA}
\affiliation{School of Natural Sciences, Institute for Advanced Study, 1 Einstein Drive, Princeton, NJ 08540, USA}
\author[0000-0003-4413-1523]{Feryal~\"{O}zel}
\email{feryal.ozel@gatech.edu}
\affiliation{School of Physics, Georgia Institute of Technology, 837 State St NW, Atlanta, GA 30332, USA}

\begin{abstract}

We characterize the relationship between the size of the bright ring observed in simulated black hole images and the size of the analytic black hole shadow. Calibrating this relationship is crucial for mass measurements and, when independent mass measurements are available, for tests of general relativity using Event Horizon Telescope (EHT) images. We perform this calibration using a large set of high-resolution simulated images generated with different accretion-flow modeling approaches and spanning a wide range of system parameters and initial conditions. We show that the theoretical uncertainty in this relationship can be reduced significantly through future observations, improved imaging techniques, and the application of astrophysical or model-based constraints. In particular, the uncertainty decreases compared to existing measurements when \textit{(i)} observing at 345 GHz, \textit{(ii)} applying geometric image constraints, such as the ring width inference from the PRIMO image reconstruction algorithm, \textit{(iii)} incorporating astrophysical constraints such as the black hole spin axis in M87 being aligned (or anti-aligned) with the large-scale jet observed at longer radio wavelengths, and \textit{(iv)} assuming that the accretion flow can be described by a magnetically arrested field configuration. Finally, we quantify how the uncertainty is expected to decrease as additional observations are obtained in subsequent years and identify dwell-time filtering, i.e., evaluating the persistence of a geometric measurement over time, as a promising avenue for improving the precision of the calibration.
\end{abstract}
 
\section{Introduction}\label{sec:intro}

The Event Horizon Telescope (EHT) is a very long baseline interferometer capable of producing resolved radio images of supermassive black holes on event horizon scales. The EHT has produced images of two nearby supermassive black holes (the one at the center of M87, \citealt{eht_m87_1,eht_m87_2,eht_m87_3,eht_m87_4,eht_m87_5,eht_m87_6,eht_m87_7,eht_m87_8,eht_m87_9,eht_m87_2018_1,EHT_m87_2018_2,eht_m87_2025}, and Sagittarius~A$^*$, \sgra, at the center of the Milky Way \citealt{eht_sgra_1,eht_sgra_2,eht_sgra_3,eht_sgra_4,eht_sgra_5,eht_sgra_6,eht_sgra_7,eht_sgra_8}). Although the EHT has only released 1.3~mm (230 GHz) images, an initial set of observations at 0.8~mm (345 GHz) have already been performed with a subset of the global array, with the first fringe detections reported recently \citep{raymond_2024_eht345}. At these radio frequencies, black hole images are dominated by a bright ring of emission surrounding a central brightness depression. The diameter of the ring has been used to infer the masses of the black holes and, when combined with independent mass measurements, it can be used to place constraints on deviations from the Kerr metric. 

Measuring the black hole mass or testing the Kerr hypothesis relies on precisely inferring the geometry of the black hole shadow \citep{johannsen_2010_nohairimages,psaltis_2015_grtesteht}, the boundary of which is defined by the critical impact parameters between photon trajectories that fall into the black hole and those that escape as seen by an observer at infinity. Although the Kerr shadow size and shape can be calculated analytically and are not influenced by astrophysical effects, the geometry of the observed ring feature depends on the details of the accretion flow. Obtaining a robust mass measurement or test of the Kerr metric, therefore, requires quantitatively calibrating the relationship between the observed ring and the geometry of the shadow.

\citealt{eht_sgra_6} (hereafter \citetalias{eht_sgra_6}) introduced a framework that separates the calibration (or measurement error) of shadow-size inference into theoretical and observational components. The theoretical uncertainty assumes perfect data and image reconstruction and quantifies the offset between the analytic black hole shadow and the geometric extent of the emission ring in high-resolution images, which depends on the astrophysical properties of the flow. The observational uncertainty captures the additional limitations imposed by sparse interferometric coverage, atmospheric and instrumental effects, and image reconstruction choices. In this work, we focus exclusively on the theoretical uncertainty; the observational uncertainty can be characterized by simulating the full observing pipeline (as done in \citetalias{eht_sgra_6}).

The calibration between the analytically defined shadow and the observed emission ring is uncertain for two distinct reasons. First, the ring diameter fluctuates stochastically within any fixed accretion-flow model as a result of turbulent variability and a single image provides only one realization drawn from a broader population of possible morphologies. As additional, effectively independent epochs are accumulated, this \emph{intramodel variability} can be reduced and the inferred population-averaged diameter becomes more precise. Second, the population-averaged diameter itself depends on uncertain aspects of the accretion physics and viewing geometry, leading to systematic differences between otherwise plausible models. This \emph{intermodel spread} does not, in general, average away with more observations, and instead must be addressed by constraining the allowed model space using complementary information.

\begin{figure*}
    \includegraphics[trim={.8cm 0 0 0},clip,width=1\textwidth]{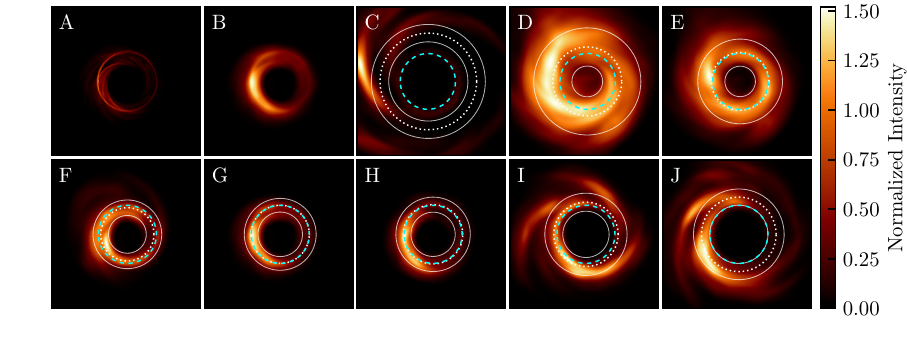}
    \caption{Example snapshots from the simulation libraries superimposed with the outputs from the \charm feature extraction algorithm. The analytically calculated black hole shadow boundary is denoted by a dashed cyan curve and the median ring diameter from \charm is denoted by a white dotted circle. 
    The median ring width from \charm is represented by two solid white circles (shown assuming that the width is symmetric about the median diameter). The particular snapshots show 
    (A) a MAD snapshot from Library B at 230~GHz at the native resolution ($\bhspin=0.5$, $R_{\mathrm{low}}=10$, $R_{\mathrm{high}}=10$), 
    (B) the same snapshot as A but blurred with a Butterworth filter with parameters $n=2$ and $r=15~\mathrm{G}\lambda$, as was done in \citetalias{eht_sgra_6} (see also \citealt{psaltis_2020_ehtfilter,medeiros_2023_primo1}, all subsequent snapshots also blurred), 
    (C) a retrograde SANE ($\bhspin=-0.9375$, $R_{\mathrm{low}}=1$, $R_{\mathrm{high}}=1$) simulation from Library B at 230~GHz ($R_{\mathrm{high}}=1$ simulations are excluded from the analysis in the main text due to their unphysically high electron temperatures, see Appendix~\ref{ap:Rhigh}),
    (D) a snapshot with high accretion rate from Library A at 230~GHz (MAD, $\bhspin=0.9$, $R_{\mathrm{high}}=20$, $n_{e} = 10^6$), 
    (E) the same GRMHD snapshot as D, but imaged at 345~GHz. Images F to J show randomly selected snapshots from Library B that span the range of fractional diameter difference values across the model ensemble. These snapshots are chosen to sample approximately the lower and upper bounds of the central $68\%$ and $95\%$ intervals as well as the median, ranging from smallest to largest in order. Snapshots B and G are the same so the reader can compare an image with and without the \charm results. \charm is able to accurately measure the ring width in most snapshots (c.f., snapshot C, see text discussion and Appendix~\ref{ap:Rhigh}). 
    }
    \label{fig:charm}
\end{figure*}

We investigate how future EHT observations and analysis strategies can reduce these two sources of theoretical uncertainty in shadow size measurements, e.g., by including observations at higher frequencies, applying astrophysical or geometric image-based constraints, and combining multiple epochs of data. In this paper, we focus on simulated images of black hole accretion in the Kerr spacetime. We use the feature extraction algorithm \charm (\citealt{eht_m87_6,ozel_2022_bhgrtests}; \citetalias{eht_sgra_6}) to measure the ring diameter in the simulated images. We use these measurements to quantify the expected theoretical uncertainty in analysis of simultaneous EHT observations at 345~GHz and 230~GHz. We further explore how this uncertainty depends on model parameters and investigate how multiple observing epochs may affect the theoretical uncertainty. We also consider how the theoretical error might be improved by applying observational image constraints such as the ring-width constraint from the \texttt{PRIMO} algorithm \citep{medeiros_2023_PRIMO_M87}.

The paper is organized as follows. In Section~\ref{sec:methods}, we describe how we characterize the theoretical uncertainty in the estimated ring diameter and width. We describe our simulation libraries in Section~\ref{sec:sims}. In Section~\ref{sec:multi-wave}, we explore how the theoretical uncertainty depends on observational frequency. We explore how constraints on the model parameter space influence the theoretical uncertainty in Section~\ref{sec:model_constraints}. Section~\ref{sec:epochs} considers how the error may be affected by combining multiple observing epochs. We summarize our work and discuss a few of its implications in Section~\ref{sec:discussion}.

\section{Methodology}\label{sec:methods}

In this section, we introduce the relevant sources of uncertainty, describe the calibration framework of \citetalias{eht_sgra_6}, and finally detail how we use \charm to perform ring feature extraction on our simulated images.

\subsection{Sources of theoretical uncertainty}

The theoretical uncertainty in the inferred ring diameter can be separated into two categories as discussed earlier. We refer to the stochastic variability within a single accretion flow model that leads to temporal fluctuations in the measured ring diameter as \emph{intramodel variability}. Even for fixed black hole and plasma parameters, individual snapshots represent different realizations of a turbulent flow, producing a distribution of diameters rather than a single value. By contrast, we define the intermodel spread as the systematic differences in the characteristic ring diameter across models that arise from uncertainties in accretion physics. These uncertainties include the black hole spin, magnetic field geometry in the accretion flow, electron thermodynamics, and observer’s inclination. These two sources of uncertainty respond differently to repeated observations: intramodel variability averages down, while intermodel spread generally does not.

In this context, it is useful to distinguish between the \emph{population} distribution of an observable and the \emph{sample} distribution derived from a finite number of observations. The population distribution is defined by the ensemble of measurements that would be obtained from an infinite number of realizations of a single model. Uncertainty arising from intramodel variability can be reduced with repeated observations: a single snapshot represents only one realization of a time-variable flow and, therefore, provides an incomplete estimate of the underlying population properties. As additional, approximately independent measurements are obtained, the sample distribution more closely approximates the population distribution, and statistical estimators like the sample mean converge toward their true values. In practice, this implies that averaging over multiple epochs can reduce the uncertainty in the inferred diameter. However, the degree of improvement depends on both the number of available measurements and the intrinsic width of the underlying distribution. As we will show, strategies such as including higher-frequency observations (which tend to produce narrower distributions) can further improve the precision of the estimated mean diameter.

Unlike intramodel variability, the uncertainty due to intermodel spread does not necessarily decrease with additional data. Repeated observations may refine the estimate of the population-averaged diameter within a given model, but they do not by themselves distinguish between models or identify which is correct. Thus, even with arbitrarily many observations, the uncertainty in the calibration factor linking the ring diameter to the shadow size may remain large when multiple models with distinct population-averaged diameters are consistent with the data. Furthermore, reducing intermodel uncertainty does not necessarily reduce the overall spread in inferred diameters: excluding a narrowly distributed but incorrect model can increase the width of the combined distribution while yielding a more accurate characterization of the remaining uncertainty.

\subsection{Calibration framework}\label{sec:alpha}
We follow the error calibration framework described in \citetalias{eht_sgra_6}. The total uncertainty in the black hole shadow diameter measurement depends on:
\begin{itemize}
\item the unknown black hole spin (the shadows of spinning black holes are never larger than the Schwarzschild shadow; the maximal shadow size difference is $\approx 7.5\%$ for all observer's inclinations),
\item the theoretical uncertainty that results in a difference between the shadow diameter and the observable ring of emission in a high-resolution image,
\item the measurement uncertainty between the inferred ring size in a high-resolution image compared to a realistic observation,
\item and, when the shadow is used to test the spacetime metric, the uncertainties in the mass over distance measurement derived from independent observations also influence the precision of the result.
\end{itemize}

Following equation 1 of \citetalias{eht_sgra_6} we write
\begin{equation}
\hat{d}_m = \frac{\hat{d}_m}{d_m} \frac{d_m}{d_{\mathrm{sh}}} d_{\mathrm{sh}} = \alpha_c \, (1+\delta) \, d_{\mathrm{sh,\,Sch}},
\end{equation}
where $\hat{d}_m$ is the ring diameter measured from the observation, $d_m$ is the ring diameter in a high-resolution image, $d_{\mathrm{sh}}$ is the shadow diameter, $\alpha_c$ is the total calibration factor ($\alpha_c = \frac{\hat{d}_m}{d_m} \frac{d_m}{d_{\mathrm{sh}}} ={\hat{d}_m}/{d_{\mathrm{sh}}}$), $\delta$ is the deviation between the inferred shadow diameter and the diameter for the Schwarzschild black hole shadow ($\delta \equiv d_{\mathrm{sh}}/d_{\mathrm{sh,\,Sch}} -1$), and $d_{\mathrm{sh,\,Sch}}$ is the diameter for a Schwarzschild black hole shadow ($d_{\mathrm{sh,\,Sch}} = 2\sqrt{27}\,M$, where we have set $G=c=1$).

We further decompose $\alpha_c$ into the theoretical uncertainty $\alpha_1 = {d_m}/{d_{\mathrm{sh}}}$ and the observational uncertainty $\alpha_2={\hat{d}_m}/ {d_m}$ such that $\alpha_c = \alpha_1\times\alpha_2$. This decomposition enables us to study the limitations of geometric constraints derived from black hole images without needing to specify a particular observing campaign or instrument configuration. As mentioned in the introduction, the present work focuses solely on the theoretical uncertainty and is agnostic to the details of the instrument and analysis software. Our results therefore serve as a lower-limit on what is theoretically possible from observations regardless of improvements to the instrument. 

In this work, we focus on simulations in the Kerr metric; however, $\alpha_1$ could in principle be different for non-Kerr metrics (see \citealt{younsi_2023_imagegrtest}). To robustly calibrate the relationship between the shadow size and the observed emission ring, particularly for constraining alternative spacetimes, a broader prior that includes non-Kerr models should be considered. \citetalias{eht_sgra_6} showed that both Kerr GRMHD simulations and non-Kerr analytic images resulted in a similar distribution of theoretical uncertainty. We leave a detailed exploration of the effects of deviations from the Kerr metric on the theoretical uncertainty to future work.

\subsection{Feature extraction with \texorpdfstring{\charm}{CHARM}}\label{sec:charm}

We use the CHaracterization Algorithm for Radius Measurements (\charm) feature extraction algorithm used in \citetalias{eht_sgra_6} (see also \citealt{eht_m87_6,ozel_2022_bhgrtests}).\footnote{\charm is available on github at \href{github.gatech.edu/fozel3/CHARM}{github.gatech.edu/fozel3/CHARM}.} Figure~\ref{fig:charm} shows several example library images (before and after blurring; see below). Each panel in the figure also shows the size of the ring (diameter and width) identified by \charm as well as the boundary of the Kerr black hole shadow.

Given a prospective ring center, \charm finds the radius of the ``best fit'' circle by measuring the distance from the center position to the peak brightness along each of 128 evenly separated radial slices. The ring radius is taken to be the median distance among the slices. To find the ring width, \charm fits an asymmetric Gaussian to the brightness profile along each of the slices and defines the width to be the median full width half max (FWHM) over all slices. In general, \charm finds the ``best fit'' ring center by searching over all possible prospective centers. However, because we are interested in $\alpha_1$, we set the center location from the approximate expression $D \approx 2 \, \bhspin \sin i,$ where $i$ is the inclination of the observer, $\bhspin$ is the dimensionless black hole spin (see Section~\ref{sec:sims}), and $D$ is the displacement from the center in the direction perpendicular to the projection of the spin axis \citep{johannsen_2010_nohairimages,medeiros_2022_asymmetry}.

Since we are only interested in structures on scales that could be observed with current or upcoming VLBI instruments, we run \charm on images that have been passed through a Butterworth filter \citep{butterworth_1930_filter}, which suppresses power on long baselines. The Butterworth filter is defined in the frequency domain as
\begin{align}
\mathcal{F}_{\rm BW}(b) = \left[ 1 + \left(\dfrac{b}{r}\right)^{2n} \right]^{-1/2},
\end{align}
where $b$ is the spatial frequency under consideration, $r$ is the characteristic frequency cutoff scale, and $n$ is the power-law index for the filter. 

We set the parameters of the filter following \citet{psaltis_2020_ehtfilter} and \citetalias{eht_sgra_6}. We choose $n = 2$ for its high-fidelity low frequency response with minimal sidelobes, and we set the frequency cutoff to $r=15\, {\rm G}\lambda$; this choice of parameters ensures that the error due to the filter in Fourier space is $\approx 1\%$ out to $\approx 8$~G$\lambda$. These parameters were originally chosen for 230~GHz EHT observations (see the left panel of Figure~8 in \citealt{medeiros_2023_primo1}), but we find that the complex error in the equivalent 345~GHz data products at $\approx 12$~G$\lambda$ (the approximate length of the longest EHT baseline at 345~GHz) is still $\approx 1\%$. We, therefore, adopt these parameters also for 345~GHz images, although less blurring could be used in the future for a full-scale 345~GHz array. The filter is applied by Fourier transforming the simulated image, multiplying it with the filter, and inverse Fourier transforming the result to recover the simulated image (i.e., the image is convolved with the kernel).\footnote{The pixel width in frequency $(uv)$ space in units of $\lambda$ is the inverse of the image field of view in radians. We pad the image by a factor of four, which correspondingly decreases the $uv$ pixel size.}

\section{Simulation Libraries}\label{sec:sims}

\begin{table*}[t]
\caption{Summary of parameter survey for Library A and B}
\label{tab:sims}
\begin{ruledtabular}
\begin{tabular}{lccccccc}
\vspace{-.7em}\\
Source& Lib. & flux & $\bhspin$ & $R_{\mathrm{low}}$ & $R_{\mathrm{high}}$ & $n_e\;({\rm cm}^{-3})$ & $i^{\dagger}$\\  
\vspace{-.5em}\\
\colrule
\vspace{-.5em}\\
M87   &  A  & MAD & 0, 0.9     & 1 & 1, 20, 80 & 1, 2.5, 5, 7.5, 10 $\;\;(\times10^5)$ & 17$^{\circ}$  \vspace{0.1em}  \\ 
      &     & SANE & 0.7, 0.9 & 1 & 1, 20, 80 & 1, 2.5, 5, 7.5, 10 $\;\;(\times10^5)$ & 17$^{\circ}$  \\
\vspace{-.9em}\\
      &  B & MAD/SANE & 0, $\pm 0.5$, $\pm 0.9375$ & 1, 10 & 1, 10, 40, 160 & set by total flux constraint & 17$^{\circ}$\\ 
\vspace{-.5em}\\
\colrule
\vspace{-.5em}\\
\sgra &  A  & MAD & 0, 0.9 & 1   & 1, 20, 80   & 1, 5, 10, 50, 100 $\;\;(\times10^6)$ & 0$^{\circ}$, 19$^{\circ}$, 42$^{\circ}$, 90$^{\circ}$ \\ 
& & SANE & 0.7, 0.9 & 1   & 1, 20, 80   & 1, 5, 10, 50, 100 $\;\;(\times10^6)$ & 0$^{\circ}$, 19$^{\circ}$, 42$^{\circ}$, 90$^{\circ}$ \\
\vspace{-.9em}\\
      &  B &  MAD/SANE & 0, $\pm 0.5$, $\pm 0.9375$ & 1 & 1, 10, 40, 160 & set by total flux constraint & 10$^{\circ}$, 50$^{\circ}$, 89$^{\circ}$ \\ 
\vspace{-.7em}\\
\end{tabular}
\vspace{1em}
\begin{tablenotes}
\item $\dagger$ For Library A \sgra simulations, the SANE, $\bhspin=0.7$, $i=0$ simulations resulted in numerical artifacts due to the pole and are therefore not included in our analysis. The highest inclination for Library B ($i=89^\circ$) was chosen to avoid numerical issues arising from tracing perfectly edge-on geodesics.
\end{tablenotes}
\end{ruledtabular}
\end{table*}

To calibrate the theoretical uncertainty, we analyze two libraries of ray-traced images generated using different simulation pipelines. Each pipeline consists of a general relativistic magnetohydrodynamics (GRMHD) fluid simulation step, which solves for the accretion system evolution, and a radiative ray tracing step, in which emission from the accretion system is propagated to an observer. For each set of simulation parameters, we produce images at $230$ and $345\,$GHz. A more detailed review of the simulation pipeline can be found in \citet{wong_2022_patoka}. In this section, we briefly overview the pipelines and describe the details of our two libraries.

\subsection{Fluid simulations and image calculation}

M87 and \sgra are typically modeled as advection dominated and radiatively inefficient accretion flows (referred to as ADAFs or RIAFs; see \citealt{yuan_2014_review}). RIAF models describe geometrically thick, optically thin disks. Numerous multiwavelength observations as well as EHT results have demonstrated significant agreement between observations of M87 and \sgra and the RIAF model \citep{yuan_2003_nonthermalriaf,narayan_2008_adafbh,ho_2008_agn,falcke_2013_sgra,eht_m87_5,eht_sgra_5}. RIAFs are often simulated using GRMHD algorithms, which produce three-dimensional time series data for the densities, internal energies, velocities, and magnetic fields in the flow. GRMHD simulations naturally recover properties of the turbulent fluid dynamics and yield a prediction for the structure of the accretion disk, winds, and relativistic jet.

At sufficiently low accretion rates (such as those expected for M87 and \sgra), radiative effects are unimportant for the fluid evolution \citep{dibi2012_radiation,ryan2017_radiation}. In this regime, the GRMHD equations are invariant under independent rescalings of both the metric length (or equivalently time) and plasma density (or equivalently accretion rate). As a result, the ideal GRMHD simulations we consider in this paper span the remaining two-dimensional parameter space over the angular momentum of the accretion system and the magnetic flux near the event horizon.

We express the black hole angular momentum $J$ in terms of the dimensionless spin parameter $\bhspin \equiv Jc / G M^2$ so that $\left| \bhspin \right| \le 1$ (hereafter we set $G=c=1$). We adopt the convention that negative values of spin signify anti-alignment between the black hole spin and the inflowing plasma far from the event horizon. Although the alignment of the angular momenta of the black hole and flow is natural, it is not required; anti-aligned systems and systems with intermediate tilt have gained widespread attention in recent years \citep{liska_2018_tilted,chatterjee_2020_tilted,white_2020_tilt,ressler_2020_madstellarwinds,white_2022_tiltvariability,liska_2023_twotempwarped,bollimpalli_2024_tiltedqpo}. In this paper, we restrict our study to the aligned/anti-aligned scenario for simplicity.

The second input parameter for the fluid simulation is the magnetic flux on the black hole event horizon. In ideal MHD, magnetic field lines are dragged toward the black hole with the plasma as accretion proceeds. If the dissipation rate is small, then magnetic flux builds up on the horizon over time until the magnetic pressure is large enough to counterbalance the inward ram pressure of the fluid, at which point the flow enters the magnetically arrested disk (MAD; \citealt{bisnovatyi_1974_madstar,igumenshchev_2003_mad,narayan_2003_mad}) state. The complementary state, when the magnetic pressure is relatively low, is known as standard and normal evolution (SANE; \citealt{narayan_2012_sane,sadowski_2013_sane}). SANE flows are turbulent but steady while MAD flows are chaotic and mediated by transient plasma accretion strands that thread between the event horizon and the sustained disk at large radius.

Although GRMHD simulations are invariant under rescalings of length and density, calculating images requires computing radiative transfer coefficients, which depend on dimensional units for quantities such as the number of electrons per volume or the magnetic field strength. When restoring units, we choose a length scale to be consistent with a particular black hole mass. The mass accretion rate (i.e., the physical electron number density) is often chosen so that the total flux produced by the system matches some observed value. The accretion rate may also be varied to study the importance of the number density, which is particularly useful in understanding systematics when the observed compact flux is poorly constrained or varies significantly.

At EHT observing frequencies, the primary emission mechanism for M87 and \sgra is synchrotron radiation, which is produced by a population of hot electrons \citep{yuan_2014_review}. The ions and electrons in the M87 and \sgra accretion flows are likely at different temperatures, owing to different heating mechanisms and the lack of an efficient coupling between the two species (see below). Nevertheless, since the electrons contribute negligibly to the total fluid energetics, GRMHD simulations typically model the plasma as a single fluid with a single total internal energy. However, the electron distribution function is required for the radiative transfer step. We assume that the electrons are thermal and assign their total energy using a prescription for the ion-to-electron temperature ratio based on the local fluid properties.

In this work, we adopt the standard ion-to-electron temperature ratio prescription of \citet[][see also \citealt{eht_m87_5}]{moscibrodzka_2016_rhigh},
\begin{align}
R = \dfrac{T_i}{T_e} &= \dfrac{R_{\rm low} + R_{\rm high} \beta^2}{1 + \beta^2},
\label{eq:Rhigh}
\end{align}
where plasma $\beta = P_{\rm gas} / P_{\rm mag}$ is the ratio between the gas pressure and the magnetic pressure $P_{\rm mag} = b^2 / 2$. Under this prescription, the temperature ratio of the ions to electrons asymptotes to $R_{\rm high}$ ($R_{\rm low}$) in the limit of large (small) plasma $\beta$. We consider multiple values for both $R_{\rm low}$ and $R_{\rm high}$ (see \citealt{wong_2022_patoka} for more detail about these choices). Both libraries include simulations with $R_{\rm high}=1$, which results in ions and electrons having the same temperature when plasma-$\beta$ is large, but expectations from theory and observations infer that both M87 and \sgra should be two temperature \citep{shapiro_1976_two_temp, rees_1982_Origin_radio_jets,narayan_Yi_1995_ADAF_underfed_BH,yuan_2014_review,dexter_2020_sgraelectrons,satapathy_2023_electronheating,satapathy_2024_alfven,chael_2025_twotempm87,salas_2025_twotempsgra}. We include the $R_{\rm high}=1$ models in our libraries but exclude them from the theoretical uncertainty analysis in the main text. Although they are unlikely to represent the M87 and \sgra systems, we consider and discuss $R_{\rm high}=1$ simulations in Appendix~\ref{ap:Rhigh} for completeness.

The time series fluid data are processed into simulated images using a general relativistic radiative transfer (GRRT) code. The observer is located at a known distance and oriented at a specified angle relative to the black hole. For M87, both libraries use a mass over distance of $M/D = 3.819\,\mu\mathrm{as}$. For \sgra, Library A uses $M/D = 5.054\,\mu\mathrm{as}$, while Library B uses $M/D = 4.98\,\mu\mathrm{as}$. Each image comprises a square grid of pixels and has a user-defined field of view. Our GRRT codes solve for the total intensity seen by the observer in two steps. First, photon trajectories are found by solving the geodesic equation backward through the simulation bulk from each pixel in the image plane. The radiative transfer equation is then solved forward along each trajectory, accounting for both emission and absorption.

\subsection{Library A}

Library A was produced using simulations that were run with the {\tt{}HARM3D} GRMHD code \citep{gammie_2003_harm} and first described in \citet{narayan_2012_sane} and \citet{sadowski_2013_sane}. The simulations were run on a hyper-logarithmic (radial) grid with a resolution of $N_r,N_\theta,N_\phi = 256,128,64$. Each simulation was initialized using the ``Polish donut'' configuration \citep{kozlowski_1978_donut} with inner edge at $r_{\rm in} = 10\,M$ and extending to $r\approx 1000\,M$. The pressure closure for the fluid evolution was provided by an ideal gas equation of state with global, uniform adiabatic index $\hat{\gamma} = 5/3$. The MAD simulations were initialized with a single magnetic field loop that results in ordered magnetic fields and high magnetic flux near the horizon and were run for a total duration of $100,000\,M$. The SANE simulations were initialized with multiple smaller magnetic loops that lead to weak and turbulent magnetic fields near the black hole and were run for a total duration of $200,000\,M$. Fluid snapshots were imaged every $10\,M$ after the system had reached steady state near the horizon ($1024$ total snapshots per simulation; for more details, see \citealt{narayan_2012_sane}). Library A contains four simulations: two MAD with black hole spins $\bhspin = 0, 0.9$ and two SANE with $\bhspin = 0.7, 0.9$.

The images in Library A were generated using the GPU-based ray-tracing code {\tt{}GRay} \citep{chan_2013_gray}. 
For M87, we set $i = 17^\circ$ and for \sgra we vary the observer inclination angle evenly in $\sin i$. We normalize the mass accretion rate for each simulation in order to match a target average electron number density ($n_e$) near the event horizon and vary this target $n_e$ across different simulations. \cite{satapathy_2022_variability} showed that the range of $n_e$ considered in the M87 subset of Library A corresponds to accretion rates between $\approx 2.9\times10^{-6}\dot{M}/\dot{M}_{\rm Edd}$ and $2.9\times10^{-5}\dot{M}/\dot{M}_{\rm Edd}$. Each image was produced with a $64\,M$ field of view at a resolution of $512\times512$ pixels. The other parameter choices have been summarized in Table~\ref{tab:sims}. 

The \texttt{CHARM} algorithm quantifies image features when the image does in fact consist primarily of a ring and a brightness depression. However, since Library A includes higher-accretion rate models, when viewed edge-on ($i=90^{\circ}$) the images in the four simulations with $\bhspin=0.9$, and $n_e = 10^8$ do not show a significant brightness depression. We exclude these models from our analysis since they are inconsistent with the observations of both M87 and Sgr~A$^*$ and a ring feature cannot be extracted.\footnote{Rerunning the analysis including the non-ring-like models does not significantly alter the results (see Section~\ref{sec:discussion}).}

\subsection{Library B}
\begin{figure*}[t!]
    \includegraphics[width=.9\textwidth]{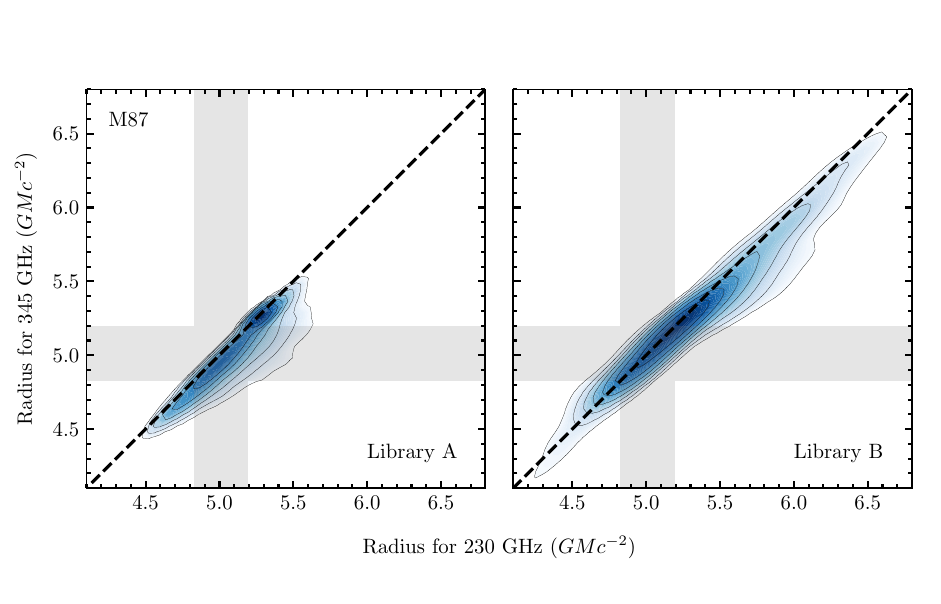}
    \caption{Correlation between the diameter calculated by \charm for a 230~GHz image vs. the same image at 345~GHz for the M87 libraries. The contours show eight logarithmically spaced levels spanning over 2 orders of magnitude (each successive contour is just less than twice the value of its neighbor). The gray shaded region corresponds to the range of possible shadow sizes for all Kerr shadows.
    }
    \label{fig:diam_correlations}
\end{figure*}

The images in Library B were produced using fluid snapshots from the GRMHD code \kharma \citep{gammie_2003_harm,prather_2021_iharm3d} that were 
raytraced with the GRRT code \ipole (\citealt{moscibrodzka_2018_ipole}, see also \citealt{wong_2022_patoka}). The models in Library B include different single-loop magnetic field initializations that lead to either the MAD or SANE accretion state. In addition to including both MAD and SANE initializations, the models in Library B also span a variety of other parameter choices, which are summarized in Table~\ref{tab:sims}. The fluid simulations were initialized with the equilibrium torus solution of \citet{fishbone_1976_torus} with inner radius $r_{\rm in}$ and pressure maximum $r_{\rm max}$ in units of $M$ set to $(20, 41)$ for MAD tori and $(10, 20)$ for SANE tori. The pressure closure was provided by an ideal gas equation of state, but with an adiabatic index $\hat{\gamma} = 4/3$. Fluid variables were saved on a (logarithmic) spherical grid with resolution of $N_r, N_\theta, N_\phi = 288, 128, 128$. The simulations were evolved for a total duration of $25,000\,M$ after an initial $\Delta t = 5000\,M$ transient, with snapshots drawn uniformly every $250\,M$ ($100$ images per simulation). 

For M87, we set $i = 163^{\circ}$ or $17^\circ$ according to the brightness asymmetry observed on the sky and choose the mass accretion rate per set of simulation parameters to reproduce the observed $230\,$GHz flux density $\approx 0.65\,$Jy \citep{eht_m87_4}. To compare with Library A, we note that the flux normalization selects an accretion rate of $5.2\times10^{-7} \,\dot{M}/\dot{M}_{\rm Edd}$ for MAD and $7.9\times10^{-6} \,\dot{M}/\dot{M}_{\rm Edd}$ for SANE (for $\bhspin=+0.9375$ and $R_{\mathrm{high}} = 40$). The electron density $n_e$ for the Library B SANE model is similar to the largest $n_e$ considered in Library A, but the $n_e$ in Library B MAD simulations is an order of magnitude smaller than the smallest value considered in Library A. For \sgra, we set the mass accretion rate per simulation to reproduce a target flux density of $2.4\,$Jy and sample over three observer inclinations $i=10^\circ, 50^\circ, 89^\circ$. We set the fields of view for our images to be $160\,\mu{\rm as}$ for M87 and $200\,\mu{\rm as}$ for \sgra (these correspond to approximately $41.88$ and $40.152\, M$, respectively, for the source distances). Following the recommendations in \citet{psaltis_2020_ehtfilter}, we set the image resolution to be two pixels per $\mu{\rm as}$.

\begin{figure*}[t!]
    \includegraphics[width=.9\textwidth]{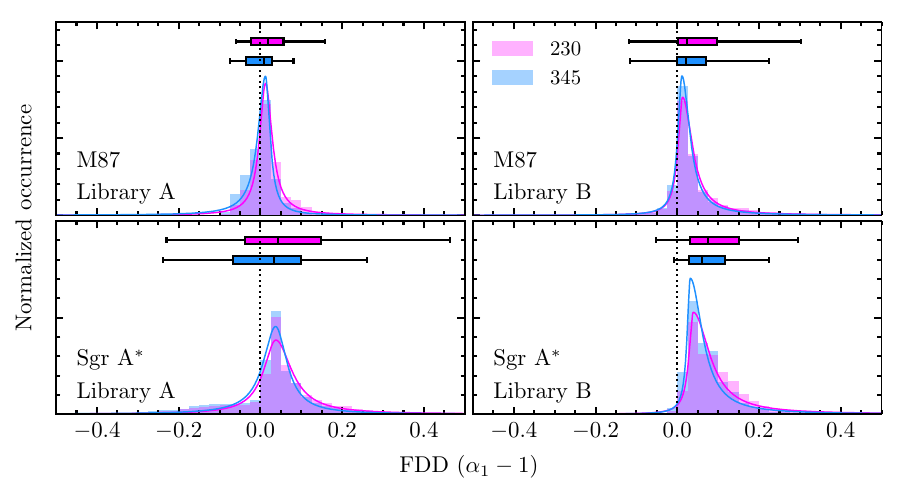}
    \includegraphics[width=.9\textwidth]{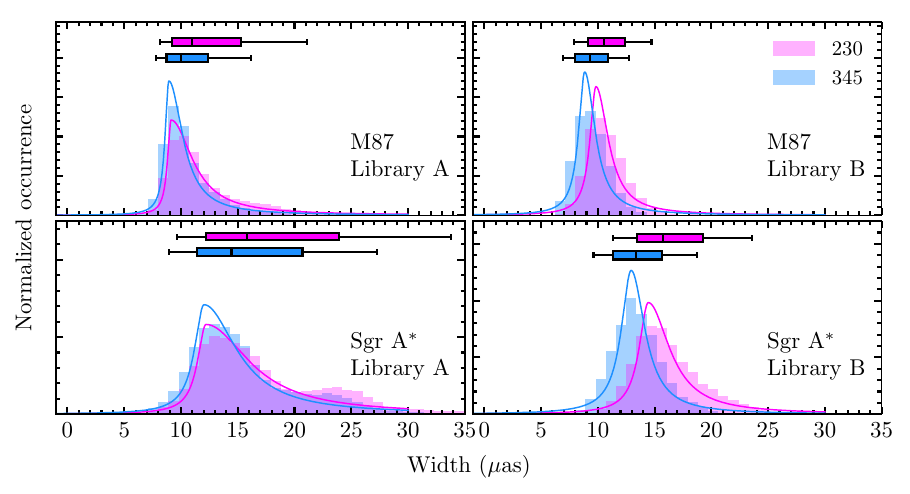}
    \caption{Fractional diameter difference (FDD, top panels) and width (bottom panels) from \charm for Library A (\textit{left}) and B (\textit{right}) at both 230~GHz (magenta) and 345~GHz (blue) for both M87 and \sgra. The modified box-and-whisker plots show the central $68\%$ (box) and $95\%$ (whisker) ranges for the distributions. We also show the results of fitting a skewed Cauchy distribution as solid lines (color coded to match the histograms), to guide the reader's eye. For both libraries and both sources, the medians as well as the two central ranges decrease when measured at 345~GHz as compared to 230~GHz (see Table~\ref{tab:results}). The widths also decrease significantly for the higher frequency images.
    }
    \label{fig:diam_total}
\end{figure*}
For the time-series studies performed in Section~\ref{sec:epochs}, we also generate a higher-cadence subsample of Library B spanning over $5,000\,M$ total duration with a sampling cadence of $5\,M$ per image. These images also have twice the linear resolution of the larger library, i.e., $160\,\mu{\rm as}$ field of view at a resolution of $640\times640$ pixels.

\section{Observing at multiple frequencies}\label{sec:multi-wave}

Observing at shorter wavelengths offers several advantages, including improved angular resolution and reduced absorption (both atmospheric and at the source). Multi-frequency observations also offer the possibility of reconstructing resolved Faraday rotation measure maps from the additional observing frequencies. They may also provide a means toward increased confidence in identifying gravitational lensing signatures, since the lensing effects are achromatic, while plasma emission effects are not. In this section, we focus on one aspect of that broader motivation: how the observing wavelength affects the distribution of measured ring diameters and, consequently, the calibration between the emission ring and the underlying shadow scale.

We apply \charm\ to images from both libraries and compare the resulting diameter distributions at 230 and 345~GHz for Sgr~A$^*$ and M87. Because synchrotron emissivity depends on local fluid properties, higher-frequency emission (e.g., at $345\,{\rm GHz}$) tends to arise from more compact regions at smaller radii. This shift in the emission region, together with an increasing contribution from lensed emission that has passed around the far side of the black hole, generally produces narrower rings and ring diameters that lie closer to the shadow boundary. Figure~\ref{fig:diam_correlations} shows the correlations between ring diameters measured in 230~GHz and 345~GHz images for both M87 libraries. As expected, the diameters are correlated, with a larger dispersion at 230~GHz than at 345~GHz (especially for Library A).

\begin{table*}[t]
\caption{Summary of Feature Extraction Results}
\label{tab:results}
\begin{ruledtabular}
\begin{tabular}{lclllllrrr}
\vspace{-.7em}\\
Library & frequency  & &    & FDD &       &   &   & Width ($\mu$as) \!\!\!\!\!\!\!\!\!\!\!   &  \\ 

       &(GHz) & &   median  &  68\% &  95\% &  & median  & 68\% & 95\%\\ 
\vspace{-.5em}\\
\colrule
\vspace{-.5em}\\
A -- M87 & 230 &  & 0.017 & 0.079 & 0.218 & & 10.96 & 6.07 & 12.90 \\ 
        & 345 &  & 0.008 & 0.064 & 0.156 & & 9.98 & 3.68 & 8.41 \\ 
&&&&&&&&&\\ 
B -- M87 & 230 &  & 0.025 & 0.095 & 0.419 & & 10.54 & 3.28 & 6.84 \\ 
        & 345 &  & 0.021 & 0.073 & 0.341 & & 9.32 & 2.90 & 5.76 \\ 
 \\ 
A -- \sgra & 230 &  & 0.043 & 0.187 & 0.694 & & 15.81 & 11.76 & 24.10 \\ 
              & 345 &  & 0.033 & 0.167 & 0.499 & & 14.45 & 9.29 & 18.35 \\ 
&&&&&&&&&\\ 
B -- \sgra & 230 &  & 0.076 & 0.120 & 0.349 & & 15.71 & 5.81 & 12.22 \\ 
              & 345 &  & 0.060 & 0.088 & 0.231 & & 13.33 & 4.28 & 9.09 \\ 
 \vspace{-.5em}\\ \colrule \vspace{-.5em}\\ 
A -- M87 MAD& 230 &  & 0.017 & 0.068 & 0.281 & & 11.09 & 7.49 & 13.99 \\ 
        & 345 &  & 0.008 & 0.052 & 0.156 & & 10.08 & 4.50 & 9.33 \\ 
&&&&&&&&&\\ 
B -- M87 MAD& 230 &  & 0.014 & 0.030 & 0.129 & & 9.93 & 2.43 & 5.24 \\ 
        & 345 &  & 0.011 & 0.031 & 0.137 & & 9.02 & 2.58 & 5.20 \\ 
 \vspace{-.5em}\\ \colrule \vspace{-.5em}\\ 
A -- M87 PRIMO width& 230 &  & 0.008 & 0.051 & 0.114 & & 9.27 & 1.28 & 2.28 \\ 
        & 345 &  & 0.006 & 0.061 & 0.115 & & 8.75 & 1.10 & 2.14 \\ 
&&&&&&&&&\\ 
B -- M87 PRIMO width& 230 &  & 0.011 & 0.052 & 0.278 & & 9.29 & 1.42 & 2.52 \\ 
        & 345 &  & 0.011 & 0.044 & 0.283 & & 8.28 & 1.59 & 2.83 \\ 
 \vspace{-.5em}\\ \colrule \vspace{-.5em}\\ 
A -- M87 temporal median& 230 &  & 0.016 & 0.044 & 0.136 & & 10.93 & 5.96 & 11.11 \\ 
        & 345 &  & 0.007 & 0.042 & 0.076 & & 9.88 & 3.24 & 6.92 \\ 
&&&&&&&&&\\ 
B -- M87 temporal median& 230 &  & 0.025 & 0.079 & 0.210 & & 10.58 & 2.48 & 4.33 \\ 
        & 345 &  & 0.023 & 0.054 & 0.179 & & 9.32 & 2.03 & 3.61 \\ 
 \\ 
A -- \sgra temporal median & 230 &  & 0.041 & 0.155 & 0.594 & & 15.66 & 11.44 & 21.38 \\ 
              & 345 &  & 0.032 & 0.144 & 0.457 & & 14.53 & 8.63 & 16.05 \\ 
&&&&&&&&&\\ 
B -- \sgra temporal median& 230 &  & 0.082 & 0.091 & 0.164 & & 15.35 & 4.57 & 8.53 \\ 
              & 345 &  & 0.063 & 0.061 & 0.107 & & 13.27 & 2.72 & 4.97 \\ 
 \\ 
\vspace{-.7em}\\
\end{tabular}
\vspace{1em}
\begin{tablenotes}
\item Summary of the median ($50$th percentile), average $68\%$ bound (half of the full central $68\%$ interval), and average $95\%$ bound (half of the full central $95\%$ interval) for both FDD and ring width for both Library A and B for both M87 and \sgra. 
\end{tablenotes}
\end{ruledtabular}
\end{table*}

\begin{figure*}
    \includegraphics[width=.95\textwidth]{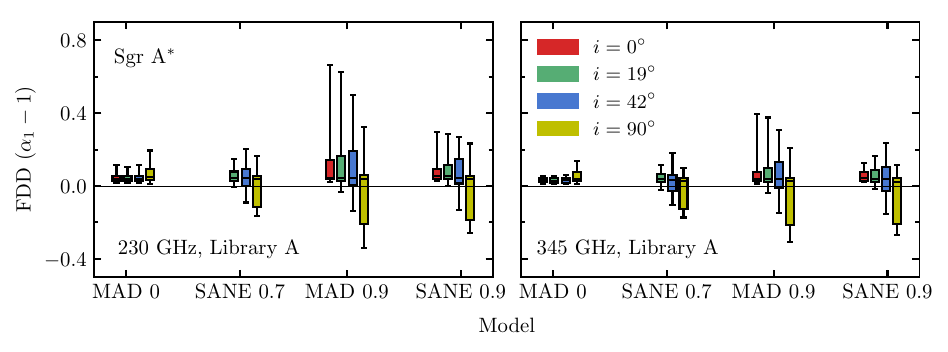}
    \includegraphics[width=.95\textwidth]{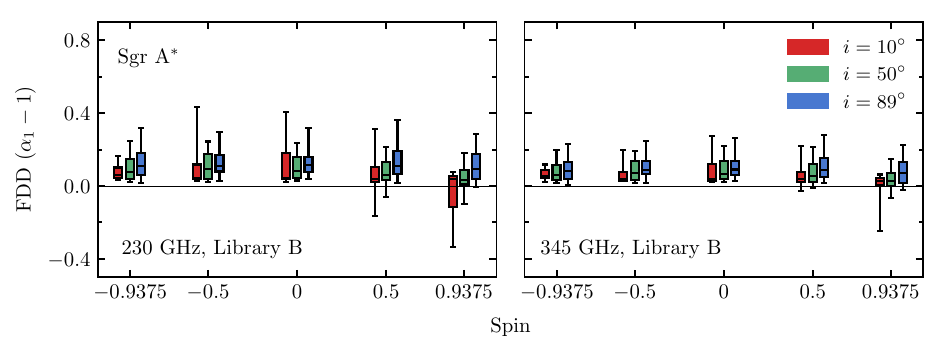}
    \caption{Modified box and whisker plots comparing the distribution of the fractional diameter difference for both libraries at both wavelengths as a function of spin ($x$-axis) and the observer's inclination (different colors) for the \sgra part of the libraries; the left column corresponds to 230~GHz and the right column corresponds to 345~GHz. As before, the medians of the distributions are shown as black lines within the boxes, the boxes show the central $68\%$ interval, and the whiskers show the central $95\%$ interval. The central $68\%$ range increases with inclination for Library A, but in Library B, there are no consistent trends with inclination in the median or central $95\%$ range.
    }
    \label{fig:inc}
\end{figure*}

We use \charm to characterize images in both simulation libraries at both wavelengths to measure the distribution of ring diameters and widths in the images. We then take the ratio of the \charm-measured diameter over the median diameter of the analytically calculated black hole shadow for each image to obtain the ``fractional diameter difference,'' or $\alpha_1-1$, where $\alpha_1\equiv d_m/d_{\mathrm{sh}}$ (see also \citetalias{eht_sgra_6}). We also note that the fractional diameter difference (FDD) is equal to the relative error in the diameter measurement, 
\begin{align}
\mathrm{FDD} &= \alpha_1-1 = \frac{d_m}{d_{\mathrm{sh}}} - 1 = \frac{d_m-d_{\mathrm{sh}}}{d_{\mathrm{sh}}}.
\end{align}
The distribution of FDD and width in both libraries, both sources, and both wavelengths is shown in Figure~\ref{fig:diam_total}. For both libraries and both sources, the median and dispersions for the FDD are significantly smaller for 345~GHz compared to 230~GHz. 

To guide the reader's eye and enable a more quantitative comparison, we overlay the best-fit skewed Cauchy function on the sampled distributions. We adopt a skew-Cauchy family (rather than a skew-normal) because the empirical distributions exhibit pronounced heavy tails that skew-normal fits fail to reproduce. The skew-Cauchy curve is used only as a visual summary of the overall shape; we do not interpret or report its fitted parameters. We instead summarize each distribution non-parametrically using the percentiles $P_{2.28}$, $P_{15.87}$, $P_{50}$, $P_{84.13}$, and $P_{97.72}$ computed directly from the samples (not from the fit). These define a median $\overline{x}$ and percentile-based scatter $s_x \equiv \frac{1}{2}\left(P_{84.13}-P_{15.87}\right)$, together with an outer interval $[P_{2.28},\,P_{97.72}]$. For a normal distribution, these percentiles coincide with $\mu\pm\sigma$ and $\mu\pm2\sigma$, so $\overline{x}$ and $s_x$ provide a familiar scale for comparison without assuming normality or relying on ill-defined moments. For clarity, we quote percentiles rounded to the nearest whole number throughout the text, although we maintain full precision in our analysis. When representing the data graphically, we use modified box-and-whisker plots showing the median of the data as a solid black line, the center $68\%$ range as a colored box, and the center $95\%$ range as whiskers.

Table~\ref{tab:results} summarizes the median and center $68\%$ and $95\%$ interval ranges for both the FDD and the width. For the M87 portion of Library A (B), the median, center $68\%$, center $95\%$ ranges improved by $53\%$, $19\%$, and $28\%$ ($16\%$, $23\%$, and $19\%$), respectively, for 345~GHz as compared to 230~GHz. Similarly, for the \sgra portion of Library A (B) the median, $68\%$, and $95\%$ ranges decreased by $23\%$, $11\%$, and $28\%$ ($21\%$, $27\%$, and $34\%$) for 345~GHz compared to 230~GHz. The differences between the M87 and the \sgra libraries are due primarily to the different observer inclinations considered, which we will discuss in more detail in Section~\ref{sec:astro_priors}. 

The lower panels of Figure~\ref{fig:diam_total} show the distribution of ring widths for both libraries, both sources, and both wavelengths. Ring width distributions are important because the shadow boundary is typically contained within the broader emission ring (see, e.g., Figure 10 of \citealt{ozel_2022_bhgrtests}). In section~\ref{sec:width}, we will discuss how ring width constraints on the parameter space can improve FDD estimates. As expected, we find that ring widths are considerably smaller: medians are lower by about $10\%$ on average and the $95\%$ ranges are lower by about $15$-$35\%$, on average for 345~GHz compared to 230~GHz. The two libraries exhibit distributions of ring widths with considerably different shapes. For Library A (both M87 and \sgra), the distribution is broader and skewed towards higher widths, likely due to the high accretion rates considered. Note that since all images were blurred by a Butterworth filter prior to \charm being applied, there is an implicit lower limit to measured ring widths that is related to observational capabilities and not to the underlying physics.

\begin{figure*}
    \includegraphics[width=.95\textwidth]{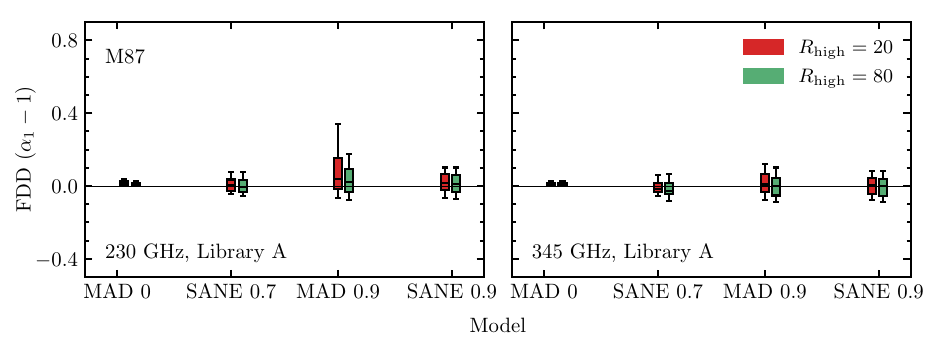}
    \includegraphics[width=.95\textwidth]{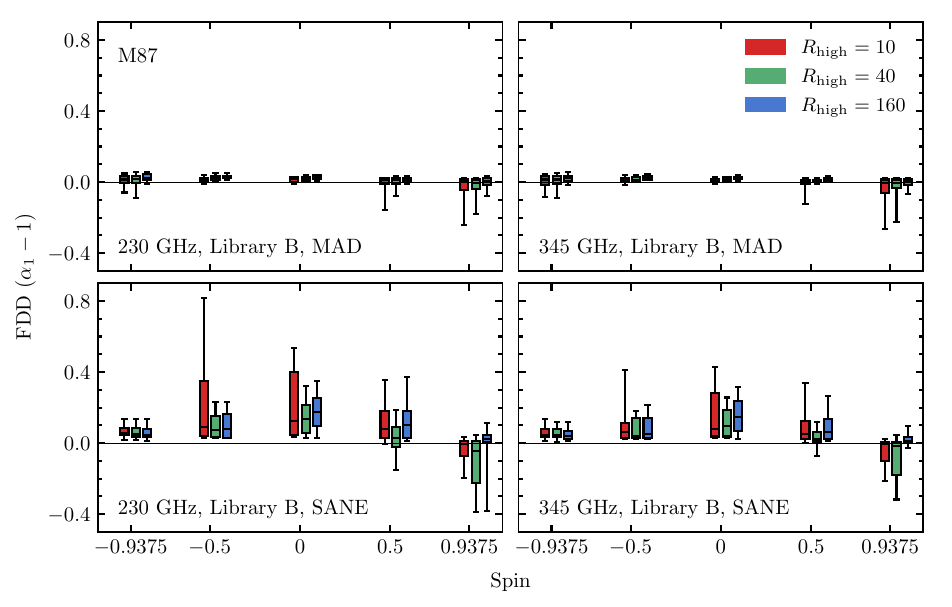}
    \caption{Comparison of the distribution of the fractional diameter difference for both M87 libraries at both wavelengths as a function of spin, magnetic flux (SANE/MAD), and the $R_{\mathrm{high}}$ parameter. For Library A (top row) the $x$-axis shows both the magnetic flux (SANE vs. MAD) as well as the spin magnitude for each model. MAD models from Library B are in the second row, while SANE are in the third row. In all panels, the different colors show different values for the $R_{\mathrm{high}}$ parameter. Due to the total flux constraint in Library B, MAD models in Library B have significantly smaller accretion rates than SANE models. This results in much smaller FDD values for MAD models for Library B. However, both MAD and SANE have the same accretion rates in Library A, resulting in similar ranges of FDD. 
    }
    \label{fig:SANE_MAD}
\end{figure*}

\begin{figure*}
    \includegraphics[width=.95\textwidth]{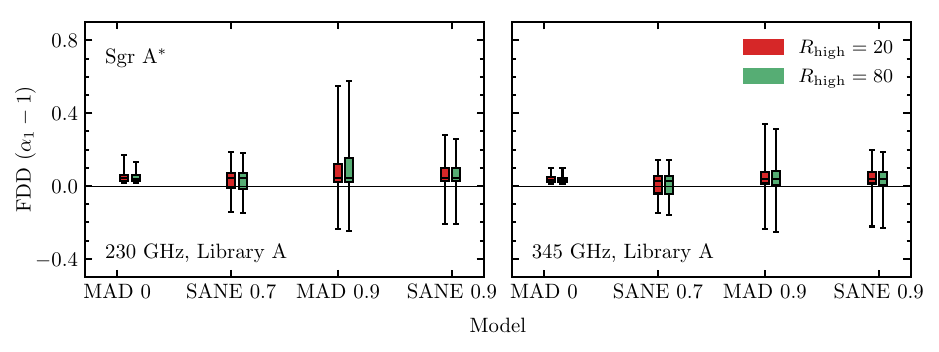}
    \includegraphics[width=.95\textwidth]{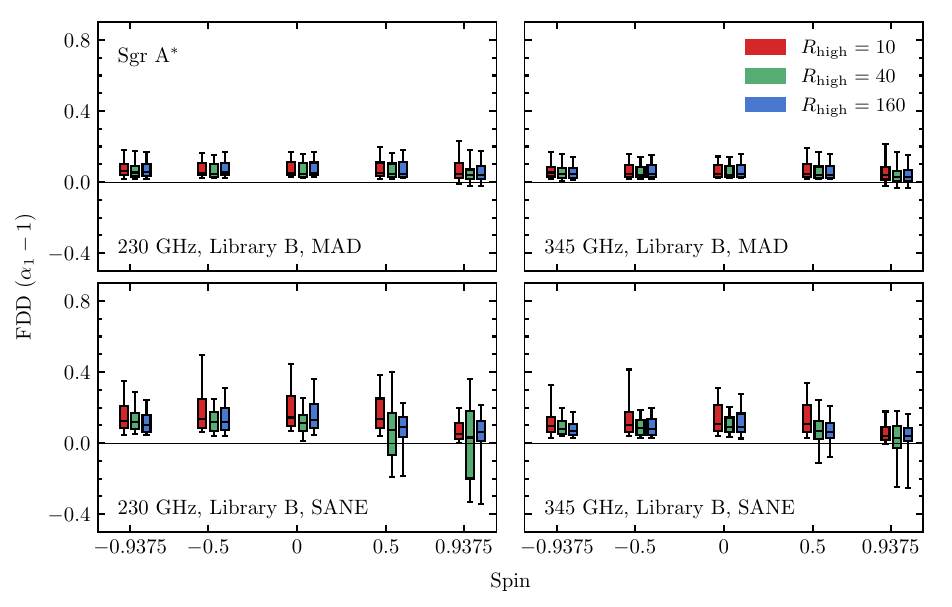}
    \caption{Same as Figure~\ref{fig:SANE_MAD} but for the \sgra parts of the libraries. 
    }
    \label{fig:SANE_MAD_SA}
\end{figure*}
 
\begin{figure*}
    \includegraphics[width=.9\textwidth]{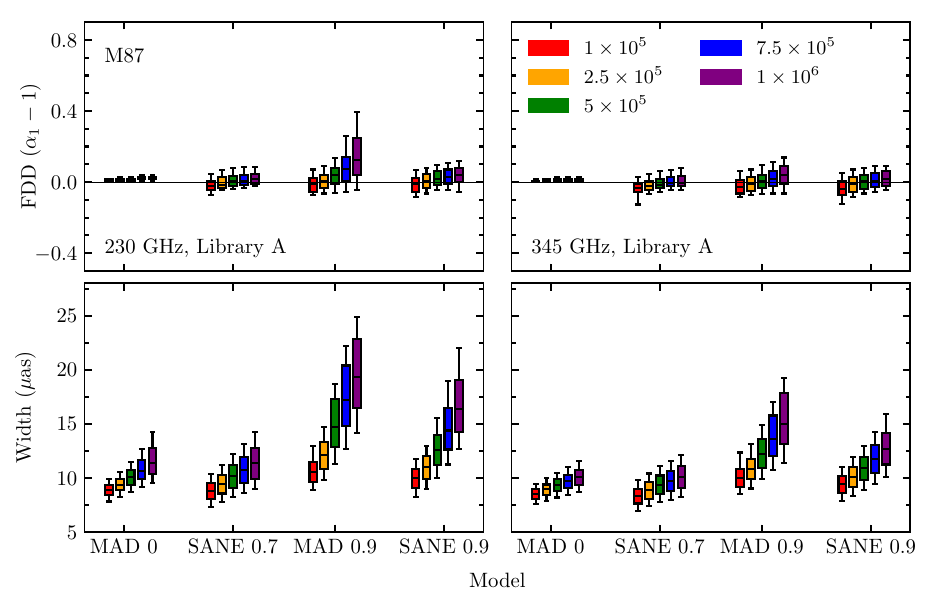}
    \caption{Distribution of FDD (\textit{top}) and ring width (\textit{bottom}) measurements for different mass accretion rates (different colors) at 230~GHz (\textit{left}) and 345~GHz (\textit{right}) for Library A as a function of spin and magnetic flux. Increasing the accretion rate consistently increases the median and spread in the width as well as the median FDD.
    }
    \label{fig:NE}
\end{figure*}
\begin{figure}
    \includegraphics[width=\columnwidth]{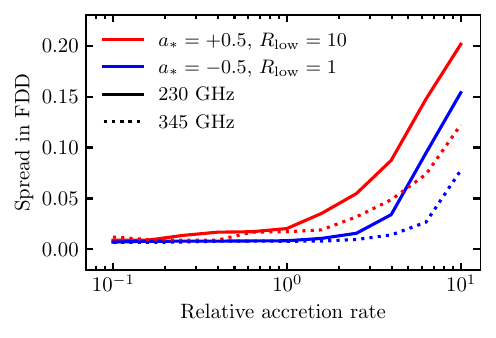}
        \caption{Dependence of the spread in FDD versus accretion rate for two ``best-bet'' models for the M87 accretion flow (based on total intensity, linear polarimetric, multiwavelength, and jet power constraints \citealt{eht_m87_8}). The images for this figure were computed from the same GRMHD simulations used to generate Library B. The spread in FDD increases with higher accretion rate for both models and wavelengths, consistent with Figure~\ref{fig:NE}.
    }
    \label{fig:golden}
\end{figure}

\begin{figure*}[htp!]
    \includegraphics[width=.9\textwidth]{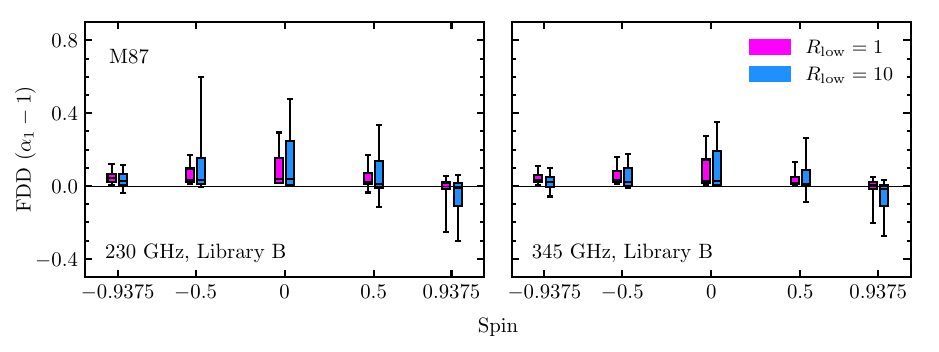}
    \caption{
    Distribution of FDD for different values of $R_{\mathrm{low}}$ (different colors) at 230~GHz (left) and 345~GHz (right) for the M87 portion of Library B. Higher $R_{\mathrm{low}}$ tends to result in higher and more variable FDD.
    }
    \label{fig:Rlow}
\end{figure*}

\section{Applying constraints to the model space}\label{sec:model_constraints}

We consider here how the FDD might improve if the model parameter space can be further constrained either through astrophysical constraints (Section~\ref{sec:astro_priors}), or through image morphology constraints (Section~\ref{sec:width}).

\subsection{Astrophysical priors}\label{sec:astro_priors}

It may be possible to constrain the observer inclination angle for M87 by associating the large-scale jet orientation with the black hole spin axis. No analogous large-scale jet prior exists for \sgra, but the observed image asymmetry can still provide a model-dependent constraint on inclination: \citet{faggert_2025_sgraasym} showed that the low observed asymmetry disfavors inclinations substantially above face-on for Keplerian flows, although \citet{medeiros_2022_asymmetry} showed that similarly low asymmetries can arise even at high inclination in strongly sub-Keplerian flows, as in some low-spin MAD models. In Figure~\ref{fig:inc}, we explore the dependence of FDD on the observer's inclination angle for both libraries. The central $68\%$ interval in FDD increases for higher inclination angles throughout Library A. The median FDD values and the central $95\%$ interval do not show a clear trend in FDD. In Library B the median value of FDD tends to increase with higher inclination, but the spread in FDD does not always increase with inclination. The higher inclination simulations in Library A also tend to have lower FDD, and a significant part of the distribution has negative FDD values (these simulations generate the left shoulder in the FDD distribution for Library A, \sgra, Figure~\ref{fig:diam_total}). 

All images included in the M87 libraries (see Figure~\ref{fig:diam_total}) assume that the spin axis of M87 is either aligned or anti-aligned with the large scale jet at $17^{\circ}$ off of the observer's line of sight \citep{walker_2018_m87jet}. These simulations also assume different values for the black hole mass and distance compared to the ones for \sgra. The FDD tends to be significantly smaller for the M87 libraries than the \sgra libraries: the median, $68\%$, and $95\%$ range values for Library A at 230~GHz are $60\%$, $58\%$, and $69\%$ smaller for the M87 library compared to \sgra (see also Table~\ref{tab:results}). For Library B the median is $67\%$ smaller and the center $68\%$ range is $21\%$ smaller. However, the $95\%$ range is $20\%$ larger for the M87 library. Therefore, if M87 is indeed at low inclination (consistent with both the jet orientation and EHT results), the theoretical error is considerably smaller in most cases (see also Figure~\ref{fig:diam_total}). It is possible, however, that the disk and jet/spin axis are tilted with respect to each other, but we leave a thorough analysis of the effect of disk tilt to future work. 

Polarized EHT observations can be used to constrain the strength of the magnetic field and distinguish between MAD and SANE flows \citep{palumbo_2020_beta2,eht_m87_8,eht_sgra_8}. In Figure~\ref{fig:SANE_MAD}, we explore how the distribution of fractional diameter difference depends on the strength of the magnetic field (SANE vs. MAD), on the black hole spin, and on the $R_{\mathrm{high}}$ parameter for the M87 libraries. We show a similar exploration for the \sgra libraries in Figure~\ref{fig:SANE_MAD_SA}. For Library A, we see a similar FDD distribution for both MAD and SANE, but the FDD increases slightly with increasing spin for both M87 and \sgra. Model fitting and parameter estimation through comparisons of theoretical simulations and observations may also constrain $R_{\mathrm{high}}$, $R_{\mathrm{low}}$, and the accretion rate. We do not see a clear trend in FDD with respect to $R_{\mathrm{high}}$ for Library A. However, we do note that the models with the highest spread in FDD for Library A, M87 at 230~GHz (MAD, $\bhspin=0.9$, and $R_{\mathrm{high}}=20$), have a significantly smaller median and spread at 345~GHz. This demonstrates that, for some regions of the model parameter space, the higher frequency observations can provide a highly significant improvement for FDD.

Library B shows a very different dependence on both $R_{\mathrm{high}}$ and SANE/MAD because the accretion rate is set by the flux constraint and because the two libraries probe different spin ranges. The spread in FDD for MAD in Library B is consistently much smaller than for SANE (this effect is most pronounced for the M87 Library). Due to the total flux constraint, the SANE simulations in Library B have higher accretion rates than the MAD simulations, so the SANE/MAD dependence we see here is strongly influenced by the dependence on accretion rate, which we will explore further below.

\begin{figure*}
    \includegraphics[width=\textwidth]{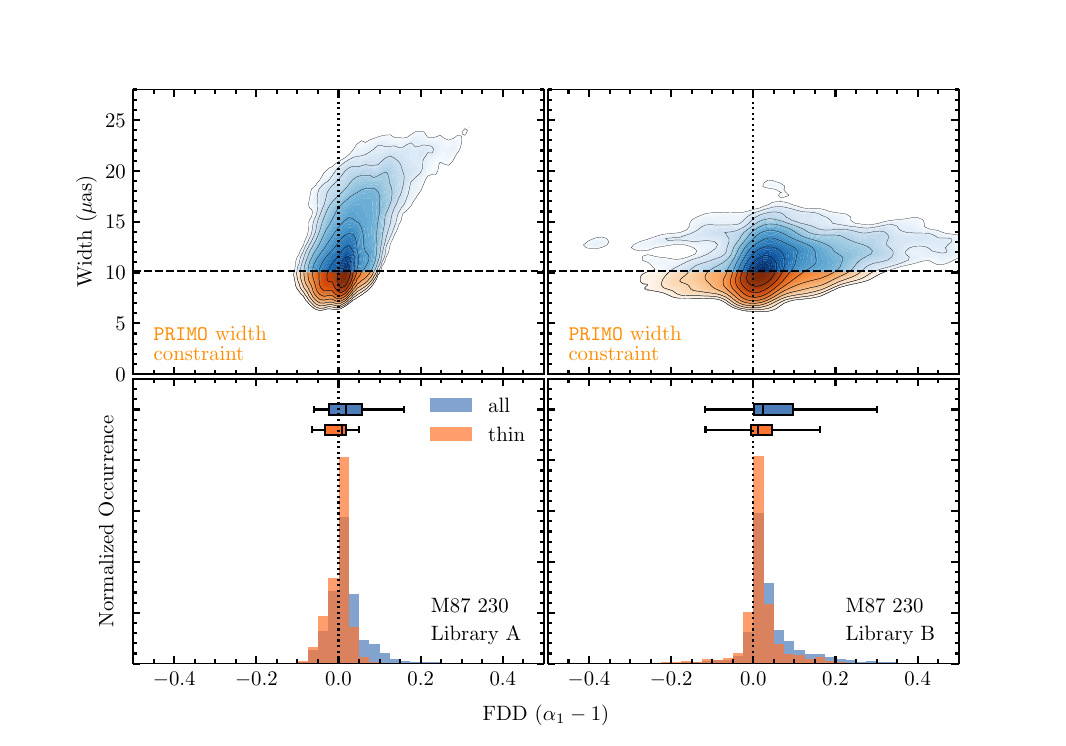}
    \caption{\textit{(top row)} The distribution of width vs. FDD for the 230~GHz images for the M87 parts of both libraries. The contours are logarithmic and span two and a half orders of magnitude. The region of the parameter space that is consistent with the \texttt{PRIMO} constraint is highlighted in orange. \textit{(bottom row)} The distribution in FDD of the full 230~GHz M87 libraries (blue) compared to the FDD distribution of the subset of the images in the library that are consistent with the \texttt{PRIMO} constraint (orange). The FDD distributions are considerably tighter for the subset of images consistent with the \texttt{PRIMO} constraints.
    }
    \label{fig:PRIMO}
\end{figure*}

The models with the highest FDD in Library B are SANE models with $R_{\mathrm{high}}=10$. Models with low $R_{\rm high}$ have hotter disk electrons and thus exhibit more extended emission out to larger radii. As with Library A, these models are considerably better behaved at 345~GHz. If M87 is in the MAD state as argued in~\citet[][but see \citealt{eht_m87_2025}]{eht_m87_8}, then the magnetic flux constraint will constrain the accretion rate to be small, and would, therefore, result in a smaller FDD. We also quote the median and central $68\%$ and $95\%$ interval FDD ranges from only the MAD models in Table~\ref{tab:results}. At 230~GHz the median, center $68\%$, and $95\%$ intervals of the MAD-only FDD are smaller by $44\%$, $68\%$, and $69\%$ respectively, compared to the full Library B. In future work, the EHT collaboration may consider how robust the MAD determination is and perhaps use this constraint together with the total flux/accretion rate constraint to considerably lower the theoretical error for both mass estimates and gravitational tests.

Since the accretion rate in Library B is determined by the requirement that the total flux match the observed value, we explore the relationship between the accretion rate, the FDD, and ring width only for Library A in Figure~\ref{fig:NE}. The median FDD increases with accretion rate and the overall spread grows. However, the detailed changes depend on model specifics.

To explore the dependence of the spread in FDD on accretion rate further, we also ran additional simulations based on the two ``best-bet'' models from EHT results: one MAD $\bhspin = 0.5$ model with $R_{\rm low} = 10$ and $R_{\rm high} = 160$ and one MAD $\bhspin = -0.5$ model with $R_{\rm low} = 1$ and $R_{\rm high} = 160$ \citep{eht_m87_5,eht_m87_8,eht_m87_9}. Images were generated from the best-bet models over 11 different accretion rates ranging between $1/10$ and $10\times$ the fiducial accretion rate found with the flux-fitting procedure. We show in Figure~\ref{fig:golden} the dependence of the standard deviation of the FDD as a function of the accretion rate. We attribute the increase in spread at higher accretion rates to the increased uncertainty and variance in diameter measurements for thick rings. This is because, for thicker rings, there is a wider range of diameters that can be preferentially ``lit-up'' for different realizations of turbulence, although these models are inconsistent with constraints on variability.

\begin{figure*}
    \includegraphics[width=.9\textwidth]{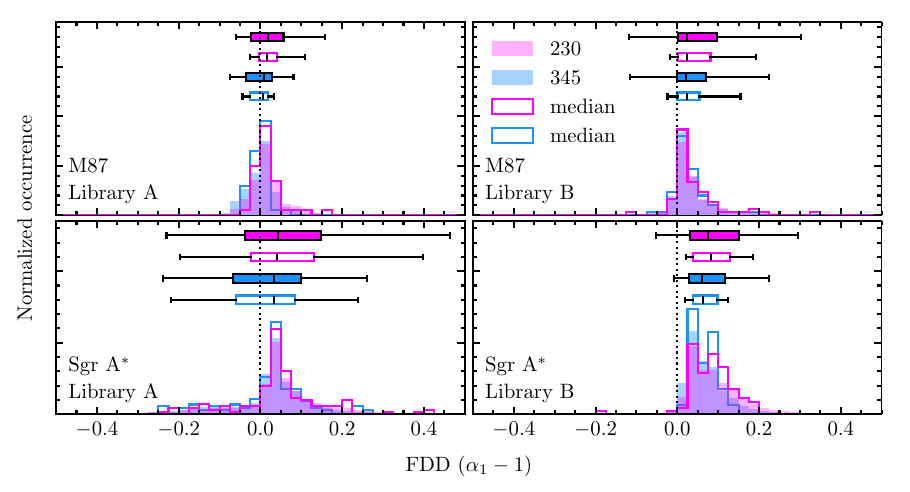}
    \caption{Filled histograms show the distribution of FDD measured from all snapshots across all simulations (identical to Figure~\ref{fig:diam_total}), incorporating both intramodel variability and intermodel spread. The unfilled step histograms show the distribution of per-model median FDDs, obtained by taking the median over time for each simulation. The step histograms therefore isolate the contribution from intermodel uncertainty. The median of the distributions remains largely unchanged, but the $68\%$ and $95\%$ ranges decrease for the step histograms. The remaining width of the step-histogram distributions reflects the intermodel spread, indicating that temporal averaging alone does not eliminate the fundamental uncertainty associated with the unknown accretion model.}
    \label{fig:models_time_avg}
\end{figure*}

We explore the dependence on the $R_{\mathrm{low}}$ parameter in Figure~\ref{fig:Rlow}. The $R_{\rm low}$ parameter controls the ion-to-electron temperature ratio in regions where the plasma $\beta$ is small (e.g., the jet and, in MAD flows, regions close to the event horizon). Increasing $R_{\rm low}$ decreases electron temperatures in low-$\beta$ regions, which results in relatively more emission being produced in the disk as well as an overall increased accretion rate (to match the observed target flux). The median and spread in FDD increase with higher $R_{\mathrm{low}}$, consistent with what one might expect from the changing emission morphology. This effect is particularly prominent for low spin magnitude models. In the future, improved constraints on the allowed model parameter space may lead to considerable improvements to FDD, which may then lead to tighter constraints on mass and/or improved gravitational tests.

\begin{figure*}
    \centering
    \includegraphics[height=.34\textwidth]{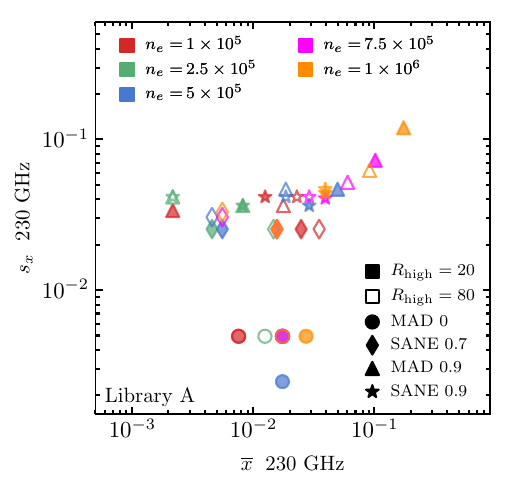}
    \includegraphics[height=.34\textwidth]{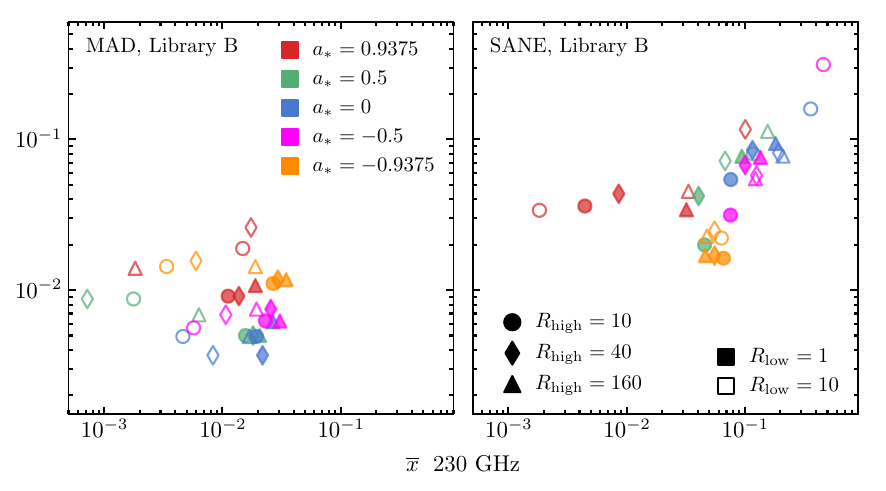}
    \caption{Fractional variability of FDD across models in Libraries A (left) and B (middle, right). Each panel shows $\left| \overline{x} \right|$ plotted against $\left|s_x\right|$ for the FDD time series of each model at $230\,$GHz. Marker shape and color identify the model. Across both libraries, models with smaller mean FDD tend to exhibit higher relative variability, consistent with the expectation that a smaller denominator amplifies the ratio.
    }
    \label{fig:sigma_mu}
\end{figure*}
\subsection{Image-based priors}\label{sec:width}
Beyond using astrophysical constraints, the potential model parameter space could also be constrained through directly observed image characteristics. As an example, we will focus on the width constraint derived by the \texttt{PRIMO} algorithm, which is a dictionary learning based algorithm for EHT image reconstruction that uses GRMHD simulations as a training set \citep{medeiros_2023_PRIMO_M87,medeiros_2023_primo1}. \texttt{PRIMO} fits a linear combination of principal components to the interferometric data directly in the Fourier domain through the Markov chain Monte Carlo (MCMC) algorithm MARCH \citep{psaltis_2022_march}. The width constraint from \texttt{PRIMO} for the 2017 EHT image of M87 is $9.6\pm0.5\,\mu$as \citep{medeiros_2023_PRIMO_M87}. The uncertainty on the ring width was derived by applying \charm to 1750 randomly sampled steps from the MCMC chains. We will use $10.1\,\mu$as as a conservative upper limit on the ring width.

In Figure~\ref{fig:PRIMO}, we compare the distributions of width vs. FDD and shows how the distribution in FDD tightens when only images with ring widths smaller than $10.1\,\mu$as are included. We also quote the median, center $68\%$, and center $95\%$ interval ranges for the width-constrained libraries in Table~\ref{tab:results} (note that the 345~GHz values quoted in the table include only 345~GHz images for which the 230~GHz image is consistent with the constraint). By including only images that are consistent with the 230~GHz \texttt{PRIMO} width constraint, the median, $68\%$, and $95\%$ interval FDD values for Library A (B) decrease by $53\%$, $35\%$, and $48\%$ ($56\%$, $45\%$, and $34\%$) respectively. 

Other image features, such as the observed brightness asymmetry, may be used to constrain the viewing geometry. For optically thin synchrotron emission, higher inclinations, i.e., more ``edge-on'' inclinations, tend to enhance the image asymmetry due to relativistic beaming and Doppler boosting of the approaching side of the flow. The low observed asymmetry of \sgra, therefore, disfavors moderately inclined or high-inclination Keplerian models \citep[$i \gtrsim 10^\circ-20^\circ$, see][but see also \citealt{medeiros_2022_asymmetry} in the context of MAD-like accretion flows]{eht_sgra_5,faggert_2025_sgraasym}.

\cite{eht_sgra_4} considered cuts on the simulation parameter space motivated by features of the interferometric EHT data for \sgra (see Section 3.1 of \citealt{eht_sgra_5} for a similar geometrically motivated set of interferometric constraints). Here, we apply analogous geometric constraints to the \sgra portions of both simulation libraries to assess how such EHT-based priors affect the theoretical uncertainty.
We retain only those images for which the first minimum in Fourier space, in either the vertical or horizontal direction, lies between 2.5 and 3.5 G$\lambda$, and whose median visibility amplitude between 6 and 8 G$\lambda$ in both directions is less than 0.057 of the zero-baseline flux, consistent with the data.
Under these constraints, the median and central 68\% and 95\% ranges of the fractional diameter difference at 230~GHz are 0.095, 0.168, and 0.442 for Library A and 0.046, 0.07, 0.476 for Library B. Comparison with Table~\ref{tab:results} shows that some summary statistics decrease while others increase. This behavior reflects the fact that excluding regions of parameter space can remove models with either small or large intrinsic dispersion.

\section{Observing over multiple epochs}\label{sec:epochs}
We now explore how observations across multiple epochs can reduce the theoretical uncertainty in inferred shadow size measurements by mitigating intramodel (temporal) variability. Because the diameter of the emission ring fluctuates stochastically over time, a single snapshot provides only one realization drawn from a broader population distribution and offers an incomplete estimate of the population-averaged ring diameter. By averaging measurements across many epochs, this intramodel source of uncertainty can be reduced, enabling more robust inference of persistent, spacetime-dependent features such as the black hole shadow size.

Multi-epoch observations do not directly resolve intermodel uncertainty: even if observations constrain the population-averaged diameter to high precision, ambiguity in the true shadow diameter would remain if multiple models with distinct mean diameters were consistent with the observations. At the same time, the structure of the temporal variability itself can be informative. Successive observations are not necessarily independent, and the variability may not be well represented as random draws from a simple stationary process. Rather, the system may linger near a median state for extended periods, punctuated by brief excursions such as hot spots, flares, or other transient phenomena. By studying this variability structure, we can thus determine both the minimum temporal spacing required for observations to be effectively independent and identify time-windowing strategies to select subsamples that may be better behaved.

\subsection{Model uncertainty}

Multi-epoch imaging suppresses intramodel variability by averaging over the stochastic fluctuations of the emission morphology. In practice, however, the accretion flow is not known \emph{a priori}, and although repeated observations sharpen the estimate of the mean ring diameter, they do not identify which model provides the correct description of the source. Because the mapping between the observed ring diameter and the underlying shadow size is generally model dependent, there remains a floor to the precision with which the calibration factor $\alpha_1$ can be inferred. This floor is determined by the dispersion among the models, i.e., the intermodel spread.

\begin{figure*}
    \centering
    \includegraphics[width=.7\textwidth]{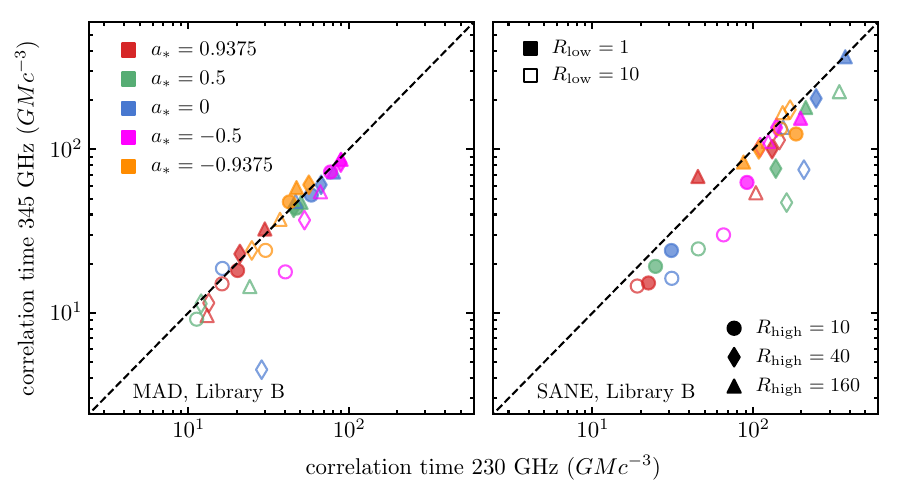}
    \caption{
    Comparison of FDD correlation times at $230\,{\rm GHz}$ and $345\,{\rm GHz}$ for MAD (left) and SANE (right) models from the high-cadence subset of Library B. Each point represents a single simulation, with the marker colors, fills, and shapes denoting spin, $R_{\rm low}$, and $R_{\rm high}$. SANE models also tend to exhibit systematically longer correlation times than MADs. Most models lie below the diagonal line, indicating that they are more correlated at lower frequencies. This reflects the fact that higher-frequency emission typically originates from the more compact regions of the flow, which tend to vary more rapidly. The MAD, $\bhspin = 0$, $R_{\rm low} = 10$, $R_{\rm high} = 40$ outlier occasionally exhibits transient emission close to the event horizon, which biases the inferred ring diameter and results in much shorter correlation times.
    }
    \label{fig:correlations}
\end{figure*}

The effectiveness of temporal averaging, therefore, depends on the relative importance of intermodel spread compared to intramodel variability. When the spread in population-averaged ring diameters across models is small compared to the intrinsic variability within a given model, the total theoretical uncertainty decreases with the number of independent measurements. Additional epochs then yield a corresponding improvement in the constraint on FDD. In contrast, if the differences among models are comparable to or exceed the characteristic intramodel variability, averaging over time cannot eliminate the calibration ambiguity. In this regime, multi-epoch observations serve a dual purpose: they reduce the stochastic uncertainty in diameter estimates and provide leverage to distinguish among competing accretion models.

In Figure~\ref{fig:models_time_avg}, we compare the distribution of FDD measured from all snapshots across all models (filled; incorporating both intra- and intermodel variability) with the distribution of per-model median FDDs (outlined; isolating intermodel spread). The narrowing of the outlined distributions relative to the filled ones demonstrates that averaging over multiple snapshots reduces the effective uncertainty within individual models. However, the intermodel spread remains substantial: the widths of the outlined distributions are comparable to those of the full distributions. Thus, for the full library of models considered here, temporal averaging improves the precision with which the ring diameter can be inferred for a given model, but does not drastically improve the accuracy of the calibration between the measured ring size and the true shadow diameter when the underlying model is unknown.

\subsection{Variability and sample size}

The filled distributions in Figure~\ref{fig:models_time_avg} can be interpreted as the convolution of the intermodel (unfilled) distribution of population-averaged FDDs with a per-model kernel that captures intramodel variability and may differ substantially across the model library. We characterize the width of this kernel by our $s_x$ proxy for the standard deviation of the FDD distribution. Figure~\ref{fig:sigma_mu} shows how this measure of intramodel variability varies, plotted against the corresponding per-model median FDD, $\overline{x}$. Models with broader per-model kernels (larger $s_x$) contribute disproportionately to the width of the combined FDD distribution, and differences in intrinsic variability therefore play a central role in determining how effectively multi-epoch observations reduce uncertainty. If future observations or theoretical considerations can be used to reject models with inconsistent levels of intrinsic variability, the allowable range of $\sigma$ would be correspondingly restricted. Such constraints would modify the effective convolution kernel and could help differentiate between otherwise similar models, indirectly reducing the impact of intermodel uncertainty.

While combining data across multiple epochs suppresses stochastic variability, the improvement depends on how the uncertainty scales with the number of independent observations. For $N$ uncorrelated epochs drawn from a fixed model distribution with standard deviation $\sigma$, the uncertainty decreases as $\mathrm{SEM} = \sigma / \sqrt{N}$, which remains a good approximation even for non-Gaussian distributions with finite variance. In practice, we take $s_x$ as a proxy for the characteristic intramodel scatter. This sets both the amplitude of stochastic variability and the normalization of the $1/\sqrt{N}$ scaling: models with smaller $s_x$ reach a given precision with fewer epochs. Since $s_x$ varies by factors of several across our models, the rate of convergence under multi-epoch averaging differs correspondingly.

\begin{figure*}
    \centering
    \includegraphics[width=\linewidth]{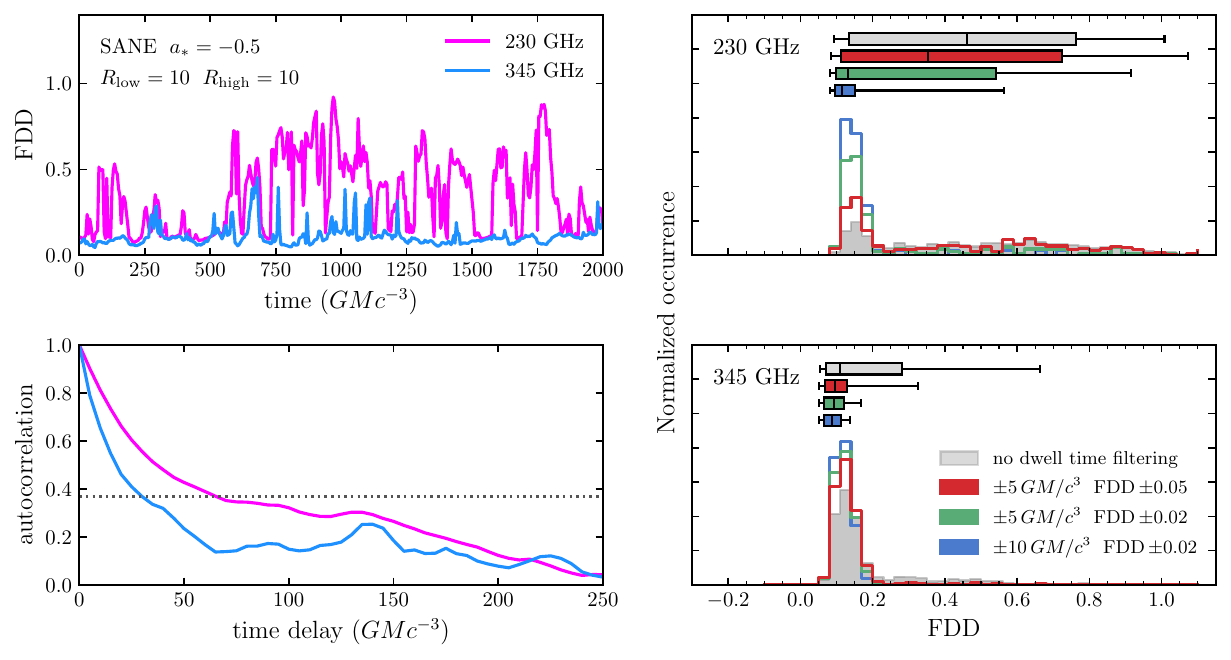}
    \caption{Effect of dwell-time-based filtering for a representative SANE model from Library B (spin $\bhspin=-0.5$, $R_{\rm low}=10$, $R_{\rm high}=10$). Left top: Time series of FDD at $230\,\mathrm{GHz}$ (magenta) and $345\,\mathrm{GHz}$ (blue). Left bottom: Normalized autocorrelation functions for both time series. The $230\,{\rm GHz}$ time series undergoes larger-amplitude excursions and has longer correlation times, while the $345\,{\rm GHz}$ data decorrelates more rapidly. Right panels: Distributions of FDD constructed from the full time series (gray) and from temporally filtered subsamples that retain only intervals persisting within specified FDD thresholds for a minimum dwell time (colored histograms; see legend). Temporally filtered subsamples exhibit substantially narrower distributions, reflecting reduced intramodel variability during extended quiescent intervals.
    }
    \label{fig:dwell_detail}
\end{figure*}

\subsection{Temporal correlations and sample independence}

The reduction of intramodel uncertainty through multi-epoch observations assumes that successive measurements are statistically independent. Consecutive snapshots of a time-variable accretion flow are not guaranteed to be statistically independent, owing to temporal correlations in the flow and to geometric lensing effects that can imprint coherence across time (see \citealt{wong_2024_vlbiechoes} and references therein). The degree to which additional observations provide independent information is therefore controlled by the correlation timescale of the variability. To quantify this, we compute the normalized autocorrelation function of the ring-diameter time series for each model,
\begin{align}
\mathcal{C}\left( \Delta t \right) = N \left< \left(x(t) - \mu\right)\left(x(t+\Delta t) - \mu\right) \right>,
\end{align}
where $x(t)$ is the time series data, $\mu$ is the time-averaged mean, and $N$ is chosen such that $\mathcal{C}(0)=1$. We define the correlation time as the lag $\Delta t$ at which $\mathcal{C}(\Delta t)$ decays to $1/e$ of its peak value and represents the minimum delay before successive observations can be treated as approximately independent.

Figure~\ref{fig:correlations} compares the correlation times measured at $230\,\mathrm{GHz}$ and $345\,\mathrm{GHz}$ for each model in the high-cadence subset of Library~B. Each point corresponds to a single simulation, with marker color, fill, and shape denoting the black hole spin and plasma parameters. Most models lie below the one-to-one line, implying systematically shorter correlation times at $345\,\mathrm{GHz}$. This is sensible given that the higher-frequency emission arises from more compact regions near the event horizon where dynamical timescales are shorter and fluctuations decorrelate more rapidly: the orbital period at $r = 3\,M$ is $25-40\,M$ for Keplerian velocities, although true dynamical periods are likely longer, especially for MAD flows, which are sub-Keplerian \citep{dhruv_2025_survey}. MAD models tend to cluster tightly and exhibit shorter correlation times ($10-100\,M$), whereas SANE models show greater scatter and systematically longer correlations ($15 - 400\,M$). These longer correlation times reflect the increased contribution of extended disk emission and intermittent large-amplitude excursions characteristic of SANE flows. The clear outlier is a MAD model with $\bhspin = 0$, $R_{\mathrm{low}} = 10$, and $R_{\mathrm{high}} = 40$. In this model, transient horizon-scale emission biases individual snapshots, leading to much shorter correlation times.

\subsection{Dwell-time filtering}

As discussed above, successive measurements cannot be assumed to be independent. Excursions away from the typical ring diameter typically occur as temporally coherent episodes embedded within longer intervals of stability, rather than isolated, uncorrelated fluctuations. This temporal structure implies that the variability contains information beyond its overall dispersion. Treating all snapshots as exchangeable realizations risks erasing physically meaningful organization in the time series. To make this explicit, we examine the dwell-time statistics of the ring-diameter time series. The dwell time measures how long the system remains in the vicinity of a given diameter before undergoing a significant excursion. In this sense it characterizes temporal persistence as a function of diameter, rather than compressing the behavior into a single global correlation timescale.

Dwell times separate flows that wander continuously through a range of diameters from those that remain near a quasi-equilibrium state and only occasionally undergo coherent excursions. This distinction is observationally useful, since intermittent excursions may dominate the tails of the distribution even though they are not representative of the typical state of the system. Because these episodes are coherent in time, they can be identified through local temporal information alone without prior knowledge of the population distribution.

Figure~\ref{fig:dwell_detail} illustrates this behavior for a representative SANE model from the high-cadence subset of Library~B. The left panels display the FDD time series and corresponding autocorrelation functions at $230\,\mathrm{GHz}$ and $345\,\mathrm{GHz}$. At $230\,\mathrm{GHz}$ the diameter exhibits larger-amplitude excursions and longer-lived correlations. At $345\,\mathrm{GHz}$ the variability is more compact and decorrelates on shorter timescales. The right panels compare the FDD distribution from the full time series with distributions derived from temporally filtered subsamples. The filtering retains only intervals that remain within specified FDD thresholds for at least a minimum dwell time.

The gain from dwell-time filtering does not come from excising isolated outliers  but instead reflects the selection of extended intervals during which the morphology remains comparatively stable. Such intervals contribute a smaller intramodel variance and, therefore, converge more rapidly under averaging. Dwell-time filtering is thus complementary to uniform multi-epoch sampling, and quantifying dwell-time behavior distinguishes flow states that are persistently stable from those dominated by intermittent activity. This distinction provides an additional diagnostic for model comparison and clarifies how observing and analysis strategies can benefit from explicitly accounting for temporal persistence.

\begin{figure*}
    \centering
    \includegraphics[width=.9\linewidth]{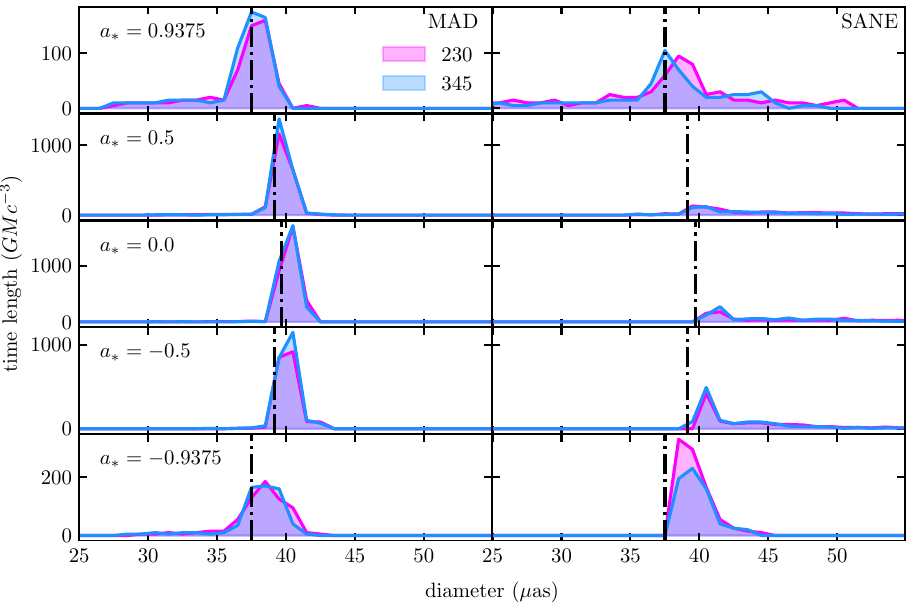}
    \caption{
    Stability of the inferred ring diameter as a function of diameter, accretion state, black hole spin, and observing frequency. Each panel shows the longest contiguous time interval over which the inferred ring diameter remains within $\pm1 \%$ of its value at the start of the interval, binned in $1\,\mu{\rm as}$ diameter intervals. Results are computed from the high-cadence subset of Library~B, with MAD models shown in the left column and SANE models in the right. Rows correspond to black hole spin. Within each panel, results at $230\,{\rm GHz}$ (magenta) and $345\,{\rm GHz}$ (blue) are shown. Larger values indicate longer temporal persistence of the inferred diameter. 
    }
    \label{fig:dwell_library}
\end{figure*}

We summarize in Figure~\ref{fig:dwell_library} dwell-time coherence across the models in the high-cadence subsample of Library~B by showing the maximum contiguous interval over which the inferred diameter remains within a fixed fractional tolerance of its initial value. Across all but the most retrograde-spin models in the library, MAD flows generally exhibit longer and more uniform stability intervals than SANE flows, consistent with their more compact and persistent emission morphology. In SANE models, stable intervals are still present but show a stronger dependence on diameter and spin, indicating greater sensitivity to flow state and viewing geometry. 

This statistic provides an upper bound on the temporal persistence at each diameter. For compactness, we display the maximum dwell time as a function of diameter, although the same qualitative trends are present in the median. Longer maximum dwell times imply that temporally filtered subsamples with reduced variance can be constructed with comparatively modest loss of data. Shorter stability times indicate either frequent or rapidly evolving excursions, so achieving a similar variance reduction would require more restrictive filtering and hence a smaller effective sample. Nevertheless, the presence of pronounced peaks at particular diameters in most models indicates that the flow often resides for extended intervals in relatively stable emission configurations. This structure suggests that dwell-time-based analyses can substantially reduce intramodel scatter even when overall variability is appreciable.

The choice of dwell-time filter parameters can be optimized for different analysis goals. One approach is to anchor the filter directly to the temporal properties of the data and use the observed time series to identify characteristic coherence intervals. In this way, dwell-time filtering could help constrain the intermodel spread by restricting the ensemble to models with similar temporal variability. Whether a given filtering window is sufficient is therefore an empirical question to be decided by the data, rather than imposed \emph{a priori} from the models alone. A more detailed discussion of observational implementation and cadence requirements is deferred to Section~\ref{sec:discussion}.
 
\section{Discussion}\label{sec:discussion}

We have considered the theoretical relation and its uncertainty between the diameter of the black hole shadow, which is fixed by the spacetime geometry, and the diameter of the bright ring of emission in simulated images. We quantified this relation using the fractional diameter difference (FDD; Section~\ref{sec:multi-wave}). We characterized its dependence on model parameters and on the observing frequency. 

Our results suggest a hierarchy of strategies for reducing the theoretical uncertainty in shadow-size inference. The most robust improvement will likely come from higher-frequency imaging: $345\,{\rm GHz}$ images have narrower rings, smaller median FDDs, and narrower FDD distributions compared to the $230\,{\rm GHz}$ images. This improvement is comparatively model agnostic in that it is not strongly sensitive to the source characteristics or theoretical modeling approaches. At the same time, our analysis indicates that the largest gains will likely come from observations that restrict the allowed model space. This is because while the intramodel variability can be averaged away, the intermodel uncertainty floor remains as the dominant uncertainty.

Image-domain constraints provide one route to reducing the intermodel floor. Because constraints on ring width, diameter stability, and asymmetry can restrict the set of viable models, they can directly reduce the intermodel scatter. The PRIMO ring-width constraint is one example: applying it substantially improves prospects for inferring the shadow diameter of M87 (i.e., the median decreases by over $50\%$, while the $68\%$ and $95\%$ ranges decrease by about $34\%$ to $48\%$ for both libraries at 230~GHz). This result demonstrates that improving image precision can play an important role in decreasing the theoretical uncertainty in shadow-size constraints. A second route is through model-aware constraints. Polarimetric and multiwavelength data may be able to distinguish MAD from SANE flows as well as constrain the accretion rate and the system geometry, all of which can improve the calibration of this theoretical relation. For example, we also demonstrated that when the accretion rate is fixed by requiring that the simulated flux matches the observed source flux, MAD models exhibit considerably smaller theoretical uncertainty, with the central $68\%$ and $95\%$ range FDD bounds smaller by $\sim 70\%$ for M87.

Finally, we find that multi-epoch observations can also serve as a highly valuable tool for improving the shadow-size calibration. Combining independent measurements decreases the intramodel variance, which reduces the uncertainty arising from stochastic variability. However, other approaches that leverage locality in time, such as dwell-time analyses, can identify intervals of stable ring geometry and provide additional model-selection constraints. As shown in Figure~\ref{fig:dwell_detail}, coherence windows of order $\pm 10\,M$ may already be informative, provided persistence can be established over the full interval. The practical viability of this approach, therefore, depends on cadence and continuity: resolving such windows requires sampling on timescales of order minutes for \sgra and days for M87 (e.g., a full $\pm 10\,M$ interval corresponds to $\approx 7$ minutes for \sgra and a week for M87).

Although we have primarily considered strategies for improving FDD constraints in isolation, future EHT analyses may be able to combine multiple constraints simultaneously. As a concrete example, suppose that the source is well represented by the MAD subset of M87 Library B. At 230~GHz, the 95\% range for the full image distribution is $0.129$. Averaging over multiple epochs and thereby reducing the stochastic intramodel contribution would then suppress the tails of the distribution and decrease this range to $0.05$.

It is important to note, however, that the behavior of combined constraints need not be simply or linearly related to the gains from each constraint applied individually: two cuts that each narrow the marginal FDD distribution can yield a broader distribution when applied together if the second cut reshapes the subset selected by the first. For example, although the \texttt{PRIMO} width and MAD constraints each reduce the FDD spread individually, combining the two constraints results in a distribution with a lower median, the same central 68\% interval, and a central 95\% interval that is larger by 20\%. The resulting FDD uncertainty is, thus, governed by the overlap geometry of the surviving model space and not simply by the tightening power of each constraint alone. Even in cases where the total spread increases, however, the allowed model space is reduced, which means that the additional independent constraints should decrease the intermodel uncertainty. With sufficiently many epochs, the remaining intramodel contribution is likewise expected to decrease.

Throughout this work, we excluded four models from Library A, because these models do not produce images that exhibit a central brightness depression. As a result, they are inconsistent with both M87 and \sgra EHT results and are not relevant for calibrating physics constraints from those observations. Nevertheless, as a deliberately permissive robustness test, we repeat the analysis by including those models that did not satisfy the physical-relevance criteria adopted in our fiducial analysis. In such cases without a clear brightness depression, the diameter of the filled image was used in the analysis, as calculated by the characterization algorithm \charm. Including these models results in only minor changes in the summary statistics, both because the image diameters are not significantly different from ring diameters and because these models comprise a small fraction of the full ensemble. Specifically, for the case of \sgra Library A, at $230\,{\rm GHz}$, the median changes from $0.043$ to $0.044$ and the central $68\%$ and $95\%$ intervals broaden modestly from $0.187$ to $0.201$ and $0.694$ to $0.726$.

Although our analysis is restricted to the Kerr spacetime, the statistical framework is not Kerr specific. The distinction between intramodel variability and intermodel spread reflects generic properties of time-variable accretion flows and of the shadow-size calibration problem. The same feature-extraction and calibration pipeline can therefore be applied to non-Kerr image libraries to test whether departures from Kerr modify the ring-shadow mapping or its uncertainty budget. Our results thus provide both a Kerr baseline and a framework for extending shadow-size inference to alternative spacetimes.

Our analysis only considered the $\alpha_1$ parameter, which quantifies the offset between the measured ring diameter and the shadow diameter under the assumption of perfect knowledge of the image and its astrometric center. In this sense, $\alpha_1$ primarily reflects the plasma astrophysics that governs the relationship between the emission morphology and the underlying spacetime geometry. For applications to mass measurements or tests of gravity with the EHT, these results must be combined with constraints on $\alpha_2$. The latter encodes instrumental and methodological effects, including measurement noise, calibration uncertainties, and biases introduced by image reconstruction. A complete error budget therefore requires joint consideration of both terms.

The methodological advances identified in this work, including multifrequency imaging, improved geometric constraints, model selection, and time-domain information, can reduce both the stochastic and intermodel components of the theoretical uncertainty. These efforts will coincide with substantial hardware improvements, such as dual band and wider band receivers, some of which are already in place or in progress. The improved calibration of the ring–shadow mapping, tighter astrophysical constraints on the model space, and enhanced instrumental performance lead to stronger mass determinations and more stringent tests of gravity in the years to come.

\begin{acknowledgements}

The authors thank S.~Cheng, D.~Psaltis, M.~Strauss, and P.~Tiede for useful discussions. L.M.\ gratefully acknowledges support from a NASA Hubble Fellowship Program, Einstein Fellowship under award number HST-HF2-51539.001-A and NSF AST-2407810.\;G.N.W.~was supported by the Taplin Fellowship and the Princeton Gravity Initiative. 
Some of the ray tracing and radiative transfer calculations were performed with the {\tt ElGato} (funded by NSF award 1228509) and {\tt Ocelote} clusters at the University of Arizona and the Texas Advanced Computing Center (TACC) at The University of Texas at Austin.
\end{acknowledgements}

\begin{appendix}
\section{Comparison of models with \texorpdfstring{$R_{\rm high} = 1$}{R_high = 1}}\label{ap:Rhigh}

The GRMHD simulations evolve the total internal energy of the fluid, but the observed synchrotron emission is produced by the electron population. In order to estimate the electron temperature from the total internal energy, we use the $R_{\rm low}$-$R_{\rm high}$ electron thermodynamics prescription described in \citet{moscibrodzka_2016_rhigh}, which models the ion-to-electron temperature ratio ($R=T_i/T_e$) as a smoothly varying function between two asymptotic values in the limits of low and high plasma $\beta = P_{\rm gas}/P_{\rm mag}$ (see equation \ref{eq:Rhigh}). This model is motivated by the physics of collisionless or weakly collisional accretion flows (see the discussion in \citealt{chan_2015_gray} relating $R$ to the collisional timescales), in which Coulomb coupling is inefficient and ions and electrons do not thermally equilibrate. In such plasmas, the partition of dissipated energy depends on the local plasma conditions, particularly plasma-$\beta$, and the heating and cooling channels for ions and electrons differ \citep[e.g.,][]{shapiro_1976_two_temp}.

The precise details of plasma heating mechanisms are not fully understood, but the relevant processes include magnetic reconnection, whistler, ion-Landau damping, and compressive cascades, as well as general scattering and instabilities in the context of velocity-space anisotropies. The net efficiency with which energy is partitioned between ions and electrons depends sensitively on the local magnetization and plasma conditions. In low-$\beta$ plasmas, which have strong magnetic fields and large Alfv\'{e}n speeds, dissipation tends to heat electrons efficiently alongside the ions. In contrast, the high-$\beta$ plasmas deeper within the accretion disk are weakly magnetized, and energy flowing along turbulent cascades reaches the ion kinetic scale first, so that ions are heated more efficiently while electrons remain relatively cooler and radiate via synchrotron and inverse Compton processes. Collectively, these mechanisms lead to an environment in which the temperature ratio at low $\beta$, i.e., $R_{\rm low}$ may be order unity while $R_{\rm high}$ at high $\beta$ is typically large (see \citealt{satapathy_2023_electronheating} and references therein).

Although $R_{\rm high}$ is greater than unity in the accretion flows observed by the EHT, the original analyses considered a range of values for the parameter in order to explore and bracket the uncertainties in the heating mechanisms \citep{eht_m87_5,eht_m87_8,eht_sgra_5,eht_sgra_8}. Qualitatively, models with low $R_{\rm high}$ have hotter electrons in their disks and therefore produce more emission at large radii. This extended emission morphology is particularly pronounced in (especially retrograde) SANE flows with small $R_{\rm high}$ (see Panel C of Figure~\ref{fig:charm}). When the \charm algorithm is applied to these models, it picks out the bright, transient emission features at large radii, which biases the ring diameter measurement to be larger, and the models with the highest FDD in Library B are consistently SANE models with $\bhspin \le 0.5$ and $R_{\mathrm{high}}=1$. Since the lowest spin value for SANE simulations in Library A is 0.7, we do not see similar behavior in Library A. 

Table~\ref{tab:Rhigh1} compares the summary statistics for the fiducial libraries used in the main analysis with those obtained when models with $R_{\mathrm{high}}=1$ are included. Including or excluding the $R_{\mathrm{high}}=1$ models has little effect on the statistics for Library~A (both for M87 and \sgra). However, since Library~B includes retrograde SANE models, the median, central $68\%$ interval, and central $95\%$ interval FDD measurements increase when the $R_{\mathrm{high}}=1$ models are included. Including $R_{\mathrm{high}}=1$ has a relatively small effect on the ring width for both libraries. Models with $R_{\mathrm{high}}=1$ typically yield large FDD, but they are disfavored both by theoretical expectations for collisionless accretion flows and by EHT constraints on ring width and other interferometric observables. They are therefore unlikely to provide realistic descriptions of the M87 and \sgra accretion systems and are excluded from the primary analysis \citep{eht_m87_5,eht_m87_8,eht_m87_9,eht_sgra_5,eht_sgra_8}. 

\begin{table*}[t]
\caption{Comparison of Feature Extraction results when $R_{\mathrm{high}}=1$ is included}
\label{tab:Rhigh1}
\begin{ruledtabular}
\begin{tabular}{lclllllrrr}
\vspace{-.7em}\\
Library & frequency  & &    & FDD &       &   &   & Width    &  \\ 
       &(GHz) & &   median  &  68\% &  95\% &  & median  & 68\% & 95\%\\ 
\vspace{-.5em}\\
\colrule
\vspace{-.5em}\\
A -- M87 & 230 &  & 0.017 & 0.079 & 0.218 & & 10.96 & 6.07 & 12.90 \\ 
        & 345 &  & 0.008 & 0.064 & 0.156 & & 9.98 & 3.68 & 8.41 \\ 
&&&&&&&&&\\ 
B -- M87 & 230 &  & 0.025 & 0.095 & 0.419 & & 10.54 & 3.28 & 6.84 \\ 
        & 345 &  & 0.021 & 0.073 & 0.341 & & 9.32 & 2.90 & 5.76 \\ 
 \\ 
A -- \sgra & 230 &  & 0.043 & 0.187 & 0.694 & & 15.81 & 11.76 & 24.10 \\ 
              & 345 &  & 0.033 & 0.167 & 0.499 & & 14.45 & 9.29 & 18.35 \\ 
&&&&&&&&&\\ 
B -- \sgra & 230 &  & 0.076 & 0.120 & 0.349 & & 15.71 & 5.81 & 12.22 \\ 
              & 345 &  & 0.060 & 0.088 & 0.231 & & 13.33 & 4.28 & 9.09 \\ 
 \vspace{-.5em}\\ \colrule \vspace{-.5em}\\ 
A -- M87 with $R_{\mathrm{high=1}}$& 230 &  & 0.017 & 0.068 & 0.203 & & 11.61 & 7.52 & 15.15 \\ 
        & 345 &  & 0.008 & 0.071 & 0.152 & & 10.43 & 4.53 & 11.12 \\ 
&&&&&&&&&\\ 
B -- M87 with $R_{\mathrm{high=1}}$& 230 &  & 0.027 & 0.141 & 0.960 & & 10.64 & 3.62 & 7.61 \\ 
        & 345 &  & 0.021 & 0.107 & 0.781 & & 9.53 & 3.24 & 6.88 \\ 
 \\ 
A -- \sgra with $R_{\mathrm{high=1}}$& 230 &  & 0.048 & 0.166 & 0.606 & & 16.59 & 11.69 & 24.54 \\ 
        & 345 &  & 0.040 & 0.154 & 0.465 & & 15.17 & 9.51 & 18.87 \\ 
&&&&&&&&&\\ 
B -- M87 with $R_{\mathrm{high=1}}$& 230 &  & 0.084 & 0.139 & 0.432 & & 15.82 & 5.81 & 12.69 \\ 
        & 345 &  & 0.068 & 0.114 & 0.360 & & 13.70 & 4.64 & 9.81 \\ 
 \\ 
\vspace{-.7em}\\
\end{tabular}
\vspace{1em}
\end{ruledtabular}
\end{table*}

\end{appendix}

\bibliography{medeiroswong}

@article{bisnovatyi_1974_madstar,
  title = {The {{Accretion}} of {{Matter}} by a {{Collapsing Star}} in the {{Presence}} of a {{Magnetic Field}}},
  author = {{Bisnovatyi-Kogan}, G. S. and Ruzmaikin, A. A.},
  year = {1974},
  month = may,
  journal = {Astrophysics and Space Science},
  volume = {28},
  pages = {45--59},
  publisher = {Springer},
  issn = {0004-640X},
  doi = {10.1007/BF00642237},
  urldate = {2024-05-02}
}

@article{bollimpalli_2024_tiltedqpo,
  title = {Truncated, Tilted Discs as a Possible Source of {{Quasi-Periodic Oscillations}}},
  author = {Bollimpalli, D. A. and Fragile, P. C. and Dewberry, J. W. and Klu{\'z}niak, W.},
  year = {2024},
  month = feb,
  journal = {Monthly Notices of the Royal Astronomical Society},
  volume = {528},
  pages = {1142--1157},
  publisher = {OUP},
  issn = {0035-8711},
  doi = {10.1093/mnras/stad3975},
  urldate = {2024-05-15}
}

@article{butterworth_1930_filter,
  title = {On the {{Theory}} of {{Filter Amplifiers}}},
  author = {Butterworth, Stephen},
  year = {1930},
  month = oct,
  journal = {Experimental Wireless \& the Wireless Engineer},
  volume = {7},
  pages = {536--541}
}

@article{chan_2013_gray,
  title = {{{GRay}}: {{A Massively Parallel GPU-based Code}} for {{Ray Tracing}} in {{Relativistic Spacetimes}}},
  shorttitle = {{{GRay}}},
  author = {Chan, Chi-kwan and Psaltis, Dimitrios and {\"O}zel, Feryal},
  year = {2013},
  month = nov,
  journal = {The Astrophysical Journal},
  volume = {777},
  pages = {13},
  publisher = {IOP},
  issn = {0004-637X},
  doi = {10.1088/0004-637X/777/1/13},
  urldate = {2024-05-24}
}

@article{chatterjee_2020_tilted,
  title = {Observational Signatures of Disc and Jet Misalignment in Images of Accreting Black Holes},
  author = {Chatterjee, K. and Younsi, Z. and Liska, M. and Tchekhovskoy, A. and Markoff, S. B. and Yoon, D. and {van Eijnatten}, D. and Hesp, C. and Ingram, A. and {van der Klis}, M. B. M.},
  year = {2020},
  month = nov,
  journal = {Monthly Notices of the Royal Astronomical Society},
  volume = {499},
  pages = {362--378},
  publisher = {OUP},
  issn = {0035-8711},
  doi = {10.1093/mnras/staa2718},
  urldate = {2024-05-15}
}

@ARTICLE{chael_2025_twotempm87,
       author = {{Chael}, Andrew},
        title = "{Survey of radiative, two-temperature magnetically arrested simulations of the black hole M87* I: turbulent electron heating}",
      journal = {\mnras},
         year = 2025,
        month = mar,
       volume = {537},
       number = {3},
        pages = {2496-2515},
          doi = {10.1093/mnras/staf200},
archivePrefix = {arXiv},
       eprint = {2501.12448},
 primaryClass = {astro-ph.HE},
       adsurl = {https://ui.adsabs.harvard.edu/abs/2025MNRAS.537.2496C}
}

@ARTICLE{salas_2025_twotempsgra,
       author = {{Salas}, L.~D.~S. and {Liska}, M.~T.~P. and {Markoff}, S.~B. and {Chatterjee}, K. and {Musoke}, G. and {Porth}, O. and {Ripperda}, B. and {Yoon}, D. and {Mulaudzi}, W.},
        title = "{Two-temperature treatments in magnetically arrested disc GRMHD simulations more accurately predict light curves of Sagittarius A*}",
      journal = {\mnras},
         year = 2025,
        month = apr,
       volume = {538},
       number = {2},
        pages = {698-710},
          doi = {10.1093/mnras/staf240},
archivePrefix = {arXiv},
       eprint = {2411.09556},
 primaryClass = {astro-ph.HE},
       adsurl = {https://ui.adsabs.harvard.edu/abs/2025MNRAS.538..698S}
}

@ARTICLE{dexter_2020_sgraelectrons,
       author = {{Dexter}, J. and {Jim{\'e}nez-Rosales}, A. and {Ressler}, S.~M. and {Tchekhovskoy}, A. and {Baub{\"o}ck}, M. and {de Zeeuw}, P.~T. and {Eisenhauer}, F. and {von Fellenberg}, S. and {Gao}, F. and {Genzel}, R. and {Gillessen}, S. and {Habibi}, M. and {Ott}, T. and {Stadler}, J. and {Straub}, O. and {Widmann}, F.},
        title = "{A parameter survey of Sgr A* radiative models from GRMHD simulations with self-consistent electron heating}",
      journal = {\mnras},
         year = 2020,
        month = may,
       volume = {494},
       number = {3},
        pages = {4168-4186},
          doi = {10.1093/mnras/staa922},
archivePrefix = {arXiv},
       eprint = {2004.00019},
 primaryClass = {astro-ph.HE},
       adsurl = {https://ui.adsabs.harvard.edu/abs/2020MNRAS.494.4168D}
}

@article{dibi2012_radiation,
  title = {General Relativistic Magnetohydrodynamic Simulations of Accretion on to {{Sgr A}}*: How Important Are Radiative Losses?},
  shorttitle = {General Relativistic Magnetohydrodynamic Simulations of Accretion on to {{Sgr A}}*},
  author = {Dibi, S. and Drappeau, S. and Fragile, P. C. and Markoff, S. and Dexter, J.},
  year = {2012},
  month = nov,
  journal = {Monthly Notices of the Royal Astronomical Society},
  volume = {426},
  pages = {1928--1939},
  publisher = {OUP},
  issn = {0035-8711},
  doi = {10.1111/j.1365-2966.2012.21857.x},
  urldate = {2024-05-01}
}

@article{eht_m87_1,
  title = {First {{M87 Event Horizon Telescope Results}}. {{I}}. {{The Shadow}} of the {{Supermassive Black Hole}}},
  author = {{Event Horizon Telescope Collaboration} and Akiyama, Kazunori and Alberdi, Antxon and Alef, Walter and Asada, Keiichi and Azulay, Rebecca and Baczko, Anne-Kathrin and Ball, David and Balokovi{\'c}, Mislav and Barrett, John and Bintley, Dan and Blackburn, Lindy and Boland, Wilfred and Bouman, Katherine L. and Bower, Geoffrey C. and Bremer, Michael and Brinkerink, Christiaan D. and Brissenden, Roger and Britzen, Silke and Broderick, Avery E. and Broguiere, Dominique and Bronzwaer, Thomas and Byun, Do-Young and Carlstrom, John E. and Chael, Andrew and Chan, Chi-kwan and Chatterjee, Shami and Chatterjee, Koushik and Chen, Ming-Tang and Chen, Yongjun and Cho, Ilje and Christian, Pierre and Conway, John E. and Cordes, James M. and Crew, Geoffrey B. and Cui, Yuzhu and Davelaar, Jordy and De Laurentis, Mariafelicia and Deane, Roger and Dempsey, Jessica and Desvignes, Gregory and Dexter, Jason and Doeleman, Sheperd S. and Eatough, Ralph P. and Falcke, Heino and Fish, Vincent L. and Fomalont, Ed and {Fraga-Encinas}, Raquel and Freeman, William T. and Friberg, Per and Fromm, Christian M. and G{\'o}mez, Jos{\'e} L. and Galison, Peter and Gammie, Charles F. and Garc{\'i}a, Roberto and Gentaz, Olivier and Georgiev, Boris and Goddi, Ciriaco and Gold, Roman and Gu, Minfeng and Gurwell, Mark and Hada, Kazuhiro and Hecht, Michael H. and Hesper, Ronald and Ho, Luis C. and Ho, Paul and Honma, Mareki and Huang, Chih-Wei L. and Huang, Lei and Hughes, David H. and Ikeda, Shiro and Inoue, Makoto and Issaoun, Sara and James, David J. and Jannuzi, Buell T. and Janssen, Michael and Jeter, Britton and Jiang, Wu and Johnson, Michael D. and Jorstad, Svetlana and Jung, Taehyun and Karami, Mansour and Karuppusamy, Ramesh and Kawashima, Tomohisa and Keating, Garrett K. and Kettenis, Mark and Kim, Jae-Young and Kim, Junhan and Kim, Jongsoo and Kino, Motoki and Koay, Jun Yi and Koch, Patrick M. and Koyama, Shoko and Kramer, Michael and Kramer, Carsten and Krichbaum, Thomas P. and Kuo, Cheng-Yu and Lauer, Tod R. and Lee, Sang-Sung and Li, Yan-Rong and Li, Zhiyuan and Lindqvist, Michael and Liu, Kuo and Liuzzo, Elisabetta and Lo, Wen-Ping and Lobanov, Andrei P. and Loinard, Laurent and Lonsdale, Colin and Lu, Ru-Sen and MacDonald, Nicholas R. and Mao, Jirong and Markoff, Sera and Marrone, Daniel P. and Marscher, Alan P. and {Mart{\'i}-Vidal}, Iv{\'a}n and Matsushita, Satoki and Matthews, Lynn D. and Medeiros, Lia and Menten, Karl M. and Mizuno, Yosuke and Mizuno, Izumi and Moran, James M. and Moriyama, Kotaro and Moscibrodzka, Monika and M{\"u}ller, Cornelia and Nagai, Hiroshi and Nagar, Neil M. and Nakamura, Masanori and Narayan, Ramesh and Narayanan, Gopal and Natarajan, Iniyan and Neri, Roberto and Ni, Chunchong and Noutsos, Aristeidis and Okino, Hiroki and Olivares, H{\'e}ctor and {Ortiz-Le{\'o}n}, Gisela N. and Oyama, Tomoaki and {\"O}zel, Feryal and Palumbo, Daniel C. M. and Patel, Nimesh and Pen, Ue-Li and Pesce, Dominic W. and Pi{\'e}tu, Vincent and Plambeck, Richard and PopStefanija, Aleksandar and Porth, Oliver and Prather, Cora and {Preciado-L{\'o}pez}, Jorge A. and Psaltis, Dimitrios and Pu, Hung-Yi and Ramakrishnan, Venkatessh and Rao, Ramprasad and Rawlings, Mark G. and Raymond, Alexander W. and Rezzolla, Luciano and Ripperda, Bart and Roelofs, Freek and Rogers, Alan and Ros, Eduardo and Rose, Mel and Roshanineshat, Arash and Rottmann, Helge and Roy, Alan L. and Ruszczyk, Chet and Ryan, Benjamin R. and Rygl, Kazi L. J. and S{\'a}nchez, Salvador and {S{\'a}nchez-Arguelles}, David and Sasada, Mahito and Savolainen, Tuomas and Schloerb, F. Peter and Schuster, Karl-Friedrich and Shao, Lijing and Shen, Zhiqiang and Small, Des and Sohn, Bong Won and SooHoo, Jason and Tazaki, Fumie and Tiede, Paul and Tilanus, Remo P. J. and Titus, Michael and Toma, Kenji and Torne, Pablo and Trent, Tyler and Trippe, Sascha and Tsuda, Shuichiro and {van Bemmel}, Ilse and {van Langevelde}, Huib Jan and {van Rossum}, Daniel R. and Wagner, Jan and Wardle, John and Weintroub, Jonathan and Wex, Norbert and Wharton, Robert and Wielgus, Maciek and Wong, George N. and Wu, Qingwen and Young, Ken and Young, Andr{\'e} and Younsi, Ziri and Yuan, Feng and Yuan, Ye-Fei and Zensus, J. Anton and Zhao, Guangyao and Zhao, Shan-Shan and Zhu, Ziyan and Algaba, Juan-Carlos and Allardi, Alexander and Amestica, Rodrigo and Anczarski, Jadyn and Bach, Uwe and Baganoff, Frederick K. and Beaudoin, Christopher and Benson, Bradford A. and Berthold, Ryan and Blanchard, Jay M. and Blundell, Ray and Bustamente, Sandra and Cappallo, Roger and {Castillo-Dom{\'i}nguez}, Edgar and Chang, Chih-Cheng and Chang, Shu-Hao and Chang, Song-Chu and Chen, Chung-Chen and Chilson, Ryan and Chuter, Tim C. and C{\'o}rdova Rosado, Rodrigo and Coulson, Iain M. and Crawford, Thomas M. and Crowley, Joseph and David, John and Derome, Mark and Dexter, Matthew and Dornbusch, Sven and Dudevoir, Kevin A. and Dzib, Sergio A. and Eckart, Andreas and Eckert, Chris and Erickson, Neal R. and Everett, Wendeline B. and Faber, Aaron and Farah, Joseph R. and Fath, Vernon and Folkers, Thomas W. and Forbes, David C. and Freund, Robert and {G{\'o}mez-Ruiz}, Arturo I. and Gale, David M. and Gao, Feng and Geertsema, Gertie and Graham, David A. and Greer, Christopher H. and Grosslein, Ronald and Gueth, Fr{\'e}d{\'e}ric and Haggard, Daryl and Halverson, Nils W. and Han, Chih-Chiang and Han, Kuo-Chang and Hao, Jinchi and Hasegawa, Yutaka and Henning, Jason W. and {Hern{\'a}ndez-G{\'o}mez}, Antonio and {Herrero-Illana}, Rub{\'e}n and Heyminck, Stefan and Hirota, Akihiko and Hoge, James and Huang, Yau-De and Impellizzeri, C. M. Violette and Jiang, Homin and Kamble, Atish and Keisler, Ryan and Kimura, Kimihiro and Kono, Yusuke and Kubo, Derek and Kuroda, John and Lacasse, Richard and Laing, Robert A. and Leitch, Erik M. and Li, Chao-Te and Lin, Lupin C. -C. and Liu, Ching-Tang and Liu, Kuan-Yu and Lu, Li-Ming and Marson, Ralph G. and {Martin-Cocher}, Pierre L. and Massingill, Kyle D. and Matulonis, Callie and McColl, Martin P. and McWhirter, Stephen R. and Messias, Hugo and {Meyer-Zhao}, Zheng and Michalik, Daniel and Monta{\~n}a, Alfredo and Montgomerie, William and {Mora-Klein}, Matias and Muders, Dirk and Nadolski, Andrew and Navarro, Santiago and Neilsen, Joseph and Nguyen, Chi H. and Nishioka, Hiroaki and Norton, Timothy and Nowak, Michael A. and Nystrom, George and Ogawa, Hideo and Oshiro, Peter and Oyama, Tomoaki and Parsons, Harriet and Paine, Scott N. and Pe{\~n}alver, Juan and Phillips, Neil M. and Poirier, Michael and Pradel, Nicolas and Primiani, Rurik A. and Raffin, Philippe A. and Rahlin, Alexandra S. and Reiland, George and Risacher, Christopher and Ruiz, Ignacio and {S{\'a}ez-Mada{\'i}n}, Alejandro F. and Sassella, Remi and Schellart, Pim and Shaw, Paul and Silva, Kevin M. and Shiokawa, Hotaka and Smith, David R. and Snow, William and Souccar, Kamal and Sousa, Don and Sridharan, T. K. and Srinivasan, Ranjani and Stahm, William and Stark, Anthony A. and Story, Kyle and Timmer, Sjoerd T. and Vertatschitsch, Laura and Walther, Craig and Wei, Ta-Shun and Whitehorn, Nathan and Whitney, Alan R. and Woody, David P. and Wouterloot, Jan G. A. and Wright, Melvin and Yamaguchi, Paul and Yu, Chen-Yu and Zeballos, Milagros and Zhang, Shuo and Ziurys, Lucy},
  year = {2019},
  month = apr,
  journal = {The Astrophysical Journal},
  volume = {875},
  pages = {L1},
  publisher = {IOP},
  issn = {0004-637X},
  doi = {10.3847/2041-8213/ab0ec7},
  urldate = {2024-05-03}
}

@article{eht_m87_2,
  title = {First {{M87 Event Horizon Telescope Results}}. {{II}}. {{Array}} and {{Instrumentation}}},
  author = {{Event Horizon Telescope Collaboration} and Akiyama, Kazunori and Alberdi, Antxon and Alef, Walter and Asada, Keiichi and Azulay, Rebecca and Baczko, Anne-Kathrin and Ball, David and Balokovi{\'c}, Mislav and Barrett, John and Bintley, Dan and Blackburn, Lindy and Boland, Wilfred and Bouman, Katherine L. and Bower, Geoffrey C. and Bremer, Michael and Brinkerink, Christiaan D. and Brissenden, Roger and Britzen, Silke and Broderick, Avery E. and Broguiere, Dominique and Bronzwaer, Thomas and Byun, Do-Young and Carlstrom, John E. and Chael, Andrew and Chan, Chi-kwan and Chatterjee, Shami and Chatterjee, Koushik and Chen, Ming-Tang and Chen, Yongjun and Cho, Ilje and Christian, Pierre and Conway, John E. and Cordes, James M. and Crew, Geoffrey B. and Cui, Yuzhu and Davelaar, Jordy and De Laurentis, Mariafelicia and Deane, Roger and Dempsey, Jessica and Desvignes, Gregory and Dexter, Jason and Doeleman, Sheperd S. and Eatough, Ralph P. and Falcke, Heino and Fish, Vincent L. and Fomalont, Ed and {Fraga-Encinas}, Raquel and Friberg, Per and Fromm, Christian M. and G{\'o}mez, Jos{\'e} L. and Galison, Peter and Gammie, Charles F. and Garc{\'i}a, Roberto and Gentaz, Olivier and Georgiev, Boris and Goddi, Ciriaco and Gold, Roman and Gu, Minfeng and Gurwell, Mark and Hada, Kazuhiro and Hecht, Michael H. and Hesper, Ronald and Ho, Luis C. and Ho, Paul and Honma, Mareki and Huang, Chih-Wei L. and Huang, Lei and Hughes, David H. and Ikeda, Shiro and Inoue, Makoto and Issaoun, Sara and James, David J. and Jannuzi, Buell T. and Janssen, Michael and Jeter, Britton and Jiang, Wu and Johnson, Michael D. and Jorstad, Svetlana and Jung, Taehyun and Karami, Mansour and Karuppusamy, Ramesh and Kawashima, Tomohisa and Keating, Garrett K. and Kettenis, Mark and Kim, Jae-Young and Kim, Junhan and Kim, Jongsoo and Kino, Motoki and Koay, Jun Yi and Koch, Patrick M. and Koyama, Shoko and Kramer, Michael and Kramer, Carsten and Krichbaum, Thomas P. and Kuo, Cheng-Yu and Lauer, Tod R. and Lee, Sang-Sung and Li, Yan-Rong and Li, Zhiyuan and Lindqvist, Michael and Liu, Kuo and Liuzzo, Elisabetta and Lo, Wen-Ping and Lobanov, Andrei P. and Loinard, Laurent and Lonsdale, Colin and Lu, Ru-Sen and MacDonald, Nicholas R. and Mao, Jirong and Markoff, Sera and Marrone, Daniel P. and Marscher, Alan P. and {Mart{\'i}-Vidal}, Iv{\'a}n and Matsushita, Satoki and Matthews, Lynn D. and Medeiros, Lia and Menten, Karl M. and Mizuno, Yosuke and Mizuno, Izumi and Moran, James M. and Moriyama, Kotaro and Moscibrodzka, Monika and M{\"u}ller, Cornelia and Nagai, Hiroshi and Nagar, Neil M. and Nakamura, Masanori and Narayan, Ramesh and Narayanan, Gopal and Natarajan, Iniyan and Neri, Roberto and Ni, Chunchong and Noutsos, Aristeidis and Okino, Hiroki and Olivares, H{\'e}ctor and {Ortiz-Le{\'o}n}, Gisela N. and Oyama, Tomoaki and {\"O}zel, Feryal and Palumbo, Daniel C. M. and Patel, Nimesh and Pen, Ue-Li and Pesce, Dominic W. and Pi{\'e}tu, Vincent and Plambeck, Richard and PopStefanija, Aleksandar and Porth, Oliver and Prather, Cora and {Preciado-L{\'o}pez}, Jorge A. and Psaltis, Dimitrios and Pu, Hung-Yi and Ramakrishnan, Venkatessh and Rao, Ramprasad and Rawlings, Mark G. and Raymond, Alexander W. and Rezzolla, Luciano and Ripperda, Bart and Roelofs, Freek and Rogers, Alan and Ros, Eduardo and Rose, Mel and Roshanineshat, Arash and Rottmann, Helge and Roy, Alan L. and Ruszczyk, Chet and Ryan, Benjamin R. and Rygl, Kazi L. J. and S{\'a}nchez, Salvador and {S{\'a}nchez-Arguelles}, David and Sasada, Mahito and Savolainen, Tuomas and Schloerb, F. Peter and Schuster, Karl-Friedrich and Shao, Lijing and Shen, Zhiqiang and Small, Des and Sohn, Bong Won and SooHoo, Jason and Tazaki, Fumie and Tiede, Paul and Tilanus, Remo P. J. and Titus, Michael and Toma, Kenji and Torne, Pablo and Trent, Tyler and Trippe, Sascha and Tsuda, Shuichiro and {van Bemmel}, Ilse and {van Langevelde}, Huib Jan and {van Rossum}, Daniel R. and Wagner, Jan and Wardle, John and Weintroub, Jonathan and Wex, Norbert and Wharton, Robert and Wielgus, Maciek and Wong, George N. and Wu, Qingwen and Young, Andr{\'e} and Young, Ken and Younsi, Ziri and Yuan, Feng and Yuan, Ye-Fei and Zensus, J. Anton and Zhao, Guangyao and Zhao, Shan-Shan and Zhu, Ziyan and Algaba, Juan-Carlos and Allardi, Alexander and Amestica, Rodrigo and Bach, Uwe and Beaudoin, Christopher and Benson, Bradford A. and Berthold, Ryan and Blanchard, Jay M. and Blundell, Ray and Bustamente, Sandra and Cappallo, Roger and {Castillo-Dom{\'i}nguez}, Edgar and Chang, Chih-Cheng and Chang, Shu-Hao and Chang, Song-Chu and Chen, Chung-Chen and Chilson, Ryan and Chuter, Tim C. and C{\'o}rdova Rosado, Rodrigo and Coulson, Iain M. and Crawford, Thomas M. and Crowley, Joseph and David, John and Derome, Mark and Dexter, Matthew and Dornbusch, Sven and Dudevoir, Kevin A. and Dzib, Sergio A. and Eckert, Chris and Erickson, Neal R. and Everett, Wendeline B. and Faber, Aaron and Farah, Joseph R. and Fath, Vernon and Folkers, Thomas W. and Forbes, David C. and Freund, Robert and {G{\'o}mez-Ruiz}, Arturo I. and Gale, David M. and Gao, Feng and Geertsema, Gertie and Graham, David A. and Greer, Christopher H. and Grosslein, Ronald and Gueth, Fr{\'e}d{\'e}ric and Halverson, Nils W. and Han, Chih-Chiang and Han, Kuo-Chang and Hao, Jinchi and Hasegawa, Yutaka and Henning, Jason W. and {Hern{\'a}ndez-G{\'o}mez}, Antonio and {Herrero-Illana}, Rub{\'e}n and Heyminck, Stefan and Hirota, Akihiko and Hoge, James and Huang, Yau-De and Impellizzeri, C. M. Violette and Jiang, Homin and Kamble, Atish and Keisler, Ryan and Kimura, Kimihiro and Kono, Yusuke and Kubo, Derek and Kuroda, John and Lacasse, Richard and Laing, Robert A. and Leitch, Erik M. and Li, Chao-Te and Lin, Lupin C. -C. and Liu, Ching-Tang and Liu, Kuan-Yu and Lu, Li-Ming and Marson, Ralph G. and {Martin-Cocher}, Pierre L. and Massingill, Kyle D. and Matulonis, Callie and McColl, Martin P. and McWhirter, Stephen R. and Messias, Hugo and {Meyer-Zhao}, Zheng and Michalik, Daniel and Monta{\~n}a, Alfredo and Montgomerie, William and {Mora-Klein}, Matias and Muders, Dirk and Nadolski, Andrew and Navarro, Santiago and Nguyen, Chi H. and Nishioka, Hiroaki and Norton, Timothy and Nystrom, George and Ogawa, Hideo and Oshiro, Peter and Oyama, Tomoaki and Padin, Stephen and Parsons, Harriet and Paine, Scott N. and Pe{\~n}alver, Juan and Phillips, Neil M. and Poirier, Michael and Pradel, Nicolas and Primiani, Rurik A. and Raffin, Philippe A. and Rahlin, Alexandra S. and Reiland, George and Risacher, Christopher and Ruiz, Ignacio and {S{\'a}ez-Mada{\'i}n}, Alejandro F. and Sassella, Remi and Schellart, Pim and Shaw, Paul and Silva, Kevin M. and Shiokawa, Hotaka and Smith, David R. and Snow, William and Souccar, Kamal and Sousa, Don and Sridharan, T. K. and Srinivasan, Ranjani and Stahm, William and Stark, Antony A. and Story, Kyle and Timmer, Sjoerd T. and Vertatschitsch, Laura and Walther, Craig and Wei, Ta-Shun and Whitehorn, Nathan and Whitney, Alan R. and Woody, David P. and Wouterloot, Jan G. A. and Wright, Melvin and Yamaguchi, Paul and Yu, Chen-Yu and Zeballos, Milagros and Ziurys, Lucy},
  year = {2019},
  month = apr,
  journal = {The Astrophysical Journal},
  volume = {875},
  pages = {L2},
  publisher = {IOP},
  issn = {0004-637X},
  doi = {10.3847/2041-8213/ab0c96},
  urldate = {2024-05-03}
}

@article{eht_m87_2018_1,
  title = {The Persistent Shadow of the Supermassive Black Hole of {{M}} 87. {{I}}. {{Observations}}, Calibration, Imaging, and Analysis},
  author = {{Event Horizon Telescope Collaboration} and Akiyama, Kazunori and Alberdi, Antxon and Alef, Walter and Algaba, Juan Carlos and Anantua, Richard and Asada, Keiichi and Azulay, Rebecca and Bach, Uwe and Baczko, Anne-Kathrin and Ball, David and Balokovi{\'c}, Mislav and Bandyopadhyay, Bidisha and Barrett, John and Baub{\"o}ck, Michi and Benson, Bradford A. and Bintley, Dan and Blackburn, Lindy and Blundell, Raymond and Bouman, Katherine L. and Bower, Geoffrey C. and Boyce, Hope and Bremer, Michael and Brissenden, Roger and Britzen, Silke and Broderick, Avery E. and Broguiere, Dominique and Bronzwaer, Thomas and Bustamante, Sandra and Carlstrom, John E. and Chael, Andrew and Chan, Chi-kwan and Chang, Dominic O. and Chatterjee, Koushik and Chatterjee, Shami and Chen, Ming-Tang and Chen, Yongjun and Cheng, Xiaopeng and Cho, Ilje and Christian, Pierre and Conroy, Nicholas S. and Conway, John E. and Crawford, Thomas M. and Crew, Geoffrey B. and {Cruz-Osorio}, Alejandro and Cui, Yuzhu and Dahale, Rohan and Davelaar, Jordy and De Laurentis, Mariafelicia and Deane, Roger and Dempsey, Jessica and Desvignes, Gregory and Dexter, Jason and Dhruv, Vedant and Dihingia, Indu K. and Doeleman, Sheperd S. and Dzib, Sergio A. and Eatough, Ralph P. and Emami, Razieh and Falcke, Heino and Farah, Joseph and Fish, Vincent L. and Fomalont, Edward and Ford, H. Alyson and Foschi, Marianna and {Fraga-Encinas}, Raquel and Freeman, William T. and Friberg, Per and Fromm, Christian M. and Fuentes, Antonio and Galison, Peter and Gammie, Charles F. and Garc{\'i}a, Roberto and Gentaz, Olivier and Georgiev, Boris and Goddi, Ciriaco and Gold, Roman and {G{\'o}mez-Ruiz}, Arturo I. and G{\'o}mez, Jos{\'e} L. and Gu, Minfeng and Gurwell, Mark and Hada, Kazuhiro and Haggard, Daryl and Hesper, Ronald and Heumann, Dirk and Ho, Luis C. and Ho, Paul and Honma, Mareki and Huang, Chih-Wei L. and Huang, Lei and Hughes, David H. and Ikeda, Shiro and Violette Impellizzeri, C. M. and Inoue, Makoto and Issaoun, Sara and James, David J. and Jannuzi, Buell T. and Janssen, Michael and Jeter, Britton and Jiang, Wu and {Jim{\'e}nez-Rosales}, Alejandra and Johnson, Michael D. and Jorstad, Svetlana and Jones, Adam C. and Joshi, Abhishek V. and Jung, Taehyun and Karuppusamy, Ramesh and Kawashima, Tomohisa and Keating, Garrett K. and Kettenis, Mark and Kim, Dong-Jin and Kim, Jae-Young and Kim, Jongsoo and Kim, Junhan and Kino, Motoki and Koay, Jun Yi and Kocherlakota, Prashant and Kofuji, Yutaro and Koch, Patrick M. and Koyama, Shoko and Kramer, Carsten and Kramer, Joana A. and Kramer, Michael and Krichbaum, Thomas P. and Kuo, Cheng-Yu and La Bella, Noemi and Lee, Sang-Sung and Levis, Aviad and Li, Zhiyuan and Lico, Rocco and Lindahl, Greg and Lindqvist, Michael and Lisakov, Mikhail and Liu, Jun and Liu, Kuo and Liuzzo, Elisabetta and Lo, Wen-Ping and Lobanov, Andrei P. and Loinard, Laurent and Lonsdale, Colin J. and Lowitz, Amy E. and Lu, Ru-Sen and MacDonald, Nicholas R. and Mao, Jirong and Marchili, Nicola and Markoff, Sera and Marrone, Daniel P. and Marscher, Alan P. and {Mart{\'i}-Vidal}, Iv{\'a}n and Matsushita, Satoki and Matthews, Lynn D. and Medeiros, Lia and Menten, Karl M. and Mizuno, Izumi and Mizuno, Yosuke and Montgomery, Joshua and Moran, James M. and Moriyama, Kotaro and Moscibrodzka, Monika and Mulaudzi, Wanga and M{\"u}ller, Cornelia and M{\"u}ller, Hendrik and Mus, Alejandro and Musoke, Gibwa and Myserlis, Ioannis and Nagai, Hiroshi and Nagar, Neil M. and Nakamura, Masanori and Narayanan, Gopal and Natarajan, Iniyan and Nathanail, Antonios and Fuentes, Santiago Navarro and Neilsen, Joey and Ni, Chunchong and Nowak, Michael A. and Oh, Junghwan and Okino, Hiroki and Olivares, H{\'e}ctor and Oyama, Tomoaki and {\"O}zel, Feryal and Palumbo, Daniel C. M. and Paraschos, Georgios Filippos and Park, Jongho and Parsons, Harriet and Patel, Nimesh and Pen, Ue-Li and Pesce, Dominic W. and Pi{\'e}tu, Vincent and PopStefanija, Aleksandar and Porth, Oliver and Prather, Cora and Psaltis, Dimitrios and Pu, Hung-Yi and Ramakrishnan, Venkatessh and Rao, Ramprasad and Rawlings, Mark G. and Raymond, Alexander W. and Rezzolla, Luciano and Ricarte, Angelo and Ripperda, Bart and Roelofs, Freek and {Romero-Ca{\~n}izales}, Cristina and Ros, Eduardo and Roshanineshat, Arash and Rottmann, Helge and Roy, Alan L. and Ruiz, Ignacio and Ruszczyk, Chet and Rygl, Kazi L. J. and S{\'a}nchez, Salvador and {S{\'a}nchez-Arg{\"u}elles}, David and {S{\'a}nchez-Portal}, Miguel and Sasada, Mahito and Satapathy, Kaushik and Savolainen, Tuomas and Schloerb, F. Peter and Schonfeld, Jonathan and Schuster, Karl-Friedrich and Shao, Lijing and Shen, Zhiqiang and Small, Des and Sohn, Bong Won and SooHoo, Jason and Salas, Le{\'o}n David Sosapanta and Souccar, Kamal and Stanway, Joshua S. and Sun, He and Tazaki, Fumie and Tetarenko, Alexandra J. and Tiede, Paul and Tilanus, Remo P. J. and Titus, Michael and Toma, Kenji and Torne, Pablo and Toscano, Teresa and Traianou, Efthalia and Trent, Tyler and Trippe, Sascha and Turk, Matthew and {van Bemmel}, Ilse and {van Langevelde}, Huib Jan and {van Rossum}, Daniel R. and Vos, Jesse and Wagner, Jan and {Ward-Thompson}, Derek and Wardle, John and Washington, Jasmin E. and Weintroub, Jonathan and Wharton, Robert and Wielgus, Maciek and Wiik, Kaj and Witzel, Gunther and Wondrak, Michael F. and Wong, George N. and Wu, Qingwen and Yadlapalli, Nitika and Yamaguchi, Paul and Yfantis, Aristomenis and Yoon, Doosoo and Young, Andr{\'e} and Younsi, Ziri and Yu, Wei and Yuan, Feng and Yuan, Ye-Fei and Anton Zensus, J. and Zhang, Shuo and Zhao, Guang-Yao and Zhao, Shan-Shan and Allardi, Alexander and Chang, Shu-Hao and Chang, Chih-Cheng and Chang, Song-Chu and Chen, Chung-Chen and Chilson, Ryan and Faber, Aaron and Gale, David M. and Han, Chih-Chiang and Han, Kuo-Chang and Hasegawa, Yutaka and {Hern{\'a}ndez-Rebollar}, Jos{\'e} Luis and Huang, Yau-De and Jiang, Homin and Jinchi, Hao and Kimura, Kimihiro and Kubo, Derek and Li, Chao-Te and Lin, Lupin C. -C. and Liu, Ching-Tang and Liu, Kuan-Yu and Lu, Li-Ming and {Martin-Cocher}, Pierre and {Meyer-Zhao}, Zheng and Monta{\~n}a, Alfredo and Moraghan, Anthony and {Moreno-Nolasco}, Marcos Emir and Nishioka, Hiroaki and Norton, Timothy J. and Nystrom, George and Ogawa, Hideo and Oshiro, Peter and Pradel, Nicolas and Principe, Giacomo and Raffin, Philippe and {Rodr{\'i}guez-Montoya}, Iv{\'a}n and Shaw, Paul and Snow, William and Sridharan, Tirupati Kumara and Srinivasan, Ranjani and Wei, Ta-Shun and Yu, Chen-Yu},
  year = {2024},
  month = jan,
  journal = {Astronomy and Astrophysics},
  volume = {681},
  pages = {A79},
  issn = {0004-6361},
  doi = {10.1051/0004-6361/202347932},
  urldate = {2024-05-03}
}

@article{eht_m87_3,
  title = {First {{M87 Event Horizon Telescope Results}}. {{III}}. {{Data Processing}} and {{Calibration}}},
  author = {{Event Horizon Telescope Collaboration} and Akiyama, Kazunori and Alberdi, Antxon and Alef, Walter and Asada, Keiichi and Azulay, Rebecca and Baczko, Anne-Kathrin and Ball, David and Balokovi{\'c}, Mislav and Barrett, John and Bintley, Dan and Blackburn, Lindy and Boland, Wilfred and Bouman, Katherine L. and Bower, Geoffrey C. and Bremer, Michael and Brinkerink, Christiaan D. and Brissenden, Roger and Britzen, Silke and Broderick, Avery E. and Broguiere, Dominique and Bronzwaer, Thomas and Byun, Do-Young and Carlstrom, John E. and Chael, Andrew and Chan, Chi-kwan and Chatterjee, Shami and Chatterjee, Koushik and Chen, Ming-Tang and Chen, Yongjun and Cho, Ilje and Christian, Pierre and Conway, John E. and Cordes, James M. and Crew, Geoffrey B. and Cui, Yuzhu and Davelaar, Jordy and De Laurentis, Mariafelicia and Deane, Roger and Dempsey, Jessica and Desvignes, Gregory and Dexter, Jason and Doeleman, Sheperd S. and Eatough, Ralph P. and Falcke, Heino and Fish, Vincent L. and Fomalont, Ed and {Fraga-Encinas}, Raquel and Friberg, Per and Fromm, Christian M. and G{\'o}mez, Jos{\'e} L. and Galison, Peter and Gammie, Charles F. and Garc{\'i}a, Roberto and Gentaz, Olivier and Georgiev, Boris and Goddi, Ciriaco and Gold, Roman and Gu, Minfeng and Gurwell, Mark and Hada, Kazuhiro and Hecht, Michael H. and Hesper, Ronald and Ho, Luis C. and Ho, Paul and Honma, Mareki and Huang, Chih-Wei L. and Huang, Lei and Hughes, David H. and Ikeda, Shiro and Inoue, Makoto and Issaoun, Sara and James, David J. and Jannuzi, Buell T. and Janssen, Michael and Jeter, Britton and Jiang, Wu and Johnson, Michael D. and Jorstad, Svetlana and Jung, Taehyun and Karami, Mansour and Karuppusamy, Ramesh and Kawashima, Tomohisa and Keating, Garrett K. and Kettenis, Mark and Kim, Jae-Young and Kim, Junhan and Kim, Jongsoo and Kino, Motoki and Koay, Jun Yi and Koch, Patrick M. and Koyama, Shoko and Kramer, Michael and Kramer, Carsten and Krichbaum, Thomas P. and Kuo, Cheng-Yu and Lauer, Tod R. and Lee, Sang-Sung and Li, Yan-Rong and Li, Zhiyuan and Lindqvist, Michael and Liu, Kuo and Liuzzo, Elisabetta and Lo, Wen-Ping and Lobanov, Andrei P. and Loinard, Laurent and Lonsdale, Colin and Lu, Ru-Sen and MacDonald, Nicholas R. and Mao, Jirong and Markoff, Sera and Marrone, Daniel P. and Marscher, Alan P. and {Mart{\'i}-Vidal}, Iv{\'a}n and Matsushita, Satoki and Matthews, Lynn D. and Medeiros, Lia and Menten, Karl M. and Mizuno, Yosuke and Mizuno, Izumi and Moran, James M. and Moriyama, Kotaro and Moscibrodzka, Monika and M{\"u}ller, Cornelia and Nagai, Hiroshi and Nagar, Neil M. and Nakamura, Masanori and Narayan, Ramesh and Narayanan, Gopal and Natarajan, Iniyan and Neri, Roberto and Ni, Chunchong and Noutsos, Aristeidis and Okino, Hiroki and Olivares, H{\'e}ctor and {Ortiz-Le{\'o}n}, Gisela N. and Oyama, Tomoaki and {\"O}zel, Feryal and Palumbo, Daniel C. M. and Patel, Nimesh and Pen, Ue-Li and Pesce, Dominic W. and Pi{\'e}tu, Vincent and Plambeck, Richard and PopStefanija, Aleksandar and Porth, Oliver and Prather, Cora and {Preciado-L{\'o}pez}, Jorge A. and Psaltis, Dimitrios and Pu, Hung-Yi and Ramakrishnan, Venkatessh and Rao, Ramprasad and Rawlings, Mark G. and Raymond, Alexander W. and Rezzolla, Luciano and Ripperda, Bart and Roelofs, Freek and Rogers, Alan and Ros, Eduardo and Rose, Mel and Roshanineshat, Arash and Rottmann, Helge and Roy, Alan L. and Ruszczyk, Chet and Ryan, Benjamin R. and Rygl, Kazi L. J. and S{\'a}nchez, Salvador and {S{\'a}nchez-Arguelles}, David and Sasada, Mahito and Savolainen, Tuomas and Schloerb, F. Peter and Schuster, Karl-Friedrich and Shao, Lijing and Shen, Zhiqiang and Small, Des and Sohn, Bong Won and SooHoo, Jason and Tazaki, Fumie and Tiede, Paul and Tilanus, Remo P. J. and Titus, Michael and Toma, Kenji and Torne, Pablo and Trent, Tyler and Trippe, Sascha and Tsuda, Shuichiro and {van Bemmel}, Ilse and {van Langevelde}, Huib Jan and {van Rossum}, Daniel R. and Wagner, Jan and Wardle, John and Weintroub, Jonathan and Wex, Norbert and Wharton, Robert and Wielgus, Maciek and Wong, George N. and Wu, Qingwen and Young, Andr{\'e} and Young, Ken and Younsi, Ziri and Yuan, Feng and Yuan, Ye-Fei and Zensus, J. Anton and Zhao, Guangyao and Zhao, Shan-Shan and Zhu, Ziyan and Cappallo, Roger and Farah, Joseph R. and Folkers, Thomas W. and {Meyer-Zhao}, Zheng and Michalik, Daniel and Nadolski, Andrew and Nishioka, Hiroaki and Pradel, Nicolas and Primiani, Rurik A. and Souccar, Kamal and Vertatschitsch, Laura and Yamaguchi, Paul},
  year = {2019},
  month = apr,
  journal = {The Astrophysical Journal},
  volume = {875},
  pages = {L3},
  publisher = {IOP},
  issn = {0004-637X},
  doi = {10.3847/2041-8213/ab0c57},
  urldate = {2024-05-03}
}

@article{eht_m87_4,
  title = {First {{M87 Event Horizon Telescope Results}}. {{IV}}. {{Imaging}} the {{Central Supermassive Black Hole}}},
  author = {{Event Horizon Telescope Collaboration} and Akiyama, Kazunori and Alberdi, Antxon and Alef, Walter and Asada, Keiichi and Azulay, Rebecca and Baczko, Anne-Kathrin and Ball, David and Balokovi{\'c}, Mislav and Barrett, John and Bintley, Dan and Blackburn, Lindy and Boland, Wilfred and Bouman, Katherine L. and Bower, Geoffrey C. and Bremer, Michael and Brinkerink, Christiaan D. and Brissenden, Roger and Britzen, Silke and Broderick, Avery E. and Broguiere, Dominique and Bronzwaer, Thomas and Byun, Do-Young and Carlstrom, John E. and Chael, Andrew and Chan, Chi-kwan and Chatterjee, Shami and Chatterjee, Koushik and Chen, Ming-Tang and Chen, Yongjun and Cho, Ilje and Christian, Pierre and Conway, John E. and Cordes, James M. and Crew, Geoffrey B. and Cui, Yuzhu and Davelaar, Jordy and De Laurentis, Mariafelicia and Deane, Roger and Dempsey, Jessica and Desvignes, Gregory and Dexter, Jason and Doeleman, Sheperd S. and Eatough, Ralph P. and Falcke, Heino and Fish, Vincent L. and Fomalont, Ed and {Fraga-Encinas}, Raquel and Freeman, William T. and Friberg, Per and Fromm, Christian M. and G{\'o}mez, Jos{\'e} L. and Galison, Peter and Gammie, Charles F. and Garc{\'i}a, Roberto and Gentaz, Olivier and Georgiev, Boris and Goddi, Ciriaco and Gold, Roman and Gu, Minfeng and Gurwell, Mark and Hada, Kazuhiro and Hecht, Michael H. and Hesper, Ronald and Ho, Luis C. and Ho, Paul and Honma, Mareki and Huang, Chih-Wei L. and Huang, Lei and Hughes, David H. and Ikeda, Shiro and Inoue, Makoto and Issaoun, Sara and James, David J. and Jannuzi, Buell T. and Janssen, Michael and Jeter, Britton and Jiang, Wu and Johnson, Michael D. and Jorstad, Svetlana and Jung, Taehyun and Karami, Mansour and Karuppusamy, Ramesh and Kawashima, Tomohisa and Keating, Garrett K. and Kettenis, Mark and Kim, Jae-Young and Kim, Junhan and Kim, Jongsoo and Kino, Motoki and Koay, Jun Yi and Koch, Patrick M. and Koyama, Shoko and Kramer, Michael and Kramer, Carsten and Krichbaum, Thomas P. and Kuo, Cheng-Yu and Lauer, Tod R. and Lee, Sang-Sung and Li, Yan-Rong and Li, Zhiyuan and Lindqvist, Michael and Liu, Kuo and Liuzzo, Elisabetta and Lo, Wen-Ping and Lobanov, Andrei P. and Loinard, Laurent and Lonsdale, Colin and Lu, Ru-Sen and MacDonald, Nicholas R. and Mao, Jirong and Markoff, Sera and Marrone, Daniel P. and Marscher, Alan P. and {Mart{\'i}-Vidal}, Iv{\'a}n and Matsushita, Satoki and Matthews, Lynn D. and Medeiros, Lia and Menten, Karl M. and Mizuno, Yosuke and Mizuno, Izumi and Moran, James M. and Moriyama, Kotaro and Moscibrodzka, Monika and M{\"u}ller, Cornelia and Nagai, Hiroshi and Nagar, Neil M. and Nakamura, Masanori and Narayan, Ramesh and Narayanan, Gopal and Natarajan, Iniyan and Neri, Roberto and Ni, Chunchong and Noutsos, Aristeidis and Okino, Hiroki and Olivares, H{\'e}ctor and Oyama, Tomoaki and {\"O}zel, Feryal and Palumbo, Daniel C. M. and Patel, Nimesh and Pen, Ue-Li and Pesce, Dominic W. and Pi{\'e}tu, Vincent and Plambeck, Richard and PopStefanija, Aleksandar and Porth, Oliver and Prather, Cora and {Preciado-L{\'o}pez}, Jorge A. and Psaltis, Dimitrios and Pu, Hung-Yi and Ramakrishnan, Venkatessh and Rao, Ramprasad and Rawlings, Mark G. and Raymond, Alexander W. and Rezzolla, Luciano and Ripperda, Bart and Roelofs, Freek and Rogers, Alan and Ros, Eduardo and Rose, Mel and Roshanineshat, Arash and Rottmann, Helge and Roy, Alan L. and Ruszczyk, Chet and Ryan, Benjamin R. and Rygl, Kazi L. J. and S{\'a}nchez, Salvador and {S{\'a}nchez-Arguelles}, David and Sasada, Mahito and Savolainen, Tuomas and Schloerb, F. Peter and Schuster, Karl-Friedrich and Shao, Lijing and Shen, Zhiqiang and Small, Des and Sohn, Bong Won and SooHoo, Jason and Tazaki, Fumie and Tiede, Paul and Tilanus, Remo P. J. and Titus, Michael and Toma, Kenji and Torne, Pablo and Trent, Tyler and Trippe, Sascha and Tsuda, Shuichiro and {van Bemmel}, Ilse and {van Langevelde}, Huib Jan and {van Rossum}, Daniel R. and Wagner, Jan and Wardle, John and Weintroub, Jonathan and Wex, Norbert and Wharton, Robert and Wielgus, Maciek and Wong, George N. and Wu, Qingwen and Young, Andr{\'e} and Young, Ken and Younsi, Ziri and Yuan, Feng and Yuan, Ye-Fei and Zensus, J. Anton and Zhao, Guangyao and Zhao, Shan-Shan and Zhu, Ziyan and Farah, Joseph R. and {Meyer-Zhao}, Zheng and Michalik, Daniel and Nadolski, Andrew and Nishioka, Hiroaki and Pradel, Nicolas and Primiani, Rurik A. and Souccar, Kamal and Vertatschitsch, Laura and Yamaguchi, Paul},
  year = {2019},
  month = apr,
  journal = {The Astrophysical Journal},
  volume = {875},
  pages = {L4},
  publisher = {IOP},
  issn = {0004-637X},
  doi = {10.3847/2041-8213/ab0e85},
  urldate = {2024-05-03}
}

@article{eht_m87_5,
  title = {First {{M87 Event Horizon Telescope Results}}. {{V}}. {{Physical Origin}} of the {{Asymmetric Ring}}},
  author = {{Event Horizon Telescope Collaboration} and Akiyama, Kazunori and Alberdi, Antxon and Alef, Walter and Asada, Keiichi and Azulay, Rebecca and Baczko, Anne-Kathrin and Ball, David and Balokovi{\'c}, Mislav and Barrett, John and Bintley, Dan and Blackburn, Lindy and Boland, Wilfred and Bouman, Katherine L. and Bower, Geoffrey C. and Bremer, Michael and Brinkerink, Christiaan D. and Brissenden, Roger and Britzen, Silke and Broderick, Avery E. and Broguiere, Dominique and Bronzwaer, Thomas and Byun, Do-Young and Carlstrom, John E. and Chael, Andrew and Chan, Chi-kwan and Chatterjee, Shami and Chatterjee, Koushik and Chen, Ming-Tang and Chen, Yongjun and Cho, Ilje and Christian, Pierre and Conway, John E. and Cordes, James M. and Crew, Geoffrey B. and Cui, Yuzhu and Davelaar, Jordy and De Laurentis, Mariafelicia and Deane, Roger and Dempsey, Jessica and Desvignes, Gregory and Dexter, Jason and Doeleman, Sheperd S. and Eatough, Ralph P. and Falcke, Heino and Fish, Vincent L. and Fomalont, Ed and {Fraga-Encinas}, Raquel and Friberg, Per and Fromm, Christian M. and G{\'o}mez, Jos{\'e} L. and Galison, Peter and Gammie, Charles F. and Garc{\'i}a, Roberto and Gentaz, Olivier and Georgiev, Boris and Goddi, Ciriaco and Gold, Roman and Gu, Minfeng and Gurwell, Mark and Hada, Kazuhiro and Hecht, Michael H. and Hesper, Ronald and Ho, Luis C. and Ho, Paul and Honma, Mareki and Huang, Chih-Wei L. and Huang, Lei and Hughes, David H. and Ikeda, Shiro and Inoue, Makoto and Issaoun, Sara and James, David J. and Jannuzi, Buell T. and Janssen, Michael and Jeter, Britton and Jiang, Wu and Johnson, Michael D. and Jorstad, Svetlana and Jung, Taehyun and Karami, Mansour and Karuppusamy, Ramesh and Kawashima, Tomohisa and Keating, Garrett K. and Kettenis, Mark and Kim, Jae-Young and Kim, Junhan and Kim, Jongsoo and Kino, Motoki and Koay, Jun Yi and Koch, Patrick M. and Koyama, Shoko and Kramer, Michael and Kramer, Carsten and Krichbaum, Thomas P. and Kuo, Cheng-Yu and Lauer, Tod R. and Lee, Sang-Sung and Li, Yan-Rong and Li, Zhiyuan and Lindqvist, Michael and Liu, Kuo and Liuzzo, Elisabetta and Lo, Wen-Ping and Lobanov, Andrei P. and Loinard, Laurent and Lonsdale, Colin and Lu, Ru-Sen and MacDonald, Nicholas R. and Mao, Jirong and Markoff, Sera and Marrone, Daniel P. and Marscher, Alan P. and {Mart{\'i}-Vidal}, Iv{\'a}n and Matsushita, Satoki and Matthews, Lynn D. and Medeiros, Lia and Menten, Karl M. and Mizuno, Yosuke and Mizuno, Izumi and Moran, James M. and Moriyama, Kotaro and Moscibrodzka, Monika and Mu{\"l}ler, Cornelia and Nagai, Hiroshi and Nagar, Neil M. and Nakamura, Masanori and Narayan, Ramesh and Narayanan, Gopal and Natarajan, Iniyan and Neri, Roberto and Ni, Chunchong and Noutsos, Aristeidis and Okino, Hiroki and Olivares, H{\'e}ctor and Oyama, Tomoaki and {\"O}zel, Feryal and Palumbo, Daniel C. M. and Patel, Nimesh and Pen, Ue-Li and Pesce, Dominic W. and Pi{\'e}tu, Vincent and Plambeck, Richard and PopStefanija, Aleksandar and Porth, Oliver and Prather, Cora and {Preciado-L{\'o}pez}, Jorge A. and Psaltis, Dimitrios and Pu, Hung-Yi and Ramakrishnan, Venkatessh and Rao, Ramprasad and Rawlings, Mark G. and Raymond, Alexander W. and Rezzolla, Luciano and Ripperda, Bart and Roelofs, Freek and Rogers, Alan and Ros, Eduardo and Rose, Mel and Roshanineshat, Arash and Rottmann, Helge and Roy, Alan L. and Ruszczyk, Chet and Ryan, Benjamin R. and Rygl, Kazi L. J. and S{\'a}nchez, Salvador and {S{\'a}nchez-Arguelles}, David and Sasada, Mahito and Savolainen, Tuomas and Schloerb, F. Peter and Schuster, Karl-Friedrich and Shao, Lijing and Shen, Zhiqiang and Small, Des and Sohn, Bong Won and SooHoo, Jason and Tazaki, Fumie and Tiede, Paul and Tilanus, Remo P. J. and Titus, Michael and Toma, Kenji and Torne, Pablo and Trent, Tyler and Trippe, Sascha and Tsuda, Shuichiro and {van Bemmel}, Ilse and {van Langevelde}, Huib Jan and {van Rossum}, Daniel R. and Wagner, Jan and Wardle, John and Weintroub, Jonathan and Wex, Norbert and Wharton, Robert and Wielgus, Maciek and Wong, George N. and Wu, Qingwen and Young, Andr{\'e} and Young, Ken and Younsi, Ziri and Yuan, Feng and Yuan, Ye-Fei and Zensus, J. Anton and Zhao, Guangyao and Zhao, Shan-Shan and Zhu, Ziyan and Anczarski, Jadyn and Baganoff, Frederick K. and Eckart, Andreas and Farah, Joseph R. and Haggard, Daryl and {Meyer-Zhao}, Zheng and Michalik, Daniel and Nadolski, Andrew and Neilsen, Joseph and Nishioka, Hiroaki and Nowak, Michael A. and Pradel, Nicolas and Primiani, Rurik A. and Souccar, Kamal and Vertatschitsch, Laura and Yamaguchi, Paul and Zhang, Shuo},
  year = {2019},
  month = apr,
  journal = {The Astrophysical Journal},
  volume = {875},
  pages = {L5},
  publisher = {IOP},
  issn = {0004-637X},
  doi = {10.3847/2041-8213/ab0f43},
  urldate = {2024-05-03}
}

@article{eht_m87_6,
  title = {First {{M87 Event Horizon Telescope Results}}. {{VI}}. {{The Shadow}} and {{Mass}} of the {{Central Black Hole}}},
  author = {{Event Horizon Telescope Collaboration} and Akiyama, Kazunori and Alberdi, Antxon and Alef, Walter and Asada, Keiichi and Azulay, Rebecca and Baczko, Anne-Kathrin and Ball, David and Balokovi{\'c}, Mislav and Barrett, John and Bintley, Dan and Blackburn, Lindy and Boland, Wilfred and Bouman, Katherine L. and Bower, Geoffrey C. and Bremer, Michael and Brinkerink, Christiaan D. and Brissenden, Roger and Britzen, Silke and Broderick, Avery E. and Broguiere, Dominique and Bronzwaer, Thomas and Byun, Do-Young and Carlstrom, John E. and Chael, Andrew and Chan, Chi-kwan and Chatterjee, Shami and Chatterjee, Koushik and Chen, Ming-Tang and Chen, Yongjun and Cho, Ilje and Christian, Pierre and Conway, John E. and Cordes, James M. and Crew, Geoffrey B. and Cui, Yuzhu and Davelaar, Jordy and De Laurentis, Mariafelicia and Deane, Roger and Dempsey, Jessica and Desvignes, Gregory and Dexter, Jason and Doeleman, Sheperd S. and Eatough, Ralph P. and Falcke, Heino and Fish, Vincent L. and Fomalont, Ed and {Fraga-Encinas}, Raquel and Friberg, Per and Fromm, Christian M. and G{\'o}mez, Jos{\'e} L. and Galison, Peter and Gammie, Charles F. and Garc{\'i}a, Roberto and Gentaz, Olivier and Georgiev, Boris and Goddi, Ciriaco and Gold, Roman and Gu, Minfeng and Gurwell, Mark and Hada, Kazuhiro and Hecht, Michael H. and Hesper, Ronald and Ho, Luis C. and Ho, Paul and Honma, Mareki and Huang, Chih-Wei L. and Huang, Lei and Hughes, David H. and Ikeda, Shiro and Inoue, Makoto and Issaoun, Sara and James, David J. and Jannuzi, Buell T. and Janssen, Michael and Jeter, Britton and Jiang, Wu and Johnson, Michael D. and Jorstad, Svetlana and Jung, Taehyun and Karami, Mansour and Karuppusamy, Ramesh and Kawashima, Tomohisa and Keating, Garrett K. and Kettenis, Mark and Kim, Jae-Young and Kim, Junhan and Kim, Jongsoo and Kino, Motoki and Koay, Jun Yi and Koch, Patrick M. and Koyama, Shoko and Kramer, Michael and Kramer, Carsten and Krichbaum, Thomas P. and Kuo, Cheng-Yu and Lauer, Tod R. and Lee, Sang-Sung and Li, Yan-Rong and Li, Zhiyuan and Lindqvist, Michael and Liu, Kuo and Liuzzo, Elisabetta and Lo, Wen-Ping and Lobanov, Andrei P. and Loinard, Laurent and Lonsdale, Colin and Lu, Ru-Sen and MacDonald, Nicholas R. and Mao, Jirong and Markoff, Sera and Marrone, Daniel P. and Marscher, Alan P. and {Mart{\'i}-Vidal}, Iv{\'a}n and Matsushita, Satoki and Matthews, Lynn D. and Medeiros, Lia and Menten, Karl M. and Mizuno, Yosuke and Mizuno, Izumi and Moran, James M. and Moriyama, Kotaro and Moscibrodzka, Monika and M{\"u}ller, Cornelia and Nagai, Hiroshi and Nagar, Neil M. and Nakamura, Masanori and Narayan, Ramesh and Narayanan, Gopal and Natarajan, Iniyan and Neri, Roberto and Ni, Chunchong and Noutsos, Aristeidis and Okino, Hiroki and Olivares, H{\'e}ctor and Oyama, Tomoaki and {\"O}zel, Feryal and Palumbo, Daniel C. M. and Patel, Nimesh and Pen, Ue-Li and Pesce, Dominic W. and Pi{\'e}tu, Vincent and Plambeck, Richard and PopStefanija, Aleksandar and Porth, Oliver and Prather, Cora and {Preciado-L{\'o}pez}, Jorge A. and Psaltis, Dimitrios and Pu, Hung-Yi and Ramakrishnan, Venkatessh and Rao, Ramprasad and Rawlings, Mark G. and Raymond, Alexander W. and Rezzolla, Luciano and Ripperda, Bart and Roelofs, Freek and Rogers, Alan and Ros, Eduardo and Rose, Mel and Roshanineshat, Arash and Rottmann, Helge and Roy, Alan L. and Ruszczyk, Chet and Ryan, Benjamin R. and Rygl, Kazi L. J. and S{\'a}nchez, Salvador and {S{\'a}nchez-Arguelles}, David and Sasada, Mahito and Savolainen, Tuomas and Schloerb, F. Peter and Schuster, Karl-Friedrich and Shao, Lijing and Shen, Zhiqiang and Small, Des and Sohn, Bong Won and SooHoo, Jason and Tazaki, Fumie and Tiede, Paul and Tilanus, Remo P. J. and Titus, Michael and Toma, Kenji and Torne, Pablo and Trent, Tyler and Trippe, Sascha and Tsuda, Shuichiro and {van Bemmel}, Ilse and {van Langevelde}, Huib Jan and {van Rossum}, Daniel R. and Wagner, Jan and Wardle, John and Weintroub, Jonathan and Wex, Norbert and Wharton, Robert and Wielgus, Maciek and Wong, George N. and Wu, Qingwen and Young, Andr{\'e} and Young, Ken and Younsi, Ziri and Yuan, Feng and Yuan, Ye-Fei and Zensus, J. Anton and Zhao, Guangyao and Zhao, Shan-Shan and Zhu, Ziyan and Farah, Joseph R. and {Meyer-Zhao}, Zheng and Michalik, Daniel and Nadolski, Andrew and Nishioka, Hiroaki and Pradel, Nicolas and Primiani, Rurik A. and Souccar, Kamal and Vertatschitsch, Laura and Yamaguchi, Paul},
  year = {2019},
  month = apr,
  journal = {The Astrophysical Journal},
  volume = {875},
  pages = {L6},
  publisher = {IOP},
  issn = {0004-637X},
  doi = {10.3847/2041-8213/ab1141},
  urldate = {2024-05-03}
}

@article{eht_m87_7,
  title = {First {{M87 Event Horizon Telescope Results}}. {{VII}}. {{Polarization}} of the {{Ring}}},
  author = {{Event Horizon Telescope Collaboration} and Akiyama, Kazunori and Algaba, Juan Carlos and Alberdi, Antxon and Alef, Walter and Anantua, Richard and Asada, Keiichi and Azulay, Rebecca and Baczko, Anne-Kathrin and Ball, David and Balokovi{\'c}, Mislav and Barrett, John and Benson, Bradford A. and Bintley, Dan and Blackburn, Lindy and Blundell, Raymond and Boland, Wilfred and Bouman, Katherine L. and Bower, Geoffrey C. and Boyce, Hope and Bremer, Michael and Brinkerink, Christiaan D. and Brissenden, Roger and Britzen, Silke and Broderick, Avery E. and Broguiere, Dominique and Bronzwaer, Thomas and Byun, Do-Young and Carlstrom, John E. and Chael, Andrew and Chan, Chi-kwan and Chatterjee, Shami and Chatterjee, Koushik and Chen, Ming-Tang and Chen, Yongjun and Chesler, Paul M. and Cho, Ilje and Christian, Pierre and Conway, John E. and Cordes, James M. and Crawford, Thomas M. and Crew, Geoffrey B. and {Cruz-Osorio}, Alejandro and Cui, Yuzhu and Davelaar, Jordy and De Laurentis, Mariafelicia and Deane, Roger and Dempsey, Jessica and Desvignes, Gregory and Dexter, Jason and Doeleman, Sheperd S. and Eatough, Ralph P. and Falcke, Heino and Farah, Joseph and Fish, Vincent L. and Fomalont, Ed and Ford, H. Alyson and {Fraga-Encinas}, Raquel and Freeman, William T. and Friberg, Per and Fromm, Christian M. and Fuentes, Antonio and Galison, Peter and Gammie, Charles F. and Garc{\'i}a, Roberto and Gentaz, Olivier and Georgiev, Boris and Goddi, Ciriaco and Gold, Roman and G{\'o}mez, Jos{\'e} L. and {G{\'o}mez-Ruiz}, Arturo I. and Gu, Minfeng and Gurwell, Mark and Hada, Kazuhiro and Haggard, Daryl and Hecht, Michael H. and Hesper, Ronald and Ho, Luis C. and Ho, Paul and Honma, Mareki and Huang, Chih-Wei L. and Huang, Lei and Hughes, David H. and Ikeda, Shiro and Inoue, Makoto and Issaoun, Sara and James, David J. and Jannuzi, Buell T. and Janssen, Michael and Jeter, Britton and Jiang, Wu and {Jimenez-Rosales}, Alejandra and Johnson, Michael D. and Jorstad, Svetlana and Jung, Taehyun and Karami, Mansour and Karuppusamy, Ramesh and Kawashima, Tomohisa and Keating, Garrett K. and Kettenis, Mark and Kim, Dong-Jin and Kim, Jae-Young and Kim, Jongsoo and Kim, Junhan and Kino, Motoki and Koay, Jun Yi and Kofuji, Yutaro and Koch, Patrick M. and Koyama, Shoko and Kramer, Michael and Kramer, Carsten and Krichbaum, Thomas P. and Kuo, Cheng-Yu and Lauer, Tod R. and Lee, Sang-Sung and Levis, Aviad and Li, Yan-Rong and Li, Zhiyuan and Lindqvist, Michael and Lico, Rocco and Lindahl, Greg and Liu, Jun and Liu, Kuo and Liuzzo, Elisabetta and Lo, Wen-Ping and Lobanov, Andrei P. and Loinard, Laurent and Lonsdale, Colin and Lu, Ru-Sen and MacDonald, Nicholas R. and Mao, Jirong and Marchili, Nicola and Markoff, Sera and Marrone, Daniel P. and Marscher, Alan P. and {Mart{\'i}-Vidal}, Iv{\'a}n and Matsushita, Satoki and Matthews, Lynn D. and Medeiros, Lia and Menten, Karl M. and Mizuno, Izumi and Mizuno, Yosuke and Moran, James M. and Moriyama, Kotaro and Moscibrodzka, Monika and M{\"u}ller, Cornelia and Musoke, Gibwa and Mej{\'i}as, Alejandro Mus and Michalik, Daniel and Nadolski, Andrew and Nagai, Hiroshi and Nagar, Neil M. and Nakamura, Masanori and Narayan, Ramesh and Narayanan, Gopal and Natarajan, Iniyan and Nathanail, Antonios and Neilsen, Joey and Neri, Roberto and Ni, Chunchong and Noutsos, Aristeidis and Nowak, Michael A. and Okino, Hiroki and Olivares, H{\'e}ctor and {Ortiz-Le{\'o}n}, Gisela N. and Oyama, Tomoaki and {\"O}zel, Feryal and Palumbo, Daniel C. M. and Park, Jongho and Patel, Nimesh and Pen, Ue-Li and Pesce, Dominic W. and Pi{\'e}tu, Vincent and Plambeck, Richard and PopStefanija, Aleksandar and Porth, Oliver and P{\"o}tzl, Felix M. and Prather, Cora and {Preciado-L{\'o}pez}, Jorge A. and Psaltis, Dimitrios and Pu, Hung-Yi and Ramakrishnan, Venkatessh and Rao, Ramprasad and Rawlings, Mark G. and Raymond, Alexander W. and Rezzolla, Luciano and Ricarte, Angelo and Ripperda, Bart and Roelofs, Freek and Rogers, Alan and Ros, Eduardo and Rose, Mel and Roshanineshat, Arash and Rottmann, Helge and Roy, Alan L. and Ruszczyk, Chet and Rygl, Kazi L. J. and S{\'a}nchez, Salvador and {S{\'a}nchez-Arguelles}, David and Sasada, Mahito and Savolainen, Tuomas and Schloerb, F. Peter and Schuster, Karl-Friedrich and Shao, Lijing and Shen, Zhiqiang and Small, Des and Sohn, Bong Won and SooHoo, Jason and Sun, He and Tazaki, Fumie and Tetarenko, Alexandra J. and Tiede, Paul and Tilanus, Remo P. J. and Titus, Michael and Toma, Kenji and Torne, Pablo and Trent, Tyler and Traianou, Efthalia and Trippe, Sascha and {van Bemmel}, Ilse and {van Langevelde}, Huib Jan and {van Rossum}, Daniel R. and Wagner, Jan and {Ward-Thompson}, Derek and Wardle, John and Weintroub, Jonathan and Wex, Norbert and Wharton, Robert and Wielgus, Maciek and Wong, George N. and Wu, Qingwen and Yoon, Doosoo and Young, Andr{\'e} and Young, Ken and Younsi, Ziri and Yuan, Feng and Yuan, Ye-Fei and Zensus, J. Anton and Zhao, Guang-Yao and Zhao, Shan-Shan},
  year = {2021},
  month = mar,
  journal = {The Astrophysical Journal},
  volume = {910},
  pages = {L12},
  publisher = {IOP},
  issn = {0004-637X},
  doi = {10.3847/2041-8213/abe71d},
  urldate = {2024-05-03}
}

@article{eht_m87_8,
  title = {First {{M87 Event Horizon Telescope Results}}. {{VIII}}. {{Magnetic Field Structure}} near {{The Event Horizon}}},
  author = {{Event Horizon Telescope Collaboration} and Akiyama, Kazunori and Algaba, Juan Carlos and Alberdi, Antxon and Alef, Walter and Anantua, Richard and Asada, Keiichi and Azulay, Rebecca and Baczko, Anne-Kathrin and Ball, David and Balokovi{\'c}, Mislav and Barrett, John and Benson, Bradford A. and Bintley, Dan and Blackburn, Lindy and Blundell, Raymond and Boland, Wilfred and Bouman, Katherine L. and Bower, Geoffrey C. and Boyce, Hope and Bremer, Michael and Brinkerink, Christiaan D. and Brissenden, Roger and Britzen, Silke and Broderick, Avery E. and Broguiere, Dominique and Bronzwaer, Thomas and Byun, Do-Young and Carlstrom, John E. and Chael, Andrew and Chan, Chi-kwan and Chatterjee, Shami and Chatterjee, Koushik and Chen, Ming-Tang and Chen, Yongjun and Chesler, Paul M. and Cho, Ilje and Christian, Pierre and Conway, John E. and Cordes, James M. and Crawford, Thomas M. and Crew, Geoffrey B. and {Cruz-Osorio}, Alejandro and Cui, Yuzhu and Davelaar, Jordy and De Laurentis, Mariafelicia and Deane, Roger and Dempsey, Jessica and Desvignes, Gregory and Dexter, Jason and Doeleman, Sheperd S. and Eatough, Ralph P. and Falcke, Heino and Farah, Joseph and Fish, Vincent L. and Fomalont, Ed and Ford, H. Alyson and {Fraga-Encinas}, Raquel and Friberg, Per and Fromm, Christian M. and Fuentes, Antonio and Galison, Peter and Gammie, Charles F. and Garc{\'i}a, Roberto and Gelles, Zachary and Gentaz, Olivier and Georgiev, Boris and Goddi, Ciriaco and Gold, Roman and G{\'o}mez, Jos{\'e} L. and {G{\'o}mez-Ruiz}, Arturo I. and Gu, Minfeng and Gurwell, Mark and Hada, Kazuhiro and Haggard, Daryl and Hecht, Michael H. and Hesper, Ronald and Himwich, Elizabeth and Ho, Luis C. and Ho, Paul and Honma, Mareki and Huang, Chih-Wei L. and Huang, Lei and Hughes, David H. and Ikeda, Shiro and Inoue, Makoto and Issaoun, Sara and James, David J. and Jannuzi, Buell T. and Janssen, Michael and Jeter, Britton and Jiang, Wu and {Jimenez-Rosales}, Alejandra and Johnson, Michael D. and Jorstad, Svetlana and Jung, Taehyun and Karami, Mansour and Karuppusamy, Ramesh and Kawashima, Tomohisa and Keating, Garrett K. and Kettenis, Mark and Kim, Dong-Jin and Kim, Jae-Young and Kim, Jongsoo and Kim, Junhan and Kino, Motoki and Koay, Jun Yi and Kofuji, Yutaro and Koch, Patrick M. and Koyama, Shoko and Kramer, Michael and Kramer, Carsten and Krichbaum, Thomas P. and Kuo, Cheng-Yu and Lauer, Tod R. and Lee, Sang-Sung and Levis, Aviad and Li, Yan-Rong and Li, Zhiyuan and Lindqvist, Michael and Lico, Rocco and Lindahl, Greg and Liu, Jun and Liu, Kuo and Liuzzo, Elisabetta and Lo, Wen-Ping and Lobanov, Andrei P. and Loinard, Laurent and Lonsdale, Colin and Lu, Ru-Sen and MacDonald, Nicholas R. and Mao, Jirong and Marchili, Nicola and Markoff, Sera and Marrone, Daniel P. and Marscher, Alan P. and {Mart{\'i}-Vidal}, Iv{\'a}n and Matsushita, Satoki and Matthews, Lynn D. and Medeiros, Lia and Menten, Karl M. and Mizuno, Izumi and Mizuno, Yosuke and Moran, James M. and Moriyama, Kotaro and Moscibrodzka, Monika and M{\"u}ller, Cornelia and Musoke, Gibwa and Mus Mej{\'i}as, Alejandro and Michalik, Daniel and Nadolski, Andrew and Nagai, Hiroshi and Nagar, Neil M. and Nakamura, Masanori and Narayan, Ramesh and Narayanan, Gopal and Natarajan, Iniyan and Nathanail, Antonios and Neilsen, Joey and Neri, Roberto and Ni, Chunchong and Noutsos, Aristeidis and Nowak, Michael A. and Okino, Hiroki and Olivares, H{\'e}ctor and {Ortiz-Le{\'o}n}, Gisela N. and Oyama, Tomoaki and {\"O}zel, Feryal and Palumbo, Daniel C. M. and Park, Jongho and Patel, Nimesh and Pen, Ue-Li and Pesce, Dominic W. and Pi{\'e}tu, Vincent and Plambeck, Richard and PopStefanija, Aleksandar and Porth, Oliver and P{\"o}tzl, Felix M. and Prather, Cora and {Preciado-L{\'o}pez}, Jorge A. and Psaltis, Dimitrios and Pu, Hung-Yi and Ramakrishnan, Venkatessh and Rao, Ramprasad and Rawlings, Mark G. and Raymond, Alexander W. and Rezzolla, Luciano and Ricarte, Angelo and Ripperda, Bart and Roelofs, Freek and Rogers, Alan and Ros, Eduardo and Rose, Mel and Roshanineshat, Arash and Rottmann, Helge and Roy, Alan L. and Ruszczyk, Chet and Rygl, Kazi L. J. and S{\'a}nchez, Salvador and {S{\'a}nchez-Arguelles}, David and Sasada, Mahito and Savolainen, Tuomas and Schloerb, F. Peter and Schuster, Karl-Friedrich and Shao, Lijing and Shen, Zhiqiang and Small, Des and Sohn, Bong Won and SooHoo, Jason and Sun, He and Tazaki, Fumie and Tetarenko, Alexandra J. and Tiede, Paul and Tilanus, Remo P. J. and Titus, Michael and Toma, Kenji and Torne, Pablo and Trent, Tyler and Traianou, Efthalia and Trippe, Sascha and {van Bemmel}, Ilse and {van Langevelde}, Huib Jan and {van Rossum}, Daniel R. and Wagner, Jan and {Ward-Thompson}, Derek and Wardle, John and Weintroub, Jonathan and Wex, Norbert and Wharton, Robert and Wielgus, Maciek and Wong, George N. and Wu, Qingwen and Yoon, Doosoo and Young, Andr{\'e} and Young, Ken and Younsi, Ziri and Yuan, Feng and Yuan, Ye-Fei and Zensus, J. Anton and Zhao, Guang-Yao and Zhao, Shan-Shan},
  year = {2021},
  month = mar,
  journal = {The Astrophysical Journal},
  volume = {910},
  pages = {L13},
  publisher = {IOP},
  issn = {0004-637X},
  doi = {10.3847/2041-8213/abe4de},
  urldate = {2024-05-03}
}

@article{eht_m87_9,
  title = {First {{M87 Event Horizon Telescope Results}}. {{IX}}. {{Detection}} of {{Near-horizon Circular Polarization}}},
  author = {{Event Horizon Telescope Collaboration} and Akiyama, Kazunori and Alberdi, Antxon and Alef, Walter and Algaba, Juan Carlos and Anantua, Richard and Asada, Keiichi and Azulay, Rebecca and Bach, Uwe and Baczko, Anne-Kathrin and Ball, David and Balokovi{\'c}, Mislav and Barrett, John and Baub{\"o}ck, Michi and Benson, Bradford A. and Bintley, Dan and Blackburn, Lindy and Blundell, Raymond and Bouman, Katherine L. and Bower, Geoffrey C. and Boyce, Hope and Bremer, Michael and Brinkerink, Christiaan D. and Brissenden, Roger and Britzen, Silke and Broderick, Avery E. and Broguiere, Dominique and Bronzwaer, Thomas and Bustamante, Sandra and Byun, Do-Young and Carlstrom, John E. and Ceccobello, Chiara and Chael, Andrew and Chan, Chi-kwan and Chang, Dominic O. and Chatterjee, Koushik and Chatterjee, Shami and Chen, Ming-Tang and Chen, Yongjun and Cheng, Xiaopeng and Cho, Ilje and Christian, Pierre and Conroy, Nicholas S. and Conway, John E. and Cordes, James M. and Crawford, Thomas M. and Crew, Geoffrey B. and {Cruz-Osorio}, Alejandro and Cui, Yuzhu and Dahale, Rohan and Davelaar, Jordy and De Laurentis, Mariafelicia and Deane, Roger and Dempsey, Jessica and Desvignes, Gregory and Dexter, Jason and Dhruv, Vedant and Doeleman, Sheperd S. and Dougal, Sean and Dzib, Sergio A. and Eatough, Ralph P. and Emami, Razieh and Falcke, Heino and Farah, Joseph and Fish, Vincent L. and Fomalont, Ed and Ford, H. Alyson and Foschi, Marianna and {Fraga-Encinas}, Raquel and Freeman, William T. and Friberg, Per and Fromm, Christian M. and Fuentes, Antonio and Galison, Peter and Gammie, Charles F. and Garc{\'i}a, Roberto and Gentaz, Olivier and Georgiev, Boris and Goddi, Ciriaco and Gold, Roman and {G{\'o}mez-Ruiz}, Arturo I. and G{\'o}mez, Jos{\'e} L. and Gu, Minfeng and Gurwell, Mark and Hada, Kazuhiro and Haggard, Daryl and Haworth, Kari and Hecht, Michael H. and Hesper, Ronald and Heumann, Dirk and Ho, Luis C. and Ho, Paul and Honma, Mareki and Huang, Chih-Wei L. and Huang, Lei and Hughes, David H. and Ikeda, Shiro and Impellizzeri, C. M. Violette and Inoue, Makoto and Issaoun, Sara and James, David J. and Jannuzi, Buell T. and Janssen, Michael and Jeter, Britton and Jiang, Wu and {Jim{\'e}nez-Rosales}, Alejandra and Johnson, Michael D. and Jorstad, Svetlana and Joshi, Abhishek V. and Jung, Taehyun and Karami, Mansour and Karuppusamy, Ramesh and Kawashima, Tomohisa and Keating, Garrett K. and Kettenis, Mark and Kim, Dong-Jin and Kim, Jae-Young and Kim, Jongsoo and Kim, Junhan and Kino, Motoki and Koay, Jun Yi and Kocherlakota, Prashant and Kofuji, Yutaro and Koch, Patrick M. and Koyama, Shoko and Kramer, Carsten and Kramer, Joana A. and Kramer, Michael and Krichbaum, Thomas P. and Kuo, Cheng-Yu and La Bella, Noemi and Lauer, Tod R. and Lee, Daeyoung and Lee, Sang-Sung and Leung, Po Kin and Levis, Aviad and Li, Zhiyuan and Lico, Rocco and Lindahl, Greg and Lindqvist, Michael and Lisakov, Mikhail and Liu, Jun and Liu, Kuo and Liuzzo, Elisabetta and Lo, Wen-Ping and Lobanov, Andrei P. and Loinard, Laurent and Lonsdale, Colin J. and Lowitz, Amy E. and Lu, Ru-Sen and MacDonald, Nicholas R. and Mao, Jirong and Marchili, Nicola and Markoff, Sera and Marrone, Daniel P. and Marscher, Alan P. and {Mart{\'i}-Vidal}, Iv{\'a}n and Matsushita, Satoki and Matthews, Lynn D. and Medeiros, Lia and Menten, Karl M. and Michalik, Daniel and Mizuno, Izumi and Mizuno, Yosuke and Moran, James M. and Moriyama, Kotaro and Moscibrodzka, Monika and Mulaudzi, Wanga and M{\"u}ller, Cornelia and M{\"u}ller, Hendrik and Mus, Alejandro and Musoke, Gibwa and Myserlis, Ioannis and Nadolski, Andrew and Nagai, Hiroshi and Nagar, Neil M. and Nakamura, Masanori and Narayan, Ramesh and Narayanan, Gopal and Natarajan, Iniyan and Nathanail, Antonios and Fuentes, Santiago Navarro and Neilsen, Joey and Neri, Roberto and Ni, Chunchong and Noutsos, Aristeidis and Nowak, Michael A. and Oh, Junghwan and Okino, Hiroki and Olivares, H{\'e}ctor and {Ortiz-Le{\'o}n}, Gisela N. and Oyama, Tomoaki and {\"O}zel, Feryal and Palumbo, Daniel C. M. and Paraschos, Georgios Filippos and Park, Jongho and Parsons, Harriet and Patel, Nimesh and Pen, Ue-Li and Pesce, Dominic W. and Pi{\'e}tu, Vincent and Plambeck, Richard and PopStefanija, Aleksandar and Porth, Oliver and P{\"o}tzl, Felix M. and Prather, Cora and {Preciado-L{\'o}pez}, Jorge A. and Psaltis, Dimitrios and Pu, Hung-Yi and Ramakrishnan, Venkatessh and Rao, Ramprasad and Rawlings, Mark G. and Raymond, Alexander W. and Rezzolla, Luciano and Ricarte, Angelo and Ripperda, Bart and Roelofs, Freek and Rogers, Alan and {Romero-Ca{\~n}izales}, Cristina and Ros, Eduardo and Roshanineshat, Arash and Rottmann, Helge and Roy, Alan L. and Ruiz, Ignacio and Ruszczyk, Chet and Rygl, Kazi L. J. and S{\'a}nchez, Salvador and {S{\'a}nchez-Arg{\"u}elles}, David and {S{\'a}nchez-Portal}, Miguel and Sasada, Mahito and Satapathy, Kaushik and Savolainen, Tuomas and Schloerb, F. Peter and Schonfeld, Jonathan and Schuster, Karl-Friedrich and Shao, Lijing and Shen, Zhiqiang and Small, Des and Sohn, Bong Won and SooHoo, Jason and Sosapanta Salas, Le{\'o}n David and Souccar, Kamal and Sun, He and Tazaki, Fumie and Tetarenko, Alexandra J. and Tiede, Paul and Tilanus, Remo P. J. and Titus, Michael and Torne, Pablo and Toscano, Teresa and Traianou, Efthalia and Trent, Tyler and Trippe, Sascha and Turk, Matthew and {van Bemmel}, Ilse and {van Langevelde}, Huib Jan and {van Rossum}, Daniel R. and Vos, Jesse and Wagner, Jan and {Ward-Thompson}, Derek and Wardle, John and Washington, Jasmin E. and Weintroub, Jonathan and Wharton, Robert and Wielgus, Maciek and Wiik, Kaj and Witzel, Gunther and Wondrak, Michael F. and Wong, George N. and Wu, Qingwen and Yadlapalli, Nitika and Yamaguchi, Paul and Yfantis, Aristomenis and Yoon, Doosoo and Young, Andr{\'e} and Young, Ken and Younsi, Ziri and Yu, Wei and Yuan, Feng and Yuan, Ye-Fei and Zensus, J. Anton and Zhang, Shuo and Zhao, Guang-Yao and Zhao, Shan-Shan},
  year = {2023},
  month = nov,
  journal = {The Astrophysical Journal},
  volume = {957},
  pages = {L20},
  publisher = {IOP},
  issn = {0004-637X},
  doi = {10.3847/2041-8213/acff70},
  urldate = {2024-05-03}
}

@article{eht_sgra_1,
  title = {First {{Sagittarius A}}* {{Event Horizon Telescope Results}}. {{I}}. {{The Shadow}} of the {{Supermassive Black Hole}} in the {{Center}} of the {{Milky Way}}},
  author = {{Event Horizon Telescope Collaboration} and Akiyama, Kazunori and Alberdi, Antxon and Alef, Walter and Algaba, Juan Carlos and Anantua, Richard and Asada, Keiichi and Azulay, Rebecca and Bach, Uwe and Baczko, Anne-Kathrin and Ball, David and Balokovi{\'c}, Mislav and Barrett, John and Baub{\"o}ck, Michi and Benson, Bradford A. and Bintley, Dan and Blackburn, Lindy and Blundell, Raymond and Bouman, Katherine L. and Bower, Geoffrey C. and Boyce, Hope and Bremer, Michael and Brinkerink, Christiaan D. and Brissenden, Roger and Britzen, Silke and Broderick, Avery E. and Broguiere, Dominique and Bronzwaer, Thomas and Bustamante, Sandra and Byun, Do-Young and Carlstrom, John E. and Ceccobello, Chiara and Chael, Andrew and Chan, Chi-kwan and Chatterjee, Koushik and Chatterjee, Shami and Chen, Ming-Tang and Chen, Yongjun and Cheng, Xiaopeng and Cho, Ilje and Christian, Pierre and Conroy, Nicholas S. and Conway, John E. and Cordes, James M. and Crawford, Thomas M. and Crew, Geoffrey B. and {Cruz-Osorio}, Alejandro and Cui, Yuzhu and Davelaar, Jordy and De Laurentis, Mariafelicia and Deane, Roger and Dempsey, Jessica and Desvignes, Gregory and Dexter, Jason and Dhruv, Vedant and Doeleman, Sheperd S. and Dougal, Sean and Dzib, Sergio A. and Eatough, Ralph P. and Emami, Razieh and Falcke, Heino and Farah, Joseph and Fish, Vincent L. and Fomalont, Ed and Ford, H. Alyson and {Fraga-Encinas}, Raquel and Freeman, William T. and Friberg, Per and Fromm, Christian M. and Fuentes, Antonio and Galison, Peter and Gammie, Charles F. and Garc{\'i}a, Roberto and Gentaz, Olivier and Georgiev, Boris and Goddi, Ciriaco and Gold, Roman and {G{\'o}mez-Ruiz}, Arturo I. and G{\'o}mez, Jos{\'e} L. and Gu, Minfeng and Gurwell, Mark and Hada, Kazuhiro and Haggard, Daryl and Haworth, Kari and Hecht, Michael H. and Hesper, Ronald and Heumann, Dirk and Ho, Luis C. and Ho, Paul and Honma, Mareki and Huang, Chih-Wei L. and Huang, Lei and Hughes, David H. and Ikeda, Shiro and Impellizzeri, C. M. Violette and Inoue, Makoto and Issaoun, Sara and James, David J. and Jannuzi, Buell T. and Janssen, Michael and Jeter, Britton and Jiang, Wu and {Jim{\'e}nez-Rosales}, Alejandra and Johnson, Michael D. and Jorstad, Svetlana and Joshi, Abhishek V. and Jung, Taehyun and Karami, Mansour and Karuppusamy, Ramesh and Kawashima, Tomohisa and Keating, Garrett K. and Kettenis, Mark and Kim, Dong-Jin and Kim, Jae-Young and Kim, Jongsoo and Kim, Junhan and Kino, Motoki and Koay, Jun Yi and Kocherlakota, Prashant and Kofuji, Yutaro and Koch, Patrick M. and Koyama, Shoko and Kramer, Carsten and Kramer, Michael and Krichbaum, Thomas P. and Kuo, Cheng-Yu and La Bella, Noemi and Lauer, Tod R. and Lee, Daeyoung and Lee, Sang-Sung and Leung, Po Kin and Levis, Aviad and Li, Zhiyuan and Lico, Rocco and Lindahl, Greg and Lindqvist, Michael and Lisakov, Mikhail and Liu, Jun and Liu, Kuo and Liuzzo, Elisabetta and Lo, Wen-Ping and Lobanov, Andrei P. and Loinard, Laurent and Lonsdale, Colin J. and Lu, Ru-Sen and Mao, Jirong and Marchili, Nicola and Markoff, Sera and Marrone, Daniel P. and Marscher, Alan P. and {Mart{\'i}-Vidal}, Iv{\'a}n and Matsushita, Satoki and Matthews, Lynn D. and Medeiros, Lia and Menten, Karl M. and Michalik, Daniel and Mizuno, Izumi and Mizuno, Yosuke and Moran, James M. and Moriyama, Kotaro and Moscibrodzka, Monika and M{\"u}ller, Cornelia and Mus, Alejandro and Musoke, Gibwa and Myserlis, Ioannis and Nadolski, Andrew and Nagai, Hiroshi and Nagar, Neil M. and Nakamura, Masanori and Narayan, Ramesh and Narayanan, Gopal and Natarajan, Iniyan and Nathanail, Antonios and Fuentes, Santiago Navarro and Neilsen, Joey and Neri, Roberto and Ni, Chunchong and Noutsos, Aristeidis and Nowak, Michael A. and Oh, Junghwan and Okino, Hiroki and Olivares, H{\'e}ctor and {Ortiz-Le{\'o}n}, Gisela N. and Oyama, Tomoaki and {\"O}zel, Feryal and Palumbo, Daniel C. M. and Paraschos, Georgios Filippos and Park, Jongho and Parsons, Harriet and Patel, Nimesh and Pen, Ue-Li and Pesce, Dominic W. and Pi{\'e}tu, Vincent and Plambeck, Richard and PopStefanija, Aleksandar and Porth, Oliver and P{\"o}tzl, Felix M. and Prather, Cora and {Preciado-L{\'o}pez}, Jorge A. and Psaltis, Dimitrios and Pu, Hung-Yi and Ramakrishnan, Venkatessh and Rao, Ramprasad and Rawlings, Mark G. and Raymond, Alexander W. and Rezzolla, Luciano and Ricarte, Angelo and Ripperda, Bart and Roelofs, Freek and Rogers, Alan and Ros, Eduardo and {Romero-Ca{\~n}izales}, Cristina and Roshanineshat, Arash and Rottmann, Helge and Roy, Alan L. and Ruiz, Ignacio and Ruszczyk, Chet and Rygl, Kazi L. J. and S{\'a}nchez, Salvador and {S{\'a}nchez-Arg{\"u}elles}, David and {S{\'a}nchez-Portal}, Miguel and Sasada, Mahito and Satapathy, Kaushik and Savolainen, Tuomas and Schloerb, F. Peter and Schonfeld, Jonathan and Schuster, Karl-Friedrich and Shao, Lijing and Shen, Zhiqiang and Small, Des and Sohn, Bong Won and SooHoo, Jason and Souccar, Kamal and Sun, He and Tazaki, Fumie and Tetarenko, Alexandra J. and Tiede, Paul and Tilanus, Remo P. J. and Titus, Michael and Torne, Pablo and Traianou, Efthalia and Trent, Tyler and Trippe, Sascha and Turk, Matthew and {van Bemmel}, Ilse and {van Langevelde}, Huib Jan and {van Rossum}, Daniel R. and Vos, Jesse and Wagner, Jan and {Ward-Thompson}, Derek and Wardle, John and Weintroub, Jonathan and Wex, Norbert and Wharton, Robert and Wielgus, Maciek and Wiik, Kaj and Witzel, Gunther and Wondrak, Michael F. and Wong, George N. and Wu, Qingwen and Yamaguchi, Paul and Yoon, Doosoo and Young, Andr{\'e} and Young, Ken and Younsi, Ziri and Yuan, Feng and Yuan, Ye-Fei and Zensus, J. Anton and Zhang, Shuo and Zhao, Guang-Yao and Zhao, Shan-Shan and Agurto, Claudio and Allardi, Alexander and Amestica, Rodrigo and Araneda, Juan Pablo and Arriagada, Oriel and Berghuis, Jennie L. and Bertarini, Alessandra and Berthold, Ryan and Blanchard, Jay and Brown, Ken and C{\'a}rdenas, Mauricio and Cantzler, Michael and Caro, Patricio and {Castillo-Dom{\'i}nguez}, Edgar and Chan, Tin Lok and Chang, Chih-Cheng and Chang, Dominic O. and Chang, Shu-Hao and Chang, Song-Chu and Chen, Chung-Chen and Chilson, Ryan and Chuter, Tim C. and Ciechanowicz, Miroslaw and {Colin-Beltran}, Edgar and Coulson, Iain M. and Crowley, Joseph and Degenaar, Nathalie and Dornbusch, Sven and Dur{\'a}n, Carlos A. and Everett, Wendeline B. and Faber, Aaron and Forster, Karl and Fuchs, Miriam M. and Gale, David M. and Geertsema, Gertie and Gonz{\'a}lez, Edouard and Graham, Dave and Gueth, Fr{\'e}d{\'e}ric and Halverson, Nils W. and Han, Chih-Chiang and Han, Kuo-Chang and Hasegawa, Yutaka and {Hern{\'a}ndez-Rebollar}, Jos{\'e} Luis and Herrera, Cristian and {Herrero-Illana}, Ruben and Heyminck, Stefan and Hirota, Akihiko and Hoge, James and Hostler Schimpf, Shelbi R. and Howie, Ryan E. and Huang, Yau-De and Jiang, Homin and Jinchi, Hao and John, David and Kimura, Kimihiro and Klein, Thomas and Kubo, Derek and Kuroda, John and Kwon, Caleb and Lacasse, Richard and Laing, Robert and Leitch, Erik M. and Li, Chao-Te and Liu, Ching-Tang and Liu, Kuan-Yu and Lin, Lupin C. -C. and Lu, Li-Ming and {Mac-Auliffe}, Felipe and {Martin-Cocher}, Pierre and Matulonis, Callie and Maute, John K. and Messias, Hugo and {Meyer-Zhao}, Zheng and Monta{\~n}a, Alfredo and {Montenegro-Montes}, Francisco and Montgomerie, William and Moreno Nolasco, Marcos Emir and Muders, Dirk and Nishioka, Hiroaki and Norton, Timothy J. and Nystrom, George and Ogawa, Hideo and Olivares, Rodrigo and Oshiro, Peter and {P{\'e}rez-Beaupuits}, Juan Pablo and Parra, Rodrigo and Phillips, Neil M. and Poirier, Michael and Pradel, Nicolas and Qiu, Richard and Raffin, Philippe A. and Rahlin, Alexandra S. and Ram{\'i}rez, Jorge and Ressler, Sean and Reynolds, Mark and {Rodr{\'i}guez-Montoya}, Iv{\'a}n and {Saez-Madain}, Alejandro F. and Santana, Jorge and Shaw, Paul and Shirkey, Leslie E. and Silva, Kevin M. and Snow, William and Sousa, Don and Sridharan, T. K. and Stahm, William and Stark, Anthony A. and Test, John and Torstensson, Karl and Venegas, Paulina and Walther, Craig and Wei, Ta-Shun and White, Chris and Wieching, Gundolf and Wijnands, Rudy and Wouterloot, Jan G. A. and Yu, Chen-Yu and Yu (于威), Wei and Zeballos, Milagros},
  year = {2022},
  month = may,
  journal = {The Astrophysical Journal},
  volume = {930},
  pages = {L12},
  publisher = {IOP},
  issn = {0004-637X},
  doi = {10.3847/2041-8213/ac6674},
  urldate = {2024-05-03}
}

@article{eht_sgra_2,
  title = {First {{Sagittarius A}}* {{Event Horizon Telescope Results}}. {{II}}. {{EHT}} and {{Multiwavelength Observations}}, {{Data Processing}}, and {{Calibration}}},
  author = {{Event Horizon Telescope Collaboration} and Akiyama, Kazunori and Alberdi, Antxon and Alef, Walter and Algaba, Juan Carlos and Anantua, Richard and Asada, Keiichi and Azulay, Rebecca and Bach, Uwe and Baczko, Anne-Kathrin and Ball, David and Balokovi{\'c}, Mislav and Barrett, John and Baub{\"o}ck, Michi and Benson, Bradford A. and Bintley, Dan and Blackburn, Lindy and Blundell, Raymond and Bouman, Katherine L. and Bower, Geoffrey C. and Boyce, Hope and Bremer, Michael and Brinkerink, Christiaan D. and Brissenden, Roger and Britzen, Silke and Broderick, Avery E. and Broguiere, Dominique and Bronzwaer, Thomas and Bustamante, Sandra and Byun, Do-Young and Carlstrom, John E. and Ceccobello, Chiara and Chael, Andrew and Chan, Chi-kwan and Chatterjee, Koushik and Chatterjee, Shami and Chen, Ming-Tang and Chen, Yongjun and Cheng, Xiaopeng and Cho, Ilje and Christian, Pierre and Conroy, Nicholas S. and Conway, John E. and Cordes, James M. and Crawford, Thomas M. and Crew, Geoffrey B. and {Cruz-Osorio}, Alejandro and Cui, Yuzhu and Davelaar, Jordy and De Laurentis, Mariafelicia and Deane, Roger and Dempsey, Jessica and Desvignes, Gregory and Dexter, Jason and Dhruv, Vedant and Doeleman, Sheperd S. and Dougal, Sean and Dzib, Sergio A. and Eatough, Ralph P. and Emami, Razieh and Falcke, Heino and Farah, Joseph and Fish, Vincent L. and Fomalont, Ed and Ford, H. Alyson and {Fraga-Encinas}, Raquel and Freeman, William T. and Friberg, Per and Fromm, Christian M. and Fuentes, Antonio and Galison, Peter and Gammie, Charles F. and Garc{\'i}a, Roberto and Gentaz, Olivier and Georgiev, Boris and Goddi, Ciriaco and Gold, Roman and {G{\'o}mez-Ruiz}, Arturo I. and G{\'o}mez, Jos{\'e} L. and Gu, Minfeng and Gurwell, Mark and Hada, Kazuhiro and Haggard, Daryl and Haworth, Kari and Hecht, Michael H. and Hesper, Ronald and Heumann, Dirk and Ho, Luis C. and Ho, Paul and Honma, Mareki and Huang, Chih-Wei L. and Huang, Lei and Hughes, David H. and Ikeda, Shiro and Impellizzeri, C. M. Violette and Inoue, Makoto and Issaoun, Sara and James, David J. and Jannuzi, Buell T. and Janssen, Michael and Jeter, Britton and Jiang, Wu and {Jim{\'e}nez-Rosales}, Alejandra and Johnson, Michael D. and Jorstad, Svetlana and Joshi, Abhishek V. and Jung, Taehyun and Karami, Mansour and Karuppusamy, Ramesh and Kawashima, Tomohisa and Keating, Garrett K. and Kettenis, Mark and Kim, Dong-Jin and Kim, Jae-Young and Kim, Jongsoo and Kim, Junhan and Kino, Motoki and Koay, Jun Yi and Kocherlakota, Prashant and Kofuji, Yutaro and Koch, Patrick M. and Koyama, Shoko and Kramer, Carsten and Kramer, Michael and Krichbaum, Thomas P. and Kuo, Cheng-Yu and La Bella, Noemi and Lauer, Tod R. and Lee, Daeyoung and Lee, Sang-Sung and Leung, Po Kin and Levis, Aviad and Li, Zhiyuan and Lico, Rocco and Lindahl, Greg and Lindqvist, Michael and Lisakov, Mikhail and Liu, Jun and Liu, Kuo and Liuzzo, Elisabetta and Lo, Wen-Ping and Lobanov, Andrei P. and Loinard, Laurent and Lonsdale, Colin J. and Lu, Ru-Sen and Mao, Jirong and Marchili, Nicola and Markoff, Sera and Marrone, Daniel P. and Marscher, Alan P. and {Mart{\'i}-Vidal}, Iv{\'a}n and Matsushita, Satoki and Matthews, Lynn D. and Medeiros, Lia and Menten, Karl M. and Michalik, Daniel and Mizuno, Izumi and Mizuno, Yosuke and Moran, James M. and Moriyama, Kotaro and Moscibrodzka, Monika and M{\"u}ller, Cornelia and Mus, Alejandro and Musoke, Gibwa and Myserlis, Ioannis and Nadolski, Andrew and Nagai, Hiroshi and Nagar, Neil M. and Nakamura, Masanori and Narayan, Ramesh and Narayanan, Gopal and Natarajan, Iniyan and Nathanail, Antonios and Fuentes, Santiago Navarro and Neilsen, Joey and Neri, Roberto and Ni, Chunchong and Noutsos, Aristeidis and Nowak, Michael A. and Oh, Junghwan and Okino, Hiroki and Olivares, H{\'e}ctor and {Ortiz-Le{\'o}n}, Gisela N. and Oyama, Tomoaki and {\"O}zel, Feryal and Palumbo, Daniel C. M. and Paraschos, Georgios Filippos and Park, Jongho and Parsons, Harriet and Patel, Nimesh and Pen, Ue-Li and Pesce, Dominic W. and Pi{\'e}tu, Vincent and Plambeck, Richard and PopStefanija, Aleksandar and Porth, Oliver and P{\"o}tzl, Felix M. and Prather, Cora and {Preciado-L{\'o}pez}, Jorge A. and Psaltis, Dimitrios and Pu, Hung-Yi and Ramakrishnan, Venkatessh and Rao, Ramprasad and Rawlings, Mark G. and Raymond, Alexander W. and Rezzolla, Luciano and Ricarte, Angelo and Ripperda, Bart and Roelofs, Freek and Rogers, Alan and Ros, Eduardo and {Romero-Ca{\~n}izales}, Cristina and Roshanineshat, Arash and Rottmann, Helge and Roy, Alan L. and Ruiz, Ignacio and Ruszczyk, Chet and Rygl, Kazi L. J. and S{\'a}nchez, Salvador and {S{\'a}nchez-Arg{\"u}elles}, David and {S{\'a}nchez-Portal}, Miguel and Sasada, Mahito and Satapathy, Kaushik and Savolainen, Tuomas and Schloerb, F. Peter and Schonfeld, Jonathan and Schuster, Karl-Friedrich and Shao, Lijing and Shen, Zhiqiang and Small, Des and Sohn, Bong Won and SooHoo, Jason and Souccar, Kamal and Sun, He and Tazaki, Fumie and Tetarenko, Alexandra J. and Tiede, Paul and Tilanus, Remo P. J. and Titus, Michael and Torne, Pablo and Traianou, Efthalia and Trent, Tyler and Trippe, Sascha and Turk, Matthew and {van Bemmel}, Ilse and {van Langevelde}, Huib Jan and {van Rossum}, Daniel R. and Vos, Jesse and Wagner, Jan and {Ward-Thompson}, Derek and Wardle, John and Weintroub, Jonathan and Wex, Norbert and Wharton, Robert and Wielgus, Maciek and Wiik, Kaj and Witzel, Gunther and Wondrak, Michael F. and Wong, George N. and Wu, Qingwen and Yamaguchi, Paul and Yoon, Doosoo and Young, Andr{\'e} and Young, Ken and Younsi, Ziri and Yuan, Feng and Yuan, Ye-Fei and Zensus, J. Anton and Zhang, Shuo and Zhao, Guang-Yao and Zhao, Shan-Shan and Agurto, Claudio and Araneda, Juan Pablo and Arriagada, Oriel and Bertarini, Alessandra and Berthold, Ryan and Blanchard, Jay and Brown, Ken and C{\'a}rdenas, Mauricio and Cantzler, Michael and Caro, Patricio and Chuter, Tim C. and Ciechanowicz, Miroslaw and Coulson, Iain M. and Crowley, Joseph and Degenaar, Nathalie and Dornbusch, Sven and Dur{\'a}n, Carlos A. and Forster, Karl and Geertsema, Gertie and Gonz{\'a}lez, Edouard and Graham, Dave and Gueth, Fr{\'e}d{\'e}ric and Han, Chih-Chiang and Herrera, Cristian and {Herrero-Illana}, Ruben and Heyminck, Stefan and Hoge, James and Huang, Yau-De and Jiang, Homin and John, David and Klein, Thomas and Kubo, Derek and Kuroda, John and Kwon, Caleb and Laing, Robert and Liu, Ching-Tang and Liu, Kuan-Yu and {Mac-Auliffe}, Felipe and {Martin-Cocher}, Pierre and Matulonis, Callie and Messias, Hugo and {Meyer-Zhao}, Zheng and {Montenegro-Montes}, Francisco and Montgomerie, William and Muders, Dirk and Nishioka, Hiroaki and Norton, Timothy J. and Olivares, Rodrigo and {P{\'e}rez-Beaupuits}, Juan Pablo and Parra, Rodrigo and Poirier, Michael and Pradel, Nicolas and Raffin, Philippe A. and Ram{\'i}rez, Jorge and Reynolds, Mark and {Saez-Madain}, Alejandro F. and Santana, Jorge and Silva, Kevin M. and Sousa, Don and Stahm, William and Torstensson, Karl and Venegas, Paulina and Walther, Craig and Wieching, Gundolf and Wijnands, Rudy and Wouterloot, Jan G. A.},
  year = {2022},
  month = may,
  journal = {The Astrophysical Journal},
  volume = {930},
  pages = {L13},
  publisher = {IOP},
  issn = {0004-637X},
  doi = {10.3847/2041-8213/ac6675},
  urldate = {2024-05-03}
}

@article{eht_sgra_3,
  title = {First {{Sagittarius A}}* {{Event Horizon Telescope Results}}. {{III}}. {{Imaging}} of the {{Galactic Center Supermassive Black Hole}}},
  author = {{Event Horizon Telescope Collaboration} and Akiyama, Kazunori and Alberdi, Antxon and Alef, Walter and Algaba, Juan Carlos and Anantua, Richard and Asada, Keiichi and Azulay, Rebecca and Bach, Uwe and Baczko, Anne-Kathrin and Ball, David and Balokovi{\'c}, Mislav and Barrett, John and Baub{\"o}ck, Michi and Benson, Bradford A. and Bintley, Dan and Blackburn, Lindy and Blundell, Raymond and Bouman, Katherine L. and Bower, Geoffrey C. and Boyce, Hope and Bremer, Michael and Brinkerink, Christiaan D. and Brissenden, Roger and Britzen, Silke and Broderick, Avery E. and Broguiere, Dominique and Bronzwaer, Thomas and Bustamante, Sandra and Byun, Do-Young and Carlstrom, John E. and Ceccobello, Chiara and Chael, Andrew and Chan, Chi-kwan and Chatterjee, Koushik and Chatterjee, Shami and Chen, Ming-Tang and Chen, Yongjun and Cheng, Xiaopeng and Cho, Ilje and Christian, Pierre and Conroy, Nicholas S. and Conway, John E. and Cordes, James M. and Crawford, Thomas M. and Crew, Geoffrey B. and {Cruz-Osorio}, Alejandro and Cui, Yuzhu and Davelaar, Jordy and De Laurentis, Mariafelicia and Deane, Roger and Dempsey, Jessica and Desvignes, Gregory and Dexter, Jason and Dhruv, Vedant and Doeleman, Sheperd S. and Dougal, Sean and Dzib, Sergio A. and Eatough, Ralph P. and Emami, Razieh and Falcke, Heino and Farah, Joseph and Fish, Vincent L. and Fomalont, Ed and Ford, H. Alyson and {Fraga-Encinas}, Raquel and Freeman, William T. and Friberg, Per and Fromm, Christian M. and Fuentes, Antonio and Galison, Peter and Gammie, Charles F. and Garc{\'i}a, Roberto and Gentaz, Olivier and Georgiev, Boris and Goddi, Ciriaco and Gold, Roman and {G{\'o}mez-Ruiz}, Arturo I. and G{\'o}mez, Jos{\'e} L. and Gu, Minfeng and Gurwell, Mark and Hada, Kazuhiro and Haggard, Daryl and Haworth, Kari and Hecht, Michael H. and Hesper, Ronald and Heumann, Dirk and Ho, Luis C. and Ho, Paul and Honma, Mareki and Huang, Chih-Wei L. and Huang, Lei and Hughes, David H. and Ikeda, Shiro and Impellizzeri, C. M. Violette and Inoue, Makoto and Issaoun, Sara and James, David J. and Jannuzi, Buell T. and Janssen, Michael and Jeter, Britton and Jiang, Wu and {Jim{\'e}nez-Rosales}, Alejandra and Johnson, Michael D. and Jorstad, Svetlana and Joshi, Abhishek V. and Jung, Taehyun and Karami, Mansour and Karuppusamy, Ramesh and Kawashima, Tomohisa and Keating, Garrett K. and Kettenis, Mark and Kim, Dong-Jin and Kim, Jae-Young and Kim, Jongsoo and Kim, Junhan and Kino, Motoki and Koay, Jun Yi and Kocherlakota, Prashant and Kofuji, Yutaro and Koch, Patrick M. and Koyama, Shoko and Kramer, Carsten and Kramer, Michael and Krichbaum, Thomas P. and Kuo, Cheng-Yu and La Bella, Noemi and Lauer, Tod R. and Lee, Daeyoung and Lee, Sang-Sung and Leung, Po Kin and Levis, Aviad and Li, Zhiyuan and Lico, Rocco and Lindahl, Greg and Lindqvist, Michael and Lisakov, Mikhail and Liu, Jun and Liu, Kuo and Liuzzo, Elisabetta and Lo, Wen-Ping and Lobanov, Andrei P. and Loinard, Laurent and Lonsdale, Colin J. and Lu, Ru-Sen and Mao, Jirong and Marchili, Nicola and Markoff, Sera and Marrone, Daniel P. and Marscher, Alan P. and {Mart{\'i}-Vidal}, Iv{\'a}n and Matsushita, Satoki and Matthews, Lynn D. and Medeiros, Lia and Menten, Karl M. and Michalik, Daniel and Mizuno, Izumi and Mizuno, Yosuke and Moran, James M. and Moriyama, Kotaro and Moscibrodzka, Monika and M{\"u}ller, Cornelia and Mus, Alejandro and Musoke, Gibwa and Myserlis, Ioannis and Nadolski, Andrew and Nagai, Hiroshi and Nagar, Neil M. and Nakamura, Masanori and Narayan, Ramesh and Narayanan, Gopal and Natarajan, Iniyan and Nathanail, Antonios and Fuentes, Santiago Navarro and Neilsen, Joey and Neri, Roberto and Ni, Chunchong and Noutsos, Aristeidis and Nowak, Michael A. and Oh, Junghwan and Okino, Hiroki and Olivares, H{\'e}ctor and {Ortiz-Le{\'o}n}, Gisela N. and Oyama, Tomoaki and {\"O}zel, Feryal and Palumbo, Daniel C. M. and Paraschos, Georgios Filippos and Park, Jongho and Parsons, Harriet and Patel, Nimesh and Pen, Ue-Li and Pesce, Dominic W. and Pi{\'e}tu, Vincent and Plambeck, Richard and PopStefanija, Aleksandar and Porth, Oliver and P{\"o}tzl, Felix M. and Prather, Cora and {Preciado-L{\'o}pez}, Jorge A. and Psaltis, Dimitrios and Pu, Hung-Yi and Ramakrishnan, Venkatessh and Rao, Ramprasad and Rawlings, Mark G. and Raymond, Alexander W. and Rezzolla, Luciano and Ricarte, Angelo and Ripperda, Bart and Roelofs, Freek and Rogers, Alan and Ros, Eduardo and {Romero-Ca{\~n}izales}, Cristina and Roshanineshat, Arash and Rottmann, Helge and Roy, Alan L. and Ruiz, Ignacio and Ruszczyk, Chet and Rygl, Kazi L. J. and S{\'a}nchez, Salvador and {S{\'a}nchez-Arg{\"u}elles}, David and {S{\'a}nchez-Portal}, Miguel and Sasada, Mahito and Satapathy, Kaushik and Savolainen, Tuomas and Schloerb, F. Peter and Schonfeld, Jonathan and Schuster, Karl-Friedrich and Shao, Lijing and Shen, Zhiqiang and Small, Des and Sohn, Bong Won and SooHoo, Jason and Souccar, Kamal and Sun, He and Tazaki, Fumie and Tetarenko, Alexandra J. and Tiede, Paul and Tilanus, Remo P. J. and Titus, Michael and Torne, Pablo and Traianou, Efthalia and Trent, Tyler and Trippe, Sascha and Turk, Matthew and {van Bemmel}, Ilse and {van Langevelde}, Huib Jan and {van Rossum}, Daniel R. and Vos, Jesse and Wagner, Jan and {Ward-Thompson}, Derek and Wardle, John and Weintroub, Jonathan and Wex, Norbert and Wharton, Robert and Wielgus, Maciek and Wiik, Kaj and Witzel, Gunther and Wondrak, Michael F. and Wong, George N. and Wu, Qingwen and Yamaguchi, Paul and Yoon, Doosoo and Young, Andr{\'e} and Young, Ken and Younsi, Ziri and Yuan, Feng and Yuan, Ye-Fei and Zensus, J. Anton and Zhang, Shuo and Zhao, Guang-Yao and Zhao, Shan-Shan},
  year = {2022},
  month = may,
  journal = {The Astrophysical Journal},
  volume = {930},
  pages = {L14},
  publisher = {IOP},
  issn = {0004-637X},
  doi = {10.3847/2041-8213/ac6429},
  urldate = {2024-05-03}
}

@article{eht_sgra_4,
  title = {First {{Sagittarius A}}* {{Event Horizon Telescope Results}}. {{IV}}. {{Variability}}, {{Morphology}}, and {{Black Hole Mass}}},
  author = {{Event Horizon Telescope Collaboration} and Akiyama, Kazunori and Alberdi, Antxon and Alef, Walter and Algaba, Juan Carlos and Anantua, Richard and Asada, Keiichi and Azulay, Rebecca and Bach, Uwe and Baczko, Anne-Kathrin and Ball, David and Balokovi{\'c}, Mislav and Barrett, John and Baub{\"o}ck, Michi and Benson, Bradford A. and Bintley, Dan and Blackburn, Lindy and Blundell, Raymond and Bouman, Katherine L. and Bower, Geoffrey C. and Boyce, Hope and Bremer, Michael and Brinkerink, Christiaan D. and Brissenden, Roger and Britzen, Silke and Broderick, Avery E. and Broguiere, Dominique and Bronzwaer, Thomas and Bustamante, Sandra and Byun, Do-Young and Carlstrom, John E. and Ceccobello, Chiara and Chael, Andrew and Chan, Chi-kwan and Chatterjee, Koushik and Chatterjee, Shami and Chen, Ming-Tang and Chen, Yongjun and Cheng, Xiaopeng and Cho, Ilje and Christian, Pierre and Conroy, Nicholas S. and Conway, John E. and Cordes, James M. and Crawford, Thomas M. and Crew, Geoffrey B. and {Cruz-Osorio}, Alejandro and Cui, Yuzhu and Davelaar, Jordy and De Laurentis, Mariafelicia and Deane, Roger and Dempsey, Jessica and Desvignes, Gregory and Dexter, Jason and Dhruv, Vedant and Doeleman, Sheperd S. and Dougal, Sean and Dzib, Sergio A. and Eatough, Ralph P. and Emami, Razieh and Falcke, Heino and Farah, Joseph and Fish, Vincent L. and Fomalont, Ed and Ford, H. Alyson and {Fraga-Encinas}, Raquel and Freeman, William T. and Friberg, Per and Fromm, Christian M. and Fuentes, Antonio and Galison, Peter and Gammie, Charles F. and Garc{\'i}a, Roberto and Gentaz, Olivier and Georgiev, Boris and Goddi, Ciriaco and Gold, Roman and {G{\'o}mez-Ruiz}, Arturo I. and G{\'o}mez, Jos{\'e} L. and Gu, Minfeng and Gurwell, Mark and Hada, Kazuhiro and Haggard, Daryl and Haworth, Kari and Hecht, Michael H. and Hesper, Ronald and Heumann, Dirk and Ho, Luis C. and Ho, Paul and Honma, Mareki and Huang, Chih-Wei L. and Huang, Lei and Hughes, David H. and Ikeda, Shiro and Impellizzeri, C. M. Violette and Inoue, Makoto and Issaoun, Sara and James, David J. and Jannuzi, Buell T. and Janssen, Michael and Jeter, Britton and Jiang, Wu and {Jim{\'e}nez-Rosales}, Alejandra and Johnson, Michael D. and Jorstad, Svetlana and Joshi, Abhishek V. and Jung, Taehyun and Karami, Mansour and Karuppusamy, Ramesh and Kawashima, Tomohisa and Keating, Garrett K. and Kettenis, Mark and Kim, Dong-Jin and Kim, Jae-Young and Kim, Jongsoo and Kim, Junhan and Kino, Motoki and Koay, Jun Yi and Kocherlakota, Prashant and Kofuji, Yutaro and Koch, Patrick M. and Koyama, Shoko and Kramer, Carsten and Kramer, Michael and Krichbaum, Thomas P. and Kuo, Cheng-Yu and La Bella, Noemi and Lauer, Tod R. and Lee, Daeyoung and Lee, Sang-Sung and Leung, Po Kin and Levis, Aviad and Li, Zhiyuan and Lico, Rocco and Lindahl, Greg and Lindqvist, Michael and Lisakov, Mikhail and Liu, Jun and Liu, Kuo and Liuzzo, Elisabetta and Lo, Wen-Ping and Lobanov, Andrei P. and Loinard, Laurent and Lonsdale, Colin J. and Lu, Ru-Sen and Mao, Jirong and Marchili, Nicola and Markoff, Sera and Marrone, Daniel P. and Marscher, Alan P. and {Mart{\'i}-Vidal}, Iv{\'a}n and Matsushita, Satoki and Matthews, Lynn D. and Medeiros, Lia and Menten, Karl M. and Michalik, Daniel and Mizuno, Izumi and Mizuno, Yosuke and Moran, James M. and Moriyama, Kotaro and Moscibrodzka, Monika and M{\"u}ller, Cornelia and Mus, Alejandro and Musoke, Gibwa and Myserlis, Ioannis and Nadolski, Andrew and Nagai, Hiroshi and Nagar, Neil M. and Nakamura, Masanori and Narayan, Ramesh and Narayanan, Gopal and Natarajan, Iniyan and Nathanail, Antonios and Fuentes, Santiago Navarro and Neilsen, Joey and Neri, Roberto and Ni, Chunchong and Noutsos, Aristeidis and Nowak, Michael A. and Oh, Junghwan and Okino, Hiroki and Olivares, H{\'e}ctor and {Ortiz-Le{\'o}n}, Gisela N. and Oyama, Tomoaki and Palumbo, Daniel C. M. and Paraschos, Georgios Filippos and Park, Jongho and Parsons, Harriet and Patel, Nimesh and Pen, Ue-Li and Pesce, Dominic W. and Pi{\'e}tu, Vincent and Plambeck, Richard and PopStefanija, Aleksandar and Porth, Oliver and P{\"o}tzl, Felix M. and Prather, Cora and {Preciado-L{\'o}pez}, Jorge A. and Pu, Hung-Yi and Ramakrishnan, Venkatessh and Rao, Ramprasad and Rawlings, Mark G. and Raymond, Alexander W. and Rezzolla, Luciano and Ricarte, Angelo and Ripperda, Bart and Roelofs, Freek and Rogers, Alan and Ros, Eduardo and {Romero-Ca{\~n}izales}, Cristina and Roshanineshat, Arash and Rottmann, Helge and Roy, Alan L. and Ruiz, Ignacio and Ruszczyk, Chet and Rygl, Kazi L. J. and S{\'a}nchez, Salvador and {S{\'a}nchez-Arg{\"u}elles}, David and {S{\'a}nchez-Portal}, Miguel and Sasada, Mahito and Satapathy, Kaushik and Savolainen, Tuomas and Schloerb, F. Peter and Schonfeld, Jonathan and Schuster, Karl-Friedrich and Shao, Lijing and Shen, Zhiqiang and Small, Des and Sohn, Bong Won and SooHoo, Jason and Souccar, Kamal and Sun, He and Tazaki, Fumie and Tetarenko, Alexandra J. and Tiede, Paul and Tilanus, Remo P. J. and Titus, Michael and Torne, Pablo and Traianou, Efthalia and Trent, Tyler and Trippe, Sascha and Turk, Matthew and {van Bemmel}, Ilse and {van Langevelde}, Huib Jan and {van Rossum}, Daniel R. and Vos, Jesse and Wagner, Jan and {Ward-Thompson}, Derek and Wardle, John and Weintroub, Jonathan and Wex, Norbert and Wharton, Robert and Wielgus, Maciek and Wiik, Kaj and Witzel, Gunther and Wondrak, Michael F. and Wong, George N. and Wu, Qingwen and Yamaguchi, Paul and Yoon, Doosoo and Young, Andr{\'e} and Young, Ken and Younsi, Ziri and Yuan, Feng and Yuan, Ye-Fei and Zensus, J. Anton and Zhang, Shuo and Zhao, Guang-Yao and Zhao, Shan-Shan and Chang, Dominic O.},
  year = {2022},
  month = may,
  journal = {The Astrophysical Journal},
  volume = {930},
  pages = {L15},
  publisher = {IOP},
  issn = {0004-637X},
  doi = {10.3847/2041-8213/ac6736},
  urldate = {2024-05-03}
}

@article{fishbone_1976_torus,
  title = {Relativistic Fluid Disks in Orbit around {{Kerr}} Black Holes.},
  author = {Fishbone, L. G. and Moncrief, V.},
  year = {1976},
  month = aug,
  journal = {The Astrophysical Journal},
  volume = {207},
  pages = {962--976},
  publisher = {IOP},
  issn = {0004-637X},
  doi = {10.1086/154565},
  urldate = {2024-05-05}
}

@article{gammie_2003_harm,
  title = {{{HARM}}: {{A Numerical Scheme}} for {{General Relativistic Magnetohydrodynamics}}},
  shorttitle = {{{HARM}}},
  author = {Gammie, Charles F. and McKinney, Jonathan C. and T{\'o}th, G{\'a}bor},
  year = {2003},
  month = may,
  journal = {The Astrophysical Journal},
  volume = {589},
  pages = {444--457},
  publisher = {IOP},
  issn = {0004-637X},
  doi = {10.1086/374594},
  urldate = {2024-05-15}
}

@article{igumenshchev_2003_mad,
  title = {Three-Dimensional {{Magnetohydrodynamic Simulations}} of {{Radiatively Inefficient Accretion Flows}}},
  author = {Igumenshchev, Igor V. and Narayan, Ramesh and Abramowicz, Marek A.},
  year = {2003},
  month = aug,
  journal = {The Astrophysical Journal},
  volume = {592},
  pages = {1042--1059},
  publisher = {IOP},
  issn = {0004-637X},
  doi = {10.1086/375769},
  urldate = {2024-05-02}
}

@article{johannsen_2010_nohairimages,
  title = {Testing the {{No-hair Theorem}} with {{Observations}} in the {{Electromagnetic Spectrum}}. {{II}}. {{Black Hole Images}}},
  author = {Johannsen, Tim and Psaltis, Dimitrios},
  year = {2010},
  month = jul,
  journal = {The Astrophysical Journal},
  volume = {718},
  pages = {446--454},
  publisher = {IOP},
  issn = {0004-637X},
  doi = {10.1088/0004-637X/718/1/446},
  urldate = {2024-05-15}
}

@article{kozlowski_1978_donut,
  title = {The Analytic Theory of Fluid Disks Orbiting the {{Kerr}} Black Hole.},
  author = {Kozlowski, M. and Jaroszynski, M. and Abramowicz, M. A.},
  year = {1978},
  month = feb,
  journal = {Astronomy and Astrophysics},
  volume = {63},
  pages = {209--220},
  issn = {0004-6361},
  urldate = {2024-05-15}
}

@article{liska_2018_tilted,
  title = {Formation of Precessing Jets by Tilted Black Hole Discs in {{3D}} General Relativistic {{MHD}} Simulations},
  author = {Liska, M. and Hesp, C. and Tchekhovskoy, A. and Ingram, A. and {van der Klis}, M. and Markoff, S.},
  year = {2018},
  month = feb,
  journal = {Monthly Notices of the Royal Astronomical Society},
  volume = {474},
  pages = {L81-L85},
  publisher = {OUP},
  issn = {0035-8711},
  doi = {10.1093/mnrasl/slx174},
  urldate = {2024-05-15}
}

@article{liska_2023_twotempwarped,
  title = {Radiation {{Transport Two-temperature GRMHD Simulations}} of {{Warped Accretion Disks}}},
  author = {Liska, M. T. P. and Kaaz, N. and Musoke, G. and Tchekhovskoy, A. and Porth, O.},
  year = {2023},
  month = feb,
  journal = {The Astrophysical Journal},
  volume = {944},
  pages = {L48},
  publisher = {IOP},
  issn = {0004-637X},
  doi = {10.3847/2041-8213/acb6f4},
  urldate = {2024-05-15}
}

@article{medeiros_2022_asymmetry,
  title = {Brightness {{Asymmetry}} of {{Black Hole Images}} as a {{Probe}} of {{Observer Inclination}}},
  author = {Medeiros, Lia and Chan, Chi-Kwan and Narayan, Ramesh and {\"O}zel, Feryal and Psaltis, Dimitrios},
  year = {2022},
  month = jan,
  journal = {The Astrophysical Journal},
  volume = {924},
  pages = {46},
  publisher = {IOP},
  issn = {0004-637X},
  doi = {10.3847/1538-4357/ac33a7},
  urldate = {2024-05-15}
}

@article{medeiros_2023_PRIMO_M87,
  title = {The {{Image}} of the {{M87 Black Hole Reconstructed}} with {{PRIMO}}},
  author = {Medeiros, Lia and Psaltis, Dimitrios and Lauer, Tod R. and {\"O}zel, Feryal},
  year = {2023},
  month = apr,
  journal = {The Astrophysical Journal},
  volume = {947},
  pages = {L7},
  publisher = {IOP},
  issn = {0004-637X},
  doi = {10.3847/2041-8213/acc32d},
  urldate = {2025-05-27}
}

@ARTICLE{falcke_2013_sgra,
       author = {{Falcke}, H. and {Markoff}, S.~B.},
        title = "{Toward the event horizon{\textemdash}the supermassive black hole in the Galactic Center}",
      journal = {Classical and Quantum Gravity},
         year = 2013,
        month = dec,
       volume = {30},
       number = {24},
          eid = {244003},
        pages = {244003},
          doi = {10.1088/0264-9381/30/24/244003},
archivePrefix = {arXiv},
       eprint = {1311.1841},
 primaryClass = {astro-ph.HE},
       adsurl = {https://ui.adsabs.harvard.edu/abs/2013CQGra..30x4003F}
}

@ARTICLE{narayan_2008_adafbh,
       author = {{Narayan}, Ramesh and {McClintock}, Jeffrey E.},
        title = "{Advection-dominated accretion and the black hole event horizon}",
      journal = {\nar},
         year = 2008,
        month = may,
       volume = {51},
       number = {10-12},
        pages = {733-751},
          doi = {10.1016/j.newar.2008.03.002},
archivePrefix = {arXiv},
       eprint = {0803.0322},
 primaryClass = {astro-ph},
       adsurl = {https://ui.adsabs.harvard.edu/abs/2008NewAR..51..733N}
}

\end{document}